\documentclass[a4paper,11pt]{article}
\usepackage{jheppub} %
\usepackage{lineno}
\usepackage{amsmath}
\usepackage{breqn}
\usepackage{subfigure}
\usepackage{diagbox}
\usepackage{multirow}
\usepackage{xcolor}
\def\md{\mathrm{d}}

\allowdisplaybreaks

\title{\boldmath Dihadron Angular Correlations in the \texorpdfstring{$e^+e^-$}{e+e-} Collision}

\author[a]{Wan-Li Ju,}
\emailAdd{wju@ualberta.ca}
\author[b]{Zhe-Yan Shu,}
\emailAdd{2200017808@stu.pku.edu.cn}
\author[c,d,e]{Tong-Zhi Yang,}
\emailAdd{toyang@physik.uzh.ch}
\author[f,g]{Zhen-Hua Zhang}
\emailAdd{zhhzhang@pku.edu.cn}
\author[f,g]{and Hua Xing Zhu}
\emailAdd{zhuhx@pku.edu.cn}

\affiliation[a]{Department of Physics, University of Alberta, Edmonton AB T6G 2J1, Canada}
\affiliation[b]{Yuanpei College, Peking University, Beijing, 100871, China}
\affiliation[c]{State Key Laboratory of Nuclear Physics and Technology, Institute of Quantum Matter, South China Normal University, Guangzhou 510006, China}
\affiliation[d]{%
Guangdong Basic Research Center of Excellence for Structure and Fundamental Interactions of Matter, Guangdong Provincial Key Laboratory of Nuclear Science, Guangzhou 510006, China
}
\affiliation[e]{Physik-Institut, Universit\"at Z\"urich, Winterthurerstrasse 190, CH-8057 Z\"urich, Switzerland}
\affiliation[f]{School of Physics, Peking University, Beijing, 100871, China}

\affiliation[g]{Center for High Energy Physics, Peking University, Beijing 100871, China}

\abstract{
The precision of fixed-order calculations on the dihadron production in electron–positron annihilation is paramount  for probing QCD factorization and constraining non-perturbative inputs.  This paper investigates the QCD corrections to the angular separation distribution $\theta_{12}$ between two observed hadrons, $H_1$ and $H_2$, in the process $e^+e^- \to H_1 H_2 + X$ up to $\mathcal{O}(\alpha_s^2)$, with particular emphasis on the intermediate region $\theta_{12} \in (0,\pi)$.  
The partonic processes at this accuracy consist of two sorts of contributions, the real-virtual and double-real corrections. Of them, the evaluation of four-body phase space integrals in the latter case is at the core of this study.  
To address them, we first employ  the integration-by-parts (IBP) identities to reduce the number of independent integrals and   then apply the differential equations (DE) method  to recursively solve  the resulting master integrals. In kinematic regions where the invariant mass of the unresolved partons vanishes, IBP coefficients can develop divergences. To this end, we resum higher-order terms in the dimensional regulator for each master integral based on the asymptotic behavior of the   canonical DEs.
After combining the   real and virtual corrections with the counter terms from fragmentation function renormalization, we  demonstrate that the pole terms in the final analytic expressions exactly cancel out in all partonic channels, thereby providing a non-trivial validation of collinear factorization at the next-to-leading order (NLO).  Eventually, when presenting our analytic expressions of the finite partonic coefficients, we transform the transcendental functions resulting from the DE solutions  into classical (poly)logarithmic functions, in order to facilitate the implementation in event generators. 
}

\begin{document}

\maketitle

\flushbottom

\section{Introduction}

Hadron production in electron-positron annihilation plays an important  role in the modern study of perturbative QCD. Among the  processes, semi-inclusive single hadron annihilation (SIA) $e^+e^-\to H+X$ holds a particularly important place. Being free from hadronic initial states, it provides an ideal setting to extract non-perturbative fragmentation functions (FFs)~\cite{Collins:1989gx,Collins:1981ta,Amati:1978by,Ellis:1978ty,Libby:1978qf}.
Owning to recent development of fixed-order calculations~\cite{Almasy:2011eq,Rijken:1996vr,Rijken:1996npa,Rijken:1996ns,Mitov:2006wy,Xu:2024rbt,He:2025hin,Goyal:2023zdi,Bonino:2024qbh,Liu:2023fsq,Zhou:2024cyk}, the extraction of FFs has been carried out at the next-to-next-to-leading order (N$^2$LO) in \cite{Bertone:2017tyb,Soleymaninia:2018uiv}, focusing on the experimental measurements on SIA, and also in \cite{Borsa:2022vvp,AbdulKhalek:2022laj,Gao:2025hlm,Gao:2025bko} including both the datasets of SIA and semi-inclusive deep-inelastic scattering (SIDIS). 

Moreover, the dihadron production $e^+e^-\to H_1H_2+X$ has drawn both theoretical and experimental attention as well.  Especially, in the dihadron back-to-back frame where the virtual photon possesses low transverse momentum with respect to the identified hadrons, large logarithmic corrections can be developed in the fixed-order calculation as a result of the soft and collinear singularities of QCD interaction, which has since prompted a variety of  theoretical 
endeavors~\cite{Collins:1985kw,Collins:1981uk,Collins:1981va,Collins:1984kg,Catani:2000vq,Bozzi:2005wk,Bozzi:2007pn,Collins:2011zzd,Ebert:2016gcn,Monni:2016ktx,Bizon:2017rah,Bizon:2019zgf,Bizon:2018foh,Becher:2010tm,GarciaEchevarria:2011rb,Becher:2011dz,Chiu:2011qc,Chiu:2012ir,Li:2016axz,Li:2016ctv,Moult:2018jzp,Moffat:2019pci,Hautmann:2022xuc,Gao:2023ivm,Boussarie:2023izj} to probe  the transverse momentum dependent factorization and in turn the logarithmic exponentiations.
Analogous phenomena can also be found in the transversally differential spectra of SIDIS \cite{Ji:2004wu,Ji:2004xq,Bacchetta:2006tn,Bacchetta:2017gcc,Kang:2015msa,Bastami:2018xqd,Boer:2011xd,Collins:2004nx,Xue:2020xba,Sun:2014dqm,Li:2020bub,Caucal:2023nci,Scimemi:2019cmh,Bury:2021sue,Li:2021txc,Bhattacharya:2025bqa} and the Drell-Yan (DY) process \cite{Becher:2010tm,Bozzi:2010xn,Becher:2011xn,Banfi:2011dx,Banfi:2011dm,Banfi:2012du,Catani:2015vma,Scimemi:2017etj,Bizon:2018foh,Bacchetta:2019sam,Bizon:2019zgf,Becher:2020ugp,Ebert:2020dfc,Re:2021con,Camarda:2021ict,Ju:2021lah,Camarda:2023dqn,Neumann:2022lft,Moos:2023yfa,Bubanja:2023nrd,Billis:2024dqq,Hautmann:2025fkw}. 
The recent experimental detections on the back-to-back region of the dihadron production include \cite{Belle:2005dmx,Belle:2008fdv,Belle:2019nve,BaBar:2013jdt,BESIII:2015fyw}, with particular emphasis on the angular correlation of the identified hadrons, thanks to its sensitivity to the asymmetric component of transverse momentum dependent fragmentation function (TMDFF)~\cite{Boer:2008fr,Metz:2016swz,Collins:1993kq,Bianconi:1999cd}, e.g., the Collins function \cite{Collins:1992kk}.  The analog of the latter at hadron collider is the azimuthal asymmetric modulation of the hadroproduction of the top-antitop quark pair \cite{Ju:2022wia,Catani:2017tuc}.
On the other hand, the kinematic region with small opening angle made by the identified hadrons has also garnered the interest as   a probe into the substructure within the collinear splittings \cite{Konishi:1979cb,Sukhatme:1980vs,Collins:1993kq,deFlorian:2003cg,Chen:2022pdu,Chen:2022muj}. At this stage, the non-perturbative dihadron fragmentation function (DiFF) is introduced to account for the possibility that the single parent parton fragments into two hadrons. As for the transverse momentum integrated DiFF, the field operator definition and its leading order (LO) evolution equation are investigated in~\cite{Konishi:1979cb,deFlorian:2003cg,Sukhatme:1980vs,Majumder:2004wh,Majumder:2004br}, with its transverse momentum dependent extension proposed in \cite{Ceccopieri:2007ip,Bacchetta:2008wb,Bacchetta:2011ip,Courtoy:2012ry,Bacchetta:2012ty,Radici:2015mwa}. Very recently, the latter has been extracted by JAM Collaboration \cite{Cocuzza:2023vqs} according to the experimental measurement \cite{Belle:2017rwm} on the invariant-mass distributions of di-hadron production. 

In these investigations, the fixed order precision of differential spectra in the dihadron production is of crucial importance to validate the QCD factorisation formalism and in turn improve the matching accuracy of the resummation onto the non-asymptotic regime. Hitherto, only the next-to-leading order (NLO) correction to the dihadron production, i.e., $\mathcal{O}(\alpha_s)$,  is available with the integrated opening angle between the identified hadrons~\cite{deFlorian:2003cg}.  The QCD calculation on the angular dependent observables is still absent beyond the tree-level accuracy, in spite of the fact that in an analogous context, the analytic expressions of $\mathcal{O}(\alpha^2_s)$ corrections to the large transverse momentum distributions had been worked out in SIDIS \cite{Wang:2019bvb,Daleo:2004pn} and DY processes \cite{Gonsalves:1989ar,Ellis:1981nt,Ellis:1981hk} some time ago.

In this study, we are focused on the differential spectra of the opening angle $\theta_{12}$ between the two identified hadrons in the intermediate angular region  $\theta_{12}\in(0,\pi)$. In this region, the first non-trivial partonic contribution arises at  $\mathcal{O}(\alpha_s)$, consisting solely of tree-level squared amplitudes for photon decay into three partons, which will be dubbed ``LO" hereafter.  
The NLO QCD correction, at $\mathcal{O}(\alpha^2_s)$, comprises both the one-loop contributions to photon decay into three partons and the tree-level contributions to photon decay into four partons, which are referred to real-virtual and double-real corrections in this work, respectively. The former case was calculated in~\cite{Ellis:1980wv} decades ago, whereas we choose to revisit this process with  \texttt{Mathematica} packages \texttt{FeynArts}~\cite{Kublbeck:1990xc},~\texttt{FeynCalc}~\cite{Shtabovenko:2023idz,Shtabovenko:2020gxv,Shtabovenko:2016sxi,Mertig:1990an}, and~\texttt{FeynHelpers}~\cite{Shtabovenko:2016whf}. 
To evaluate the double real contribution, we apply integration-by-parts (IBP) identities~\cite{Tkachov:1981wb,Chetyrkin:1981qh,Laporta:2000dsw} to reduce the phase space integrals to a set of master integrals. These are then computed using the differential equations (DE) method~\cite{Kotikov:1990kg,Remiddi:1997ny,Gehrmann:1999as,Argeri:2007up,Henn:2013pwa}, which allows for their recursive solution. During the calculation, spurious divergences may arise from the expansion in the dimensional regulator, particularly in regions where the invariant mass of the undetected partons approaches zero. To address this, we resum the higher-order terms in the dimensional regulator for each master integral, guided by the asymptotic behavior of their differential equations in the canonical basis. Ultimately, we will demonstrate that, after summing all contributions, the pole terms in our analytic results for the partonic processes exactly cancel those in the fragmentation functions. This cancellation provides a non-trivial validation of collinear factorization up to NLO.

This paper is organized as follows. Section~\ref{sec:def:kin} introduces the collinear factorization framework for the angular differential spectra in dihadron production, along with the LO partonic coefficients. The details of the NLO calculation are presented in Section~\ref{sec:def:NLO}, including the double-real contribution in Section~\ref{sec:def:NLO:RR}, the real-virtual sector in Section~\ref{sec:def:NLO:RV}, and the renormalization of the fragmentation functions in Section~\ref{sec:def:NLO:renFFs}. Section~\ref{sec:def:finite:coeffs} is devoted to the presentation of the finite renormalized coefficient functions. Finally, we summarize our findings in Section~\ref{sec:def:recap}.

\section{Definition of the observable}\label{sec:def:kin}

\subsection{Factorization formula}

The differential observables in the dihadron production of this work's concern include the measurement of the momentum fraction  $\tau_{1(2)}$ for the detected particle $H_{1(2)}$ and their opening angle $\theta_{12}$ in the rest reference frame of the virtual photon, 
\begin{align}\label{eq:def:tau12z}
 \tau_1=\frac{2E_1}{Q}\,,\quad
 \tau_2=\frac{2E_2}{Q}\,,\quad
 \cos \theta_{12}=1-2z\,.
\end{align}
Here, $Q=\sqrt{q_{\gamma}^2}$ stands for the colliding energy of the initial leptons, with $q_{\gamma}$ the 
4-momentum of the intermediate virtual photon. $E_{1,2}$ refer to the energies of the identified hadrons $H_{1,2}$, respectively. To facilitate the later discussion, we also bring in the dimensionless variable $z$ from the cosine of $\theta_{12}$, which is related to the transverse momentum of the virtual photon in the hadron frame \cite{Moffat:2019pci}.

Throughout this work, we will focus on the perturbative region 
\begin{align}\label{eq:def:m12}
m_{12}^2\sim z Q^2\tau_1\tau_2\gg\Lambda_{\mathrm{QCD}}^2\,,
\end{align}
where the hadrons $H_1$ and $H_2$ are kinematically well separated.  In the above, we have defined $m_{12}$ as the invariant mass of the two-hadron system. In this region, the collinear splittings associated with $H_{1(2)}$  come about independently, while the soft interactions are expected to drop out after summing up the inclusive radiations \cite{Collins:1981ta}, akin to the Drell-Yan process \cite{Collins:1980ui,Collins:1984kg,Collins:1985ue,Collins:1981uk,Bodwin:1984hc}. 
Therefore, following the collinear factorization theorem, the leading contribution from the dihadron correlation admits the factorized structure~\cite{Collins:1989gx,Collins:1981ta,Amati:1978by,Ellis:1978ty,Libby:1978qf},
\begin{align}\label{eq:def:fac}
\frac{\md^3\sigma}{\md\tau_1\md\tau_2\md z}  = 
\frac{2\pi\alpha}{3Q^4}\sum_{i,j} 
  \int_{\tau_1}^1\frac{\mathrm{d}x_1}{x_1} D_{H_1/i}\left(\frac{\tau_1}{x_1},\mu\right)\,
  \int_{\tau_2}^1\frac{\mathrm{d}x_2}{x_2} D_{H_2/j}\left(\frac{\tau_2}{x_2},\mu\right)\,
  {C}_{ij}\left(x_1,x_2,z,\mu\right)\,,
\end{align} 
where $\alpha$ is the fine-structure constant. 
$N_C=3$ stands for the number of colors. $D_{H_{1(2)}/i}(x,\mu)$ denotes the parton fragmentation function for the hadron $H_{1(2)}$ with the momentum fraction $x$ from the parton $i$, encoding all the non-perturbative physics of the energy transfer$\sim\mathcal{O}(\Lambda_{\mathrm{QCD}})$. 

The perturbative contributions are collected by the Wilson coefficients $ {C}_{ij}$, where the subscripts $i$ and $j$ characterize the mother partons of FFs.
 ${C}_{ij}$ can be evaluated via the phase space integral of the squared amplitudes under the selection conditions $\Theta_{ab}$, 
\begin{align}  \label{eq:def:Cij}  
  {C}_{ij}\left(x_1,x_2,z,\mu\right)\,=&\,
      {\sum_m} \int\md\Phi_{[m]} \,\sum_{a\neq b\in X_m } \Theta_{ij}^{ab} \; 
 \sum_{\mathrm{pol,col}} \Big|\mathcal{M}(\gamma^*\to  i+j+\cdots)\Big|^2\,\nonumber\\
  =&\, 2\alpha N_CQ^2\left\{\sum_q e_q^2\; \mathcal{C}_{ij;q}
  \,+\,
  \sum_{q,q^\prime}e_qe_{q^\prime}\; \mathcal{C}_{ij;qq^\prime}\right\}\,,
\end{align}
where the set $X_m$ collects all partons in the final state. The $m$-particle phase-space measure is defined as 
\begin{align}  \label{eq:def:dps:fac}  
   \int\md\Phi_{[m]} \,\equiv\, \prod_{\rho}\frac{1}{n_{\rho}!} \int\prod_{n=1}^m\frac{1}{2E_n \,}\frac{\md^{d-1} \vec{k}_n}{(2\pi)^{d-1}  }\,(2\pi)^d\delta^d\left(q_{\gamma}-\sum_{n=1}^mk_n\right). 
\end{align}
Herein, $E_n$ and $\vec{k}_n$ represent the energy and spatial momentum of the $n$-th  emitted parton, respectively. The factorial $n_{\rho}!$ serves as the averaging factor in the case that there are $n_{\rho}$ indistinguishable particles of the same type $\rho$ in the final state. 
$\Theta_{ij}^{ab}$ characterizes the measurement, 
\begin{align}\label{eq:def:selection} 
  \Theta_{ij}^{ab}= \,
  \delta^a_i \,\delta\left(x_1 - \frac{2 q_{\gamma} \!\cdot\! k_a}{Q^2}\right)\,
  \delta^b_j\,\delta\left( x_2 - \frac{2 q_{\gamma} \!\cdot\! k_b}{Q^2}\right)\, \delta\left( z- \frac{1}{x_1 x_2}\frac{2 k_a \!\cdot\! k_b}{Q^2}\right)\,,
\end{align}
where the Kronecker delta functions $\delta^a_i$ and $\delta^b_j$ seek out the mother particles $i$ and $j$ from the real emissions $X_m$, respectively, for which the Dirac delta functions imposes the kinematic constraint. It is worth noting that, as a result of the asymptotic expansion in $(\Lambda_{\mathrm{QCD}}/m_{12})$, the variable $z$ in  eq.~\eqref{eq:def:selection} concerns the invariant mass of two {\it massless} partons $a$ and $b$, differing from the hadron level kinematics in eq.~\eqref{eq:def:m12}.  

  \begin{figure}[t!]
    \centering
    \includegraphics[width=0.9\linewidth]{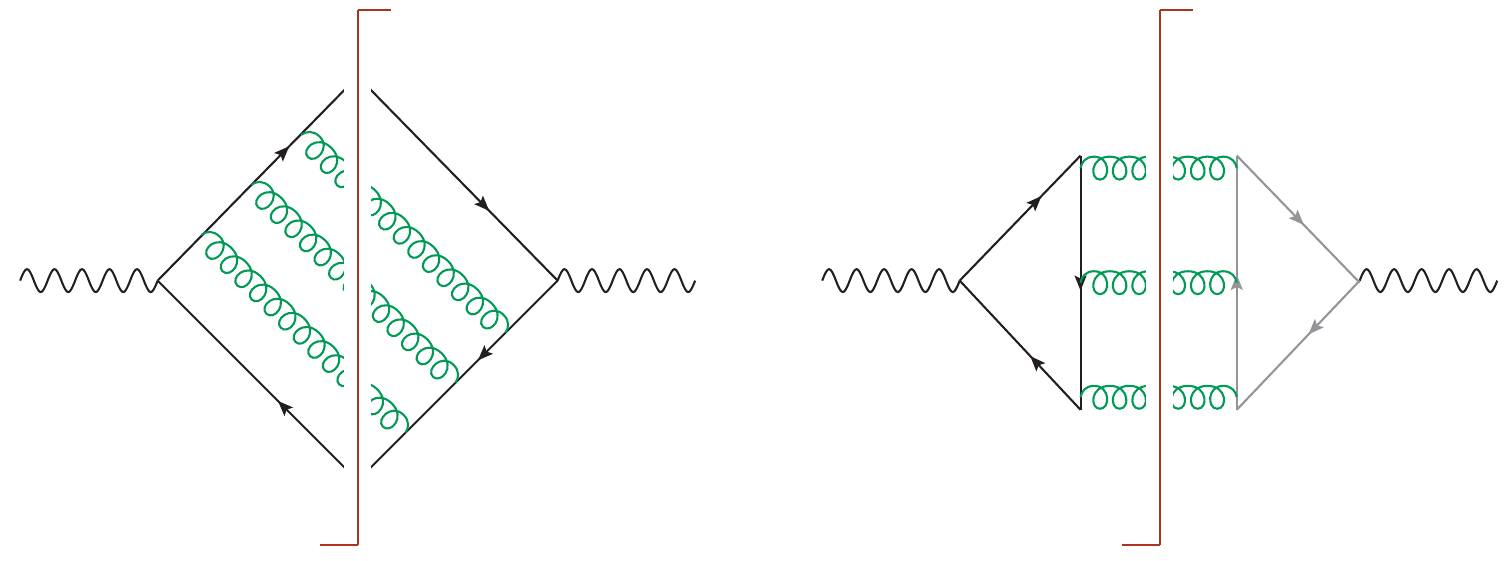}
    \caption{Characteristic Feynman diagrams contributing to $\mathcal{C}_{ij;q}$ (left panel) and $\mathcal{C}_{ij;qq^\prime}$  (right panel). The wavy line stands for the initial photon. The spring lines denote the gluons. The quarks $q$ and $q^\prime$ are illustrated in the black and gray straight lines, respectively. They start to contribute from ${\cal O}(\alpha_s^3)$ corrections to the Wilson coefficients, which are beyond the scope of this work.}
    \label{fig:feynman_eq}
\end{figure}

In the last step of eq.~\eqref{eq:def:Cij}, we further categorize the partonic contributions based on the configurations of the resulting Feynman diagrams. More specifically, $\mathcal{C}_{ij;q}$ encodes the contributions in which the prompt quarks--- directly connected to the initial photons---have the same flavor in both the amplitude and its complex conjugate, whereas $\mathcal{C}_{ij;qq^\prime}$ accounts for configurations involving prompt quarks of different flavors.
The charge for the prompt quark is denoted by $e_q$. 
Throughout this paper, our primary concern is the perturbative corrections to $\mathcal{C}_{ij;q}$ up to $\mathcal{O}(\alpha_s^2)$. The contribution of $\mathcal{C}_{ij;qq^\prime}$ begins with $\mathcal{O}(\alpha^3_s)$ (see fig.~\ref{fig:feynman_eq}) as a result of Furry's theorem and is beyond the scope of this work. 

Without loss of any generality, the perturbative expansion of $\mathcal{C}_{ij;q}$ is  defined as 
\begin{align}    
  \mathcal{C}_{ij;q}\left(x_1,x_2,z,\mu\right)=  \sum_{i} \left[\frac{\alpha_s(\mu)}{4\pi}\right]^i  \mathcal{C}^{(i)}_{ij;q}\left(x_1,x_2,z,\mu\right)\,,
\end{align}
where the scale $\mu$ is invoked during UV renormalization and collinear factorization. 

In the following, we will evaluate $\mathcal{C}_{ij;q}$  in the physical domain,  
\begin{align}\label{eq:def:phy:ps}
  0<\{x_1,x_2,z\}<1,\qquad m^2_{X}\equiv   1-x_1-x_2+x_1x_2z \ge 0\,.
\end{align}
The second inequality of $m_X$ concerning the \emph{dimensionless} invariant mass of the undetected final particles  is derived by the energy-momentum conservation in eq.~\eqref{eq:def:Cij} and also the energy positivity of the real emissions.  The inequalities in eq.~\eqref{eq:def:phy:ps} can reshape the boundary terms in eq.~\eqref{eq:def:fac} when convoluting with FFs, more specifically, 
\begin{align} 
 \tau_{1(2)}\le x_{1(2)}\le\frac{1-\tau_{2(1)}}{1-z\,\tau_{2(1)}}<1\,\qquad  \mathrm{and}\,\qquad \tau_{2(1)}\le x_{2(1)}\le\frac{1-x_{1(2)}}{1-z\, x_{1(2)}}<1\,.
\end{align}
These relations indicate that  within the physical domain in eq.~\eqref{eq:def:phy:ps},  the integral of momentum fractions $x_{1(2)}$ in eq.~\eqref{eq:def:fac} are never able to strike the end points $x_{1(2)}=1$.  Hence, in the subsequent calculations, we can process the denominators, e.g.,  $(1-x_{1(2)})$,  as regular functions, without recasting them into distributions. Moreover, it is worth noting that this conclusion is generic: it holds not only at NLO (as demonstrated in this work) but also remains valid at N$^2$LO and higher orders.

\subsection{LO results}

In the kinematic region defined by eq.~\eqref{eq:def:phy:ps}, the first non-vanishing contribution to dihadron correlation comes from the tree-level partonic process $\gamma^*\to q\bar{q}g$, which is referred to as LO precision.  

According to eq.~\eqref{eq:def:Cij}, the calculation on the coefficients $\mathcal{C}_{ij;q}$ at LO entails the phase space integrals over the tree-level squared amplitudes $|\mathcal{M}_B(\gamma^*\to  q\bar{q}g)|^2$. 
Therein, the kinematics of the inclusive emission are fully constrained by the momentum conservation condition given in eq.~\eqref{eq:def:Cij}, i.e., 
\begin{align}\label{eq:def:momX:LO}
\int
\frac{1}{2E_c \,}\frac{\md^{d-1} \vec{k}_c}{(2\pi)^{d-1}  }
(2\pi)^d\delta^d\left(q_{\gamma}-k_a-k_b-k_c\right)=\frac{(2\pi)\delta\left(m_X^2\right)}{Q^2}\,.
\end{align}
Moreover, the phase space of the identified partons can be integrated out via the measurement function in eq.~\eqref{eq:def:selection}, more explicitly, 
\begin{align}
\label{eq:def:kappa}
&\int
\frac{1}{2E_a \,}\frac{\md^{d-1} \vec{k}_a}{(2\pi)^{d-1}  }
\frac{1}{2E_b \,}\frac{\md^{d-1} \vec{k}_b}{(2\pi)^{d-1}  }\,\Theta_{ij}^{ab}\nonumber\\
=&\delta^a_i \,\delta^b_j \,\frac{\Omega_{d-1}\,\Omega_{d-2}}{2^{3d-1}\pi^{2d-2} (Q^2)^{2-d}}[(1-z)z]^{\frac{d-4}{2}} \left( x_1x_2\right)^{d-3}\notag\\
\equiv&\delta^a_i \,\delta^b_j\,\kappa(z;\varepsilon)\,\left( x_1x_2\right)^{1-2\varepsilon}\,,
\end{align}
where $\varepsilon\equiv(4-d)/2$ denotes the dimensional regulator.
The total solid angle for a $d$-dimensional hypersphere reads,
\begin{align}
\Omega_{d} = \frac{2 \pi^{d/2}}{\Gamma(d/2)}\,.
\end{align}
In deriving eq.~\eqref{eq:def:kappa}, we stick by the reference frame where $\vec{k}_a$ is aligned with $z$-axis and the component of $\vec{k}_b$ that is perpendicular to $\vec{k}_a$ points to the positive $x$ direction, from which the solid angles of $\vec{k}_a$ and azimuthal angles of $\vec{k}_b$ can be immediately integrated out.

To evaluate the squared amplitudes of the partonic process $\gamma^*\to q\bar{q}g$, we utilize the \texttt{Mathematica} packages \texttt{FeynArts}~\cite{Kublbeck:1990xc}   to create the Feynman diagrams and then apply ~\texttt{FeynCalc}~\cite{Shtabovenko:2023idz,Shtabovenko:2020gxv,Shtabovenko:2016sxi,Mertig:1990an} to generate the amplitudes and manipulate the Dirac matrices. After summing up the color indices and polarizations for the external partons, we obtain,
\begin{align} \label{eq:def:amp2:LO}   
&\left(\frac{\alpha_s}{4\pi}\right)\mathcal{H}^{(1)}(x_q,x_{\bar{q}};\varepsilon)
\equiv\sum_{\mathrm{pol,col}} \Big|\mathcal{M}_{\mathrm{B}}(\gamma^*\to q\bar{q}g)\Big|^2\,\nonumber\\
=&  \frac{64\pi^2 N_A\alpha \alpha_se_q^2(1-\varepsilon) }{(1-x_q)(1-x_{\bar{q}})}\left[(x_q^2+x_{\bar{q}}^2)-\varepsilon(2-x_q-x_{\bar{q}})^2\right]\,,
\end{align}
where $N_A=N_C^2-1$. The dimensionless variables $x_{q,{\bar{q}}}$ are defined as 
\begin{align}    \label{eq:def:xq:xqb}
  x_q=\frac{2p_q\!\cdot\!q_{\gamma}}{Q^2}\,,\qquad
  x_{\bar{q}}=&\frac{2p_{\bar{q}}\!\cdot\!q_{\gamma}}{Q^2}\,,
\end{align}
with $p_{q}$ and $p_{\bar{q}}$ the momenta of the quark and antiquark, respectively. 
We have checked eq.~\eqref{eq:def:amp2:LO} agrees with the previous calculation in \cite{Ellis:1980wv}.

Combining the expressions in eqs.~\eqref{eq:def:momX:LO}, \eqref{eq:def:kappa}, \eqref{eq:def:amp2:LO}, we arrive at the LO coefficients, 
\begin{align}    
\label{eq:def:sig:All:LO}
  \mathcal{C}^{(1)}_{ij;q}=&     \delta\left(m_X^2\right)   \mathcal{C}^{(1)}_{ij;q,\delta}\,,
\end{align}
where
\begin{align}    
\label{eq:def:sig:qqb:LO}
  \hat{\sigma}_q\,\mathcal{C}^{(1)}_{q\bar{q};q,\delta}(x_1,x_2;\varepsilon)=& \,  \frac{2\pi }{Q^2}\, \kappa(z;\varepsilon)\,\left( x_1x_2\right)^{1-2\varepsilon} \,\mathcal{H}^{(1)}(x_1,x_2;\varepsilon)   \,,
  \\
  \label{eq:def:sig:qg:LO}
\hat{\sigma}_q\,\mathcal{C}^{(1)}_{qg;q,\delta}(x_1,x_2;\varepsilon)= &\,  \frac{2\pi}{Q^2}\, \kappa(z;\varepsilon)\,\left( x_1x_2\right)^{1-2\varepsilon}  \,\mathcal{H}^{(1)}(x_1,2-x_1-x_2;\varepsilon)  \,,\\
  \label{eq:def:sig:qbg:LO}
  \mathcal{C}^{(1)}_{g\bar{q};q,\delta}(x_1,x_2;\varepsilon)=&\,\mathcal{C}^{(1)}_{qg;q,\delta}(x_2,x_1;\varepsilon).
\end{align}
Here $\hat{\sigma}_q=2\alpha N_CQ^2e_q^2$ is introduced to absorb coupling constants and dimensional variables. 
We omit the expressions of  $\mathcal{C}^{(1)}_{\bar{q}q;q}$, $\mathcal{C}^{(1)}_{gq;q}$, and $\mathcal{C}^{(1)}_{\bar{q}g;q}$ here for brevity, which can be derived by swapping the momentum fractions $x_1\leftrightarrow x_2$  in eqs.~(\ref{eq:def:sig:qqb:LO}-\ref{eq:def:sig:qbg:LO}) as appropriate. 
Expanding eqs.~(\ref{eq:def:sig:qqb:LO}-\ref{eq:def:sig:qbg:LO}) in $\varepsilon$, we obtain
\begin{align}    %
\label{eq:sig:qqb:LO:eps0}
  \mathcal{C}^{(1)}_{q\bar{q};q}(x_1,x_2;\varepsilon=0) &=\delta(m_X^2)  \frac{N_A}{N_C}\frac{x_1x_2(x_1^2+x_2^2)}{(1-x_1)(1-x_2)}  \,,
  \\
  \label{eq:sig:qg:LO:eps0}
\mathcal{C}^{(1)}_{qg;q}(x_1,x_2;\varepsilon=0) &=    \delta(m_X^2)  \frac{N_A}{N_C}\frac{x_1x_2[x_1^2+(2-x_1-x_2)^2]}{(1-x_1)(x_1+x_2-1)}  \,,
\\
\label{eq:sig:qbg:LO:eps0}
  \mathcal{C}^{(1)}_{g\bar{q};q}(x_1,x_2;\varepsilon=0)&=\mathcal{C}^{(1)}_{qg;q}(x_2,x_1;\varepsilon=0).
\end{align}

\section{The method of calculation at NLO}
\label{sec:def:NLO}

The coefficients at NLO consist of three parts, the real-real emission $\mathcal{C}_{ij;q}^{\mathrm{RR}}$, the real-virtual radiation $\mathcal{C}_{ij;q}^{\mathrm{RV}}$, and the counter terms in the collinear renormalization $\mathcal{C}_{ij;q}^{\mathrm{CT}}$, namely, 
\begin{align}    
\label{eq:def:coeff:nlo:finite}
\mathcal{C}^{(2)}_{ij;q}=\mathcal{C}_{ij;q}^{\mathrm{RR}}+\mathcal{C}_{ij;q}^{\mathrm{RV}}+\mathcal{C}_{ij;q}^{\mathrm{CT}}\,.
\end{align}
In this section, we will elaborate on their evaluations individually.

\subsection{Double-real emission}
\label{sec:def:NLO:RR}
\subsubsection{Phase space integrals and IBP}
  \begin{figure}[tb]
    \centering
    \includegraphics[width=0.9\linewidth]{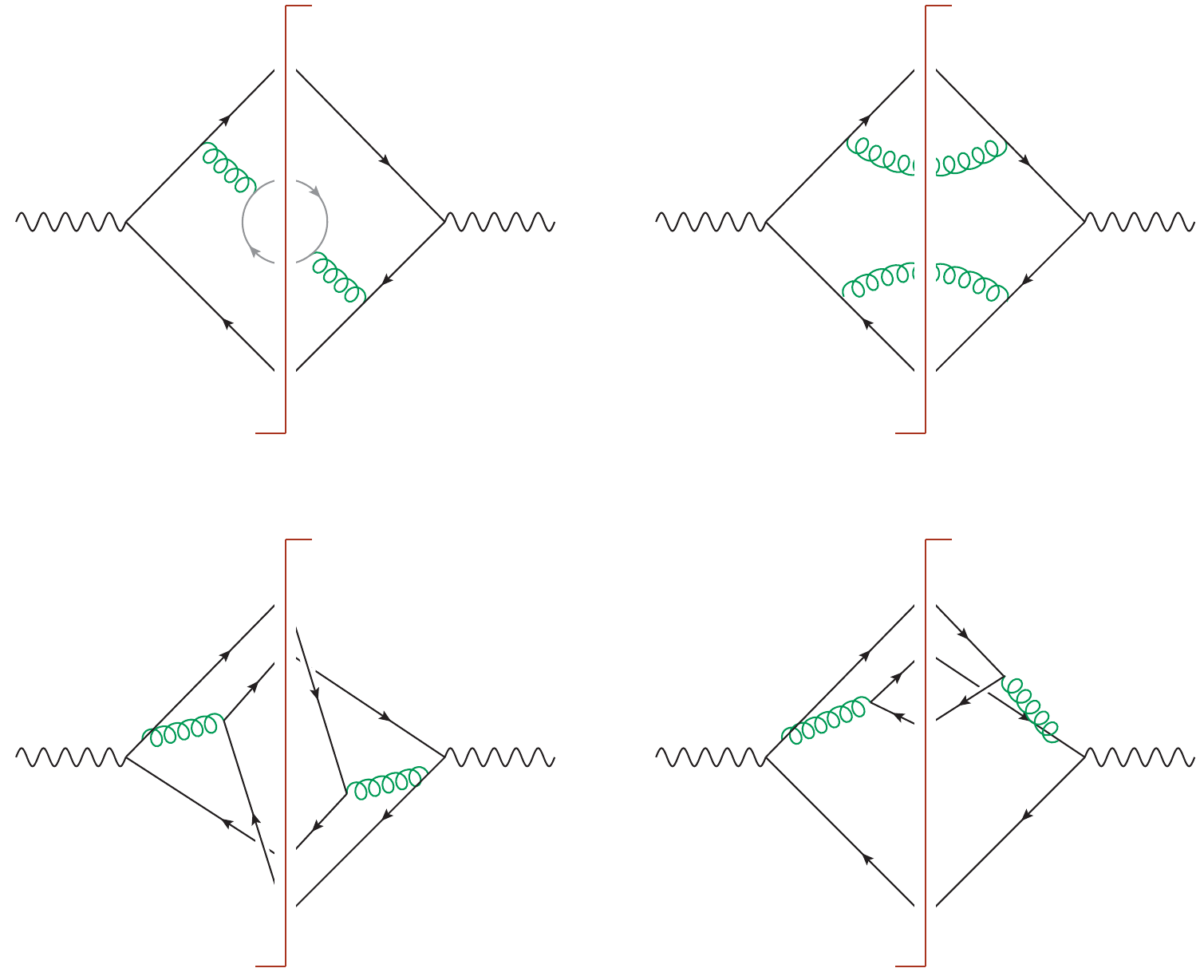}
    \caption{Characteristic Feynman diagrams for the   squared amplitudes $|\mathcal{M}_B|^2$ of $\gamma^*\to $ 4 partons. The top-left diagram shows the contribution of $\gamma^*\to q\bar{q}q^{\prime} \bar{q}^{\prime}$ with the gray lines representing the quark pair $q^{\prime}\bar{q}^{\prime}$ possessing distinct flavor from those directly connecting to the electromagnetic vertices. The contribution of the double-gluon emission $\gamma^*\to q\bar{q}gg$ is exhibited in the top-right diagram. The remaining two plots in the bottom row are for the $\gamma^*\to q\bar{q}q\bar{q}$ process.}
    \label{fig:feynman_RR}
\end{figure}
The coefficients $\mathcal{C}_{ij;q}^{\mathrm{RR}}$ for the double-real emissions are induced by the tree-level partonic processes, 
\begin{align}
\label{eq:def:processes:RR:qqp}
\gamma^* &\to q  \;\bar q  \; q^\prime  \; \bar q^\prime  \,,\\
\label{eq:def:processes:RR:gg}
\gamma^* &\to q  \;\bar q   \;g  \; g \,,\\
\label{eq:def:processes:RR:qq}
\gamma^* &\to q \; \bar q   \;q   \;\bar q  \,,
\end{align}
where $q(\bar{q})$ stands for the (anti)quark directly connecting to the electromagnetic vertices, while $q'(\bar{q}')$ refers to that produced by QCD interactions and possessing different flavors from $q(\bar{q})$. Some typical Feynman diagrams for the amplitude square of these double-real emissions are shown in fig.~\ref{fig:feynman_RR}. 
In analogy to the calculation at LO, we also employ the software \texttt{FeynArts}~\cite{Kublbeck:1990xc} and ~\texttt{FeynCalc}~\cite{Shtabovenko:2023idz,Shtabovenko:2020gxv,Shtabovenko:2016sxi,Mertig:1990an} 
to compute the squared amplitudes for each partonic channel in eqs.~(\ref{eq:def:processes:RR:qqp}-\ref{eq:def:processes:RR:qq}). 
We examine the results by comparing with those in ref.~\cite{Dixon:2018qgp}, finding full agreement.  
Plugging the resulting squared amplitudes back into eq.~\eqref{eq:def:Cij}, we arrive at %
\begin{align}
\label{eq:def:Cqqb:RR}
  \hat{\sigma}_q \, \left(\frac{\alpha_s}{4\pi}\right)^2 \mathcal{C}_{q\bar{q};q}^{\mathrm{RR}}
&= \int\mathrm{d}\widetilde{\Phi}_{cd}\,
 \Bigg\{     \Bigg|\mathcal{M}_{\mathrm{B}}\Big(\gamma^*\to q(\tilde{k}_i)+\bar q(\tilde{k}_j)  + q  (k_c)+\bar q (k_d)\Big)\Bigg|^2\notag\\
&   +  \frac{1}{2!} \, \Bigg|\mathcal{M}_{\mathrm{B}}\Big(\gamma^*\to q(\tilde{k}_i)+\bar q(\tilde{k}_j)  + g (k_c)+g(k_d)\Big)\Bigg|^2 \notag\\
&+   (N_f-1)\Bigg|\mathcal{M}_{\mathrm{B}}\Big(\gamma^*\to q(\tilde{k}_i)+\bar q(\tilde{k}_j)  + q^\prime (k_c)+\bar q^\prime(k_d)\Big)\Bigg|^2
\Bigg\}\,,
\\
    \label{eq:def:Cqqp:RR}
   \hat{\sigma}_q \, \left(\frac{\alpha_s}{4\pi}\right)^2  \mathcal{C}_{qq^{\prime};q}^{\mathrm{RR}}
&= \int\mathrm{d}\widetilde{\Phi}_{cd}\,
     \Bigg|\mathcal{M}_{\mathrm{B}}\Big(\gamma^*\to q(\tilde{k}_i)+\bar q(k_c)  + q^\prime (\tilde{k}_j)+\bar q^\prime(k_d)\Big)\Bigg|^2\,,
     \\
      \label{eq:def:Cqpqbp:RR}
   \hat{\sigma}_q \,  \left(\frac{\alpha_s}{4\pi}\right)^2  \mathcal{C}_{q^{\prime}{\bar{q}}^\prime;q}^{\mathrm{RR}}
&= \int\mathrm{d}\widetilde{\Phi}_{cd}\,
     \Bigg|\mathcal{M}_{\mathrm{B}}\Big(\gamma^*\to q( k_c)+\bar q(k_d)  + q^\prime (\tilde{k}_i)+\bar{q}^\prime(\tilde{k}_j)\Big)\Bigg|^2\,,
     \\
\label{eq:def:Cqg:RR}
   \hat{\sigma}_q \,\left(\frac{\alpha_s}{4\pi}\right)^2    \mathcal{C}_{qg;q}^{\mathrm{RR}}
&=\int\mathrm{d}\widetilde{\Phi}_{cd}\,
    \Bigg|\mathcal{M}_{\mathrm{B}}\Big(\gamma^*\to q(\tilde{k}_i)+\bar q(k_c)  + g (k_d)+ g(\tilde{k}_j)\Big)\Bigg|^2\,,    \\
    \label{eq:def:Cgg:RR}
   \hat{\sigma}_q \,  \left(\frac{\alpha_s}{4\pi}\right)^2  \mathcal{C}_{gg;q}^{\mathrm{RR}}
&= \int\mathrm{d}\widetilde{\Phi}_{cd}\,
     \Bigg|\mathcal{M}_{\mathrm{B}}\Big(\gamma^*\to q( k_c)+\bar q(k_d)  + g (\tilde{k}_i)+  g(\tilde{k}_j)\Big)\Bigg|^2\,,
   \\
\label{eq:def:Cqq:RR}
  \hat{\sigma}_q \,  \left(\frac{\alpha_s}{4\pi}\right)^2   \mathcal{C}_{qq;q}^{\mathrm{RR}}
&=\frac{1}{2!} \,\int\mathrm{d}\widetilde{\Phi}_{cd}\,
     \Bigg|\mathcal{M}_{\mathrm{B}}\Big(\gamma^*\to q(\tilde{k}_i)+\bar q(k_c)  + q ( \tilde{k}_j)+ \bar q(k_d )\Big)\Bigg|^2\,,    
\end{align}
where $N_f$ denotes the total number of the active quark flavors in question. 
The factorials in eqs.~\eqref{eq:def:Cqqb:RR} and \eqref{eq:def:Cqq:RR} account for the average factors in eq.~\eqref{eq:def:dps:fac}  for the indistinguishable particles in the final states.
The differential $\mathrm{d}\widetilde{\Phi}_{cd}$ for the undetected partons is defined as,
\begin{align}
    \int\mathrm{d}\widetilde{\Phi}_{cd}\,\equiv\,&  \kappa(z;\varepsilon)(x_1x_2)^{1-2\varepsilon}\,\int \frac{\mathrm{d}^{d-1}\vec{k}_c}{(2\pi)^{d-1}}\,\frac{1}{2E_c}\,
    \frac{\mathrm{d}^{d-1}\vec{k}_d}{(2\pi)^{d-1}}\,
    \frac{1}{2E_d}\,\nonumber\\
    &\times
    (2\pi)^{d}\,\delta^d\left(q_{\gamma}-\tilde{k}_i-\tilde{k}_j-k_c-k_d\right)\,.
\end{align}
In eqs.~(\ref{eq:def:Cqqb:RR}-\ref{eq:def:Cqq:RR}), we have completed the integration over the momenta of the identified partons, $k^{\mu}_a$ and $k^{\mu}_b$, leaving the same kinematic prefactor $\kappa(z;\varepsilon)$ as in eq.~\eqref{eq:def:kappa} and also fixing the momenta,
\begin{align}
   k^{\mu}_a\to \tilde{k}^{\mu}_i=\frac{Qx_1}{2}n_a^{\mu}\,,
    \qquad
  k^{\mu}_b\to  \tilde{k}^{\mu}_j=\frac{Qx_2}{2}n_b^{\mu}\,.
\end{align}
Herein, $n_a^{\mu}$ and $n_b^{\mu}$ are two light-like vectors pointing at the flight directions of the identified hadrons $H_1$ and $H_2$, respectively.
In eqs.~(\ref{eq:def:Cqqb:RR}-\ref{eq:def:Cqq:RR}), we only spell out the expressions for $\mathcal{C}_{ij;q}^{\mathrm{RR}}$ where 
$\{ij\}\in\{q\bar{q},qq^{\prime},q^{\prime}{\bar{q}}
^\prime,qg,gg,qq\}$ are referred as six sectors in the following. 
Since QCD interactions are insensitive to quark charges, other combinations can be derived from the results in these six sectors, namely, 
\begin{align}
\label{eq:def:Cgqb}
\mathcal{C}_{g\bar{q};q}^{\mathrm{RR}}(x_1,x_2,z;\varepsilon;\mu)&=  \mathcal{C}_{qg;q}^{\mathrm{RR}}(x_2,x_1,z;\varepsilon;\mu)\,,\\
\label{eq:def:Cqbqb}
\mathcal{C}_{\bar{q}\bar{q};q }^{\mathrm{RR}}(x_1,x_2,z;\varepsilon;\mu)&=\mathcal{C}_{qq;q }^{\mathrm{RR}}(x_1,x_2,z;\varepsilon;\mu)\,,\\
\label{eq:def:Cqqbp}
\mathcal{C}_{ {q}\bar{q}^{\prime};q }^{\mathrm{RR}}(x_1,x_2,z;\varepsilon;\mu)&=\mathcal{C}_{qq^{\prime};q }^{\mathrm{RR}}(x_1,x_2,z;\varepsilon;\mu)\,,\\
\label{eq:def:Cqbqp}
\mathcal{C}_{\bar{q}q^{\prime};q }^{\mathrm{RR}}(x_1,x_2,z;\varepsilon;\mu)&=\mathcal{C}_{qq^{\prime} ;q}^{\mathrm{RR}}(x_1,x_2,z;\varepsilon;\mu)\,,\\
\label{eq:def:Cqbqbp}
\mathcal{C}_{ \bar{q}\bar{q}^{\prime};q }^{\mathrm{RR}}(x_1,x_2,z;\varepsilon;\mu)&=\mathcal{C}_{qq^{\prime};q  }^{\mathrm{RR}}(x_1,x_2,z;\varepsilon;\mu)\,.
\end{align}
The remaining task is to evaluate the phase-space integrals in eqs.~(\ref{eq:def:Cqqb:RR}-\ref{eq:def:Cqq:RR}) over the unmeasured momenta $k_c$ and $k_d$. In contrast to that of the identified particles, the calculation at this stage is generally more involved and   in the following, we resort to IBP and DE approaches to tackle them. 

First, we apply reverse unitarity~\cite{Anastasiou:2002yz,Anastasiou:2003yy} to recast on-shell conditions of the external particles as differences of Feynman propagators with opposite $i\epsilon$ prescriptions,
  \begin{align}\label{eq:def:rev:uni}
 \int\frac{\md^{d-1} \vec{k}_l}{2E_l  }=\int\md^{d} k_l \delta_+(k^2_l)\to\int\frac{\md^{d} k_l}{2\pi i} \left\{\frac{1}{k_l^2-i\epsilon}-\frac{1}{k_l^2+i\epsilon}\right\}\,,
\end{align}
where $k_l\in\{k_c,k_d\}$. We then calculate the product of the Feynman propagators in eq.~\eqref{eq:def:rev:uni} and the squared amplitudes of the partonic processes in eqs.~(\ref{eq:def:Cqqb:RR}-\ref{eq:def:Cqq:RR}).
To facilitate the later IBP process, the package \texttt{Apart}~\cite{Feng:2012iq} is utilized here to convert the expressions possessing linearly dependent propagators into the irreducible ones.
Eventually, owning to the patterns of resulting denominators,  the integrals over $k_{c(d)}$ fall into $6$ independent topologies
\begin{align}
\label{eq:def:ind:tops1}
\mathbf{topo1}&:\,\{k_c^2,k_d^2,(\tilde{k}_i+k_d)^2, (\tilde{k}_j+k_c)^2\}\,,\\
\label{eq:def:ind:tops2}
\mathbf{topo2}&:\,\{k_c^2,k_d^2,(\tilde{k}_i+k_d)^2,(\tilde{k}_j+k_d)^2\}\,,\\
\label{eq:def:ind:tops3}
\mathbf{topo3}&:\,\{k_c^2,k_d^2,(\tilde{k}_i+k_c)^2,(\tilde{k}_i+\tilde{k}_j+k_c)^2\}\,,\\
\label{eq:def:ind:tops4}
\mathbf{topo4}&:\,\{k_c^2,k_d^2,(\tilde{k}_j+k_c)^2,(\tilde{k}_i+\tilde{k}_j+k_c)^2\}\,,\\
\label{eq:def:ind:tops5}
\mathbf{topo5}&:\,\{k_c^2,k_d^2,(\tilde{k}_i+k_c)^2,(\tilde{k}_i+\tilde{k}_j+k_d)^2\}\,,\\
\label{eq:def:ind:tops6}
\mathbf{topo6}&:\,\{k_c^2,k_d^2,(\tilde{k}_j+k_c)^2,(\tilde{k}_i+\tilde{k}_j+k_d)^2\}\,,
\end{align}
plus those subjecting to the exchange $k_c\leftrightarrow k_d$,
\begin{align}
\label{eq:def:ind:tops7}
\mathbf{topo7}&:\,\{k_c^2,k_d^2,(\tilde{k}_i+k_c)^2,(\tilde{k}_j+k_d)^2\}\,,\\
\label{eq:def:ind:tops8}
\mathbf{topo8}&:\,\{k_c^2,k_d^2,(\tilde{k}_i+k_c)^2,(\tilde{k}_j+k_c)^2\}\,,\\
\label{eq:def:ind:tops9}
\mathbf{topo9}&:\,\{k_c^2,k_d^2,(\tilde{k}_i+k_d)^2,(\tilde{k}_i+\tilde{k}_j+k_d)^2\}\,,\\
\label{eq:def:ind:tops10}
\mathbf{topo10}&:\, \{k_c^2,k_d^2,(\tilde{k}_j+k_d)^2,(\tilde{k}_i+\tilde{k}_j+k_d)^2\}\,,\\
\label{eq:def:ind:tops11}
\mathbf{topo11}&:\,\{k_c^2,k_d^2,(\tilde{k}_i+k_d)^2,(\tilde{k}_i+\tilde{k}_j+k_c)^2\}\,,\\
\label{eq:def:ind:tops12}
\mathbf{topo12}&:\,\{k_c^2, k_d^2, (\tilde{k}_j+k_d)^2, (\tilde{k}_i+\tilde{k}_j+k_c)^2\}\,.
\end{align}
Now we are in a position to apply the IBP relations~\cite{Tkachov:1981wb,Chetyrkin:1981qh} to reduce the number of integrals and also the tensor structures in the numerators. More specifically, we feed each of the independent topologies in eqs.~(\ref{eq:def:ind:tops1}-\ref{eq:def:ind:tops6}) into the program~\texttt{LiteRed}~\cite{Lee:2012cn,Lee:2013mka} %
to establish IBP identities and reduction rules among the ensuing phase space integrals. The integrals dictated by topologies in eqs.~(\ref{eq:def:ind:tops7}-\ref{eq:def:ind:tops12}) are all mapped onto the  independent cases in eqs.~(\ref{eq:def:ind:tops1}-\ref{eq:def:ind:tops6}) by swapping the momenta $k_c\leftrightarrow k_d$ as appropriate. 
During the calculation, all the denominators induced by reverse unitarity in eq.~\eqref{eq:def:rev:uni} are marked as the cut propagators to improve the IBP efficiency. 
It is followed by exploiting the command \texttt{Toj} in \texttt{LiteRed} for projecting the integration variables $k_{c,d}^{\mu}$ from the numerators into  linearly independent bases in the occurring topologies.  Ultimately,  the phase space integrals in the double real corrections are reforged into the linear combination of $9$ master integrals,
 \begin{align}
 \label{eq:def:IBP:coeffs}
 \mathcal{C}_{ij;q}^{\mathrm{RR}}(x_1,x_2,z;\varepsilon;\mu)=\theta(m_X^2)\,e^{-\varepsilon L_h}\,\sum_{k=1}^9\, d^{[k] }_{ij;q}(x_1,x_2,z;\varepsilon) \,\mathcal{I}_k(x_1,x_2,z;\varepsilon)\,,
\end{align}
where $\theta(m_X^2)$ is introduced to maintain the kinematic constraint in eq.~\eqref{eq:def:phy:ps}, with $L_h=\ln(Q^2/\mu^2)$. $d^{[k] }_{ij;q}$ encodes the resulting coefficients of the $k$th master integral $\mathcal{I}_k$ due to the IBP reductions.   The master integrals involved in this work include
 \begin{align}
\label{eq:def:MI1}
\mathcal{I}_1&
=
 {r_{\Gamma}}{(Q^2)^{\frac{4-d}{2}}}\int \frac{d^d k_c}{(2 \pi)^{d- 1}} \delta_+(k_c^2) \delta_+( (q_{\gamma}-\tilde{k}_i-\tilde{k}_j-k_c)^2)\,,
 \\[5pt] 
\label{eq:def:MI2}
\mathcal{I}_2
&=
r_{\Gamma}{(Q^2)^{\frac{6-d}{2}}}\int \frac{d^d k_c}{(2 \pi)^{d-1}} \frac{\delta_+(k_c^2) \delta_+( (q_{\gamma}-\tilde{k}_i-\tilde{k}_j-k_c)^2)}{(\tilde{k}_i+\tilde{k}_j+k_c)^2}\,,
\\
\label{eq:def:MI3}
\mathcal{I}_3&
=
r_{\Gamma}{(Q^2)^{\frac{8-d}{2}}}\int \frac{d^d k_c}{(2 \pi)^{d-1}} \frac{\delta_+(k_c^2) \delta_+( (q_{\gamma}-\tilde{k}_i-\tilde{k}_j-k_c)^2)}{(\tilde{k}_i+k_d)^2  (\tilde{k}_j+k_c)^2}\,,\\[5pt]
\label{eq:def:MI4}
\mathcal{I}_4&
=
r_{\Gamma}{(Q^2)^{\frac{8-d}{2}}}\int \frac{d^d k_c}{(2 \pi)^{d-1}} \frac{\delta_+(k_c^2) \delta_+( (q_{\gamma}-\tilde{k}_i-\tilde{k}_j-k_c)^2)}{(\tilde{k}_i+k_d)^2  (\tilde{k}_j+k_d)^2}\,,
\\[5pt]
\label{eq:def:MI5}
\mathcal{I}_5&
=
r_{\Gamma}{(Q^2)^{\frac{8-d}{2}}}\int \frac{d^d k_c}{(2 \pi)^{d-1}} \frac{\delta_+(k_c^2) \delta_+( (q_{\gamma}-\tilde{k}_i-\tilde{k}_j-k_c)^2)}{(\tilde{k}_i+k_c)^2 (\tilde{k}_i+\tilde{k}_j+k_c)^2}\,,\\[5pt]
\label{eq:def:MI6}
\mathcal{I}_6&
=
r_{\Gamma}{(Q^2)^{\frac{8-d}{2}}}\int \frac{d^d k_c}{(2 \pi)^{d-1}} \frac{\delta_+(k_c^2) \delta_+( (q_{\gamma}-\tilde{k}_i-\tilde{k}_j-k_c)^2)}{(\tilde{k}_i+k_c)^2 (\tilde{k}_i+\tilde{k}_j+k_d)^2}\,,
\end{align}
as well as those under the symmetry $\tilde{k}_i\leftrightarrow \tilde{k}_j$ or $k_c\leftrightarrow k_d$,
 \begin{align}
 \label{eq:def:MI7}
\mathcal{I}_7
&=r_{\Gamma}{(Q^2)^{\frac{8-d}{2}}}\int \frac{d^d k_c}{(2 \pi)^{d-1}} \frac{\delta_+(k_c^2) \delta_+( (q_{\gamma}-\tilde{k}_i-\tilde{k}_j-k_c)^2)}{(\tilde{k}_i+\tilde{k}_j+k_d)^2}=\mathcal{I}_2\,,
\\[5pt]
\label{eq:def:MI8}
\mathcal{I}_8
&=
r_{\Gamma}{(Q^2)^{\frac{8-d}{2}}}\int \frac{d^d k_c}{(2 \pi)^{d-1}} \frac{\delta_+(k_c^2) \delta_+( (q_{\gamma}-\tilde{k}_i-\tilde{k}_j-k_c)^2)}{(\tilde{k}_j+k_c)^2 (\tilde{k}_i+\tilde{k}_j+k_c)^2}=\mathcal{I}_5\Bigg|_{\tilde{k}_i\leftrightarrow \tilde{k}_j}\,,\\[5pt]
\label{eq:def:MI9}
\mathcal{I}_9
&=
r_{\Gamma}{(Q^2)^{\frac{8-d}{2}}}\int \frac{d^d k_c}{(2 \pi)^{d-1}} \frac{\delta_+(k_c^2) \delta_+( (q_{\gamma}-\tilde{k}_i-\tilde{k}_j-k_c)^2)}{(\tilde{k}_j+k_c)^2  (\tilde{k}_i+\tilde{k}_j+k_d)^2}=\mathcal{I}_6\Bigg|_{\tilde{k}_i\leftrightarrow \tilde{k}_j}\,.
\end{align}
Therein, all the master integrals are defined dimensionlessly by introducing the kinematic factor $(Q^2)^{\frac{n-d}{2}}(n=4,6,8)$ in front of the phase space integrals. We notice that the   integrals above are one-loop phase-space integrals with four external legs, i.e., $\tilde{k}_i,\tilde{k}_j, q_\gamma,q_\gamma-\tilde{k}_i-\tilde{k}_j$, where the first two legs are on-shell, the last two are off-shell. The kinematic invariants read 
\begin{align}
    q_\gamma^2= Q^2, \qquad\frac{2 q_\gamma \cdot \tilde{k}_i}{Q^2} = x_1,\qquad  \frac{2 q_\gamma \cdot \tilde{k}_j}{Q^2} = x_2, \qquad2 \tilde{k}_i \cdot \tilde{k}_j = x_1 x_2 z  Q^2\,.
\end{align}
The above calculation bears resemblance to   the calculation on the production of two off-shell vector bosons with different invariant masses~\cite{Chavez:2012kn,Gehrmann:2014bfa,Caola:2014lpa,Henn:2014lfa,Gehrmann:2015ora}. To absorb all the Euler's constant $\gamma_{\mathrm{E}}$ and $\ln \pi$ terms from  the results, we also put in place an overall factor 
\begin{align}
    r_{\Gamma}\equiv\left[\frac{\exp(\gamma_{\mathrm{E}})}{4\pi}\right]^{\frac{4-d}{2}}\,.
\end{align}
The calculation of $\mathcal{I}_1$, $\mathcal{I}_2$, and $\mathcal{I}_7$ is straightforward. At this point, it merits noting that the master integrals are Lorentz invariant, such that an appropriate choice on the reference frame can simplify our calculation substantially. Here we stick to the rest reference frame of the undetected partons $c$ and $d$, in which the energies and spatial momenta of $c$ and $d$ are all fixed by the Dirac delta functions in eqs.~(\ref{eq:def:MI1}-\ref{eq:def:MI2}) and eq.~(\ref{eq:def:MI7}), leaving at most one-fold angular integrals over the scattering angles of the identified particles $a$ and $b$. Completing these angular integrations  leads to, 
\begin{align}
\label{eq:misolution1}
\mathcal{I}_1&=\frac{2^{2 \varepsilon-5} \pi ^{ -\frac{3}{2}}    }{\Gamma \left(\frac{3}{2}-\varepsilon\right)}\,
\exp(\varepsilon\gamma_{\mathrm{E}})
\,m_{X}^{-2 \varepsilon}=\mathcal{I}_7\,,
\\
\label{eq:misolution2}
\mathcal{I}_2&=\frac{  \Gamma (1-\varepsilon) \exp(\varepsilon\gamma_{\mathrm{E}}) }{8\pi^2}  
\,_2\tilde{F}_1\left(1,1-\varepsilon;2-2 \varepsilon;\frac{2 \beta }{x_1+x_2+\beta  }\right)
\,\frac{m_{X}^{-2 \varepsilon}}{  \beta +x_1+x_2}\,.
\end{align}
Here $\varepsilon=(4-d)/2$ is again the dimensional regulator. The dimensionless variable $\beta $ characterizes  the relative velocity between the identified  and unmeasured final particles,
\begin{align}
\beta = {\sqrt{(x_1+x_2)^2-4 x_1 x_2 z}} \,.
\end{align}
 Nevertheless, the calculations on the other master integrals in eqs.~(\ref{eq:def:MI2}-\ref{eq:def:MI6}) and eqs.~(\ref{eq:def:MI8}-\ref{eq:def:MI9}) are more involved, as the corresponding integrands include additional propagators. In earlier studies, analogous integrals were computed using dispersion relations~\cite{vanNeerven:1985xr,Beenakker:1988bq} or via Mellin-Barnes representations~\cite{Somogyi:2011ir}. Partial results up to $\mathcal{O}(\varepsilon^0)$ are also available in~\cite{Devoto:1984wu}. In what follows, we revisit these integrals using the differential equations method~\cite{Kotikov:1990kg,Remiddi:1997ny,Gehrmann:1999as,Argeri:2007up,Henn:2013pwa}, which has proven to be a powerful and systematic tool in modern multi-loop Feynman integral computations. We hope the discussion here can provide an alternative strategy for handling these integrals and, in doing so, lay the groundwork for future investigations of NNLO differential distributions. 
 
\subsubsection{Differential equation and its solution}\label{sec:DEs}

Given the relations in eqs.~(\ref{eq:def:MI7}-\ref{eq:def:MI9}), the discussion below will lay stress on the master integrals $\mathcal{I}_1$-$\mathcal{I}_6$, pertaining to the topologies $\mathbf{topo1}$, $\mathbf{topo2}$, $\mathbf{topo3}$ and $\mathbf{topo5}$.  Differential equations of those master integrals with respect to the kinematic variable $$l\in\{x_1,x_2,z\}$$ can be derived by IBP reduction rules in the occurring topology.  In practice the built-in function \texttt{MakeDSystem} in~\texttt{LiteRed}~\cite{Lee:2012cn,Lee:2013mka} is utilized for this purpose. However, the resulting expression here invokes higher power polynomials in $l$ from the denominators, which hinders the further partial fraction and in turn the transformation into the canonical bases. Analogous circumstances can also be found in the two loop calculations on the production of two off-shell vector bosons~\cite{Chavez:2012kn,Gehrmann:2014bfa,Caola:2014lpa,Henn:2014lfa,Gehrmann:2015ora}.  

To this end, we bring in an alternative set of coordinates $$\xi\in\{\alpha,y,\bar y\}$$ for each topology. They are defined as
\begin{eqnarray}
\label{eq:def:para:123}
\mathbf{topo1/2/3}:\;&
\left\{
  \begin{aligned}
  &\alpha \to x_1-x_2\,,\\
 &y\to \frac{1}{2} (-{\beta }-x_1-x_2+2),\\
 &\bar{y}\to \frac{1}{2} ({\beta }-x_1-x_2+2),\\
 \end{aligned}
 \right.
 \\[1ex]
 \label{eq:def:para:5}
\mathbf{topo5}:\;&
\left\{
  \begin{aligned}
  &\alpha \to 2 x_1 \,,\\
 &y\to \frac{1}{2} (-{\beta }-x_1-x_2+2),\\
&\bar{y}\to \frac{1}{2} ({\beta }-x_1-x_2+2),\\
 \end{aligned}
\right.
\end{eqnarray}
where $y\bar y =m_X^2$ in all four topologies.  A different parametrization for $\mathbf{topo5}$ is introduced here to leave out any quadratic terms  from the alphabet.

 The differential equations with regards to $\xi\in\{\alpha,y,\bar y\}$ can be derived from those in $l\in\{x_1,x_2,z\}$ by means of a series of linear transformations. We then feed the results to the \texttt{Mathematica}
package \texttt{CANONICA} \cite{Meyer:2016slj,Meyer:2017joq} to transform the differential equations to  the canonical form~\cite{Henn:2013pwa}. %
Eventually, we obtain, 
\begin{equation}\label{eq:def:de}
\frac{\partial}{\partial \xi}  \vec{f}_{i}(\alpha,y,\bar y; \varepsilon)=\varepsilon\mathcal{A}^{[i]}_{\xi}(\alpha,y,\bar y)\, \vec{f}_{i}(\alpha,y,\bar y; \varepsilon)\,,
\end{equation}
where the subscript $i$ runs over $\{1,2,3,5\}$ characterizing the indices of the topologies. The vector $ \vec{f}_{i}$ is made up of the linear combinations of master integrals in the $i$-th topology.  $\mathcal{A}^{[i]}_{\xi}$ entails the rational functions of the dimensionless invariants $\xi\in\{\alpha,y,\bar y\}$. The set of irreducible denominators in each topology, customarily dubbed letters in alphabet, reads 
\begin{align}
\label{eq:def:letters:top12}
\mathbf{topo1(2)} &: \{ \bar{y} - 1, \bar{y}, y - 1, y,  \bar{y} - \alpha - y, -\bar{y} - \alpha + y, 2 \bar{y} y -\bar{y}-  \alpha  - y, \notag\\
&\qquad2 \bar{y} y - \bar{y} + \alpha - y \}, \\
\label{eq:def:letters:top3}
\mathbf{topo3}  &: \{ \bar{y} - 1, \bar{y}, \bar{y}-y ,y-1, y, \bar{y} - \alpha - y, -\bar{y} - \alpha + y, 2 \bar{y} y - \bar{y}-{\alpha} - y, \notag\\
&\qquad\bar{y} + \alpha + y \}\,, \\
\label{eq:def:letters:top5}
\mathbf{topo5} &: \{ \bar{y} - 1, \bar{y}, \bar{y} - y, y - 1, y, 2 \bar{y} y - 2 \bar{y} -\alpha- 2 y + 2, 2 \bar{y} + \alpha - 2, \alpha + 2 y - 2, \notag\\
&\qquad\alpha y \bar{y}+2y\bar{y} - 2 \bar{y} - \alpha - 2 y + 2 \} \,.
\end{align}
Within the physical domain in eq.~\eqref{eq:def:phy:ps}, the letters above are generally finite, with the only exception that $y$  could approach zero in the limit $m_X^2\to0$, where the final emissions form  a three-prong configuration.  Similarly, this type of singularity, i.e, $\propto 1/m_X^2$, is also found in the coefficients  $d^{[k]}_{ij}$ in front of the master integrals when we carry out the IBP reductions. In light of this phenomenon, the perturbative solution of $\vec{f}_{i}$ in the $\varepsilon$ expansion is not always sufficient when convoluting with FFs in eq.~\eqref{eq:def:fac}.  In order to regularize the ensuing inferred and collinear (IRC) divergence, a systematic resummation of the higher power $\varepsilon$ terms is still necessitated in the vicinity of $y=0$, akin to \cite{Papadopoulos:2015jft,Canko:2020gqp}.
To this end, we will deliver two kinds of solutions for $\vec{f}_{i}$ in the following. The first one focuses on the regular region $m_X^2>0$, solving eq.~\eqref{eq:def:de} via Chen iterated integrals \cite{Chen:1977oja}. 
In the second case, we take on the divergent regime $y\to0$. We will carry out the asymptotic expansion of  $\mathcal{A}^{[i]}_{\xi}$  in the limit $y\to0$,  thereby resumming all powers of $\varepsilon$ terms associated with leading singular behaviors $\propto 1/m_X^2$ from the three prong configuration.

\subsubsection*{Solving DE in the regular domain $m_X^2>0$}  

The solution of eq.~\eqref{eq:def:de} observes the form of the path-ordered integrals, 
\begin{align}\label{eq:def:de:sol}
&  \vec{f}_{i}(\alpha,y,\bar y; \varepsilon)=\mathbf{P}\exp\Big(\varepsilon\int_{\gamma(\xi_0)}^{\gamma(\xi)}  \mathrm{d}\mathbf{A}^{[i]}\Big)\,\vec{f}_{i}(\alpha_0,y_0,\bar y_0; \varepsilon)\,,
\end{align}
where $\gamma$ characterizes a path in the space of $\xi\in\{\alpha,y,\bar y\}$, starting from $\gamma(\xi_0)=(\alpha_0,y_0,\bar y_0)$ and ending at $\gamma(\xi)=(\alpha,y,\bar y)$. 
$\mathbf{P}$ stands for the path-ordered operator, always placing the integrands closer to the end point to the left,
\begin{align}
\mathbf{P}\exp\Big(\varepsilon\int_{\gamma(\xi_0)}^{\gamma(\xi)}  \mathrm{d}\mathbf{A}^{[i]}\Big)\equiv 1+\varepsilon\int_{\gamma(\xi_0)}^{\gamma(\xi)}  \mathrm{d}\mathbf{A}^{[i]}(\xi')+\varepsilon^2\int_{\gamma(\xi_0)}^{\gamma(\xi)}  \mathrm{d}\mathbf{A}^{[i]}(\xi')\int_{\gamma(\xi_0)}^{\gamma(\xi')}  \mathrm{d}\mathbf{A}^{[i]}(\xi'')+\dots \,.
\end{align}
The total differential $ \mathrm{d}\mathbf{A}^{[i]}$ situated in the exponent is defined as 
\begin{align} 
\label{eq:def:dA:tot}
 \mathrm{d}\mathbf{A}^{[i]}(\xi) \equiv  \mathcal{A}_{y}^{[i]}(\alpha,y,\bar y)\mathrm{d}y +\mathcal{A}_{\bar y}^{[i]}(\alpha,y,\bar y)\mathrm{d}\bar y+\mathcal{A}_{\alpha}^{[i]}(\alpha,y,\bar y)\mathrm{d}\alpha \,.
 \end{align}
 Considering that all the letters in eqs.~(\ref{eq:def:letters:top12}-\ref{eq:def:letters:top5}) are finite in the regular domain $m_X^2>0$, the perturbative expansion in $\varepsilon$ has been performed in the second step of eq.~\eqref{eq:def:de:sol}, leading to the Chen iterated integrals \cite{Chen:1977oja} after applying the path ordering. 
 
From eq.~\eqref{eq:def:de:sol}, we are capable of solving eq.~\eqref{eq:def:de} recursively. In practice, we choose a piecewise path ${\gamma(\xi)}$ to connect   the boundary position $\xi_0\in\{\alpha_0=0,y_0=0,\bar y_0=0\}$ and  the destination $\xi\in\{\alpha,y,\bar y\}$, more specifically, 
\begin{align}\label{eq:def:de:wt1}
 \int_{\gamma(\xi_0)}^{\gamma(\xi)}  \mathrm{d}\mathbf{A}^{[i]}(\xi')=\int_0^y\mathrm{d}y' \mathcal{A}^{[i]}_y(0,y',0)+\int_0^{\bar{y}}\mathrm{d}\bar{y}' \mathcal{A}^{[i]}_{\bar{y}}(0,y,\bar y')+\int_0^{\alpha}\mathrm{d}\alpha' \mathcal{A}^{[i]}_{\alpha}(\alpha',y,\bar y)\,.
  \end{align}
  Evaluating the one fold integrals on the r.h.s of eq.~\eqref{eq:def:de:wt1} and summing up the results from each segment, we arrive at the leading order ($\mathtt{weight}\!=\!1$)  correction in $\varepsilon$ to eq.~\eqref{eq:def:de:sol}. The second order contribution ($\mathtt{weight}\!=\!2$) can be appraised from the integration of the product of eq.~\eqref{eq:def:de:wt1} and  eq.~\eqref{eq:def:dA:tot} along the path $\gamma(\xi)$ once again, 
   \begin{align}\label{eq:def:de:wt2}
 &\int_{\gamma(\xi_0)}^{\gamma(\xi)}  \mathrm{d}\mathbf{A}^{[i]}(\xi')\int_{\gamma(\xi_0)}^{\gamma(\xi')}  \mathrm{d}\mathbf{A}^{[i]}(\xi'')\nonumber\\
=&\int_0^{\alpha}\mathrm{d}\alpha' \mathcal{A}^{[i]}_{\alpha}(\alpha',y,\bar y)\bigg[\int_0^y\mathrm{d}y' \mathcal{A}^{[i]}_y(0,y',0) +\int_0^{\bar{y}}\mathrm{d}\bar{y}' \mathcal{A}^{[i]}_{\bar{y}}(0,y,\bar y')+\int_0^{\alpha'}\mathrm{d}\alpha^{\prime\prime} \mathcal{A}^{[i]}_{\alpha}(\alpha^{\prime\prime} ,y,\bar y)\bigg] \nonumber\\
&+\int_0^{\bar{y}}\mathrm{d}\bar{y}'\mathcal{A}_{\bar y}^{[i]}(0,y,\bar y')  \bigg[\int_0^{y}\mathrm{d}y^{\prime} \mathcal{A}^{[i]}_y(0,y',0) +\int_0^{\bar{y}'}\mathrm{d}\bar{y}^{\prime\prime} \mathcal{A}^{[i]}_{\bar{y}}(0,y,\bar y^{\prime\prime})\bigg] \nonumber\\
&+\int_0^y\mathrm{d}y' \mathcal{A}^{[i]}_y(0,y',0)  \int_0^{y'}\mathrm{d}y^{\prime\prime} \mathcal{A}^{[i]}_y(0,y^{\prime\prime},0) \,.
  \end{align}
Repeating this procedure allows us to access higher weight. In this work, we  figure  out the analytic expressions up to $\mathtt{weight}\!=\!5$ and  recast them in terms of Goncharov polynomials (GPL) \cite{Goncharov:1998kja,Goncharov:2001iea} by means of the definition
   \begin{align}\label{eq:def:GPL}
G(a_1,\dots, a_n; z)=\int^z_0\frac{\mathrm{d}t}{t-a_1}G(a_2,\dots, a_n; z)\,,
  \end{align}
where   $G(z)=1$ and the arguments $a_i$ and $z$ could be complex. This process is straightforward due to the absence of the quadratic terms from the alphabets in eqs.~(\ref{eq:def:letters:top12}-\ref{eq:def:letters:top5}). During the calculation, the packages \texttt{PolyLogTools} \cite{Duhr:2019tlz} and \texttt{GiNaC} \cite{Bauer:2000cp} are utilized for the GPL manipulations and also their numeric evaluations. 
  
The remaining task is to determine the initial condition $\vec{f}_{i}(\alpha_0,y_0,\bar y_0; \varepsilon)$ in eq.~\eqref{eq:def:de:sol}. To this end,  the softwares \texttt{AMFlow} \cite{Liu:2022chg,Liu:2017jxz}
and \texttt{FiniteFlow} \cite{Peraro:2019svx} are employed to compute the master integrals $\mathcal{I}_1$-$\mathcal{I}_9$ of eqs.~(\ref{eq:def:MI1}-\ref{eq:def:MI9}) at a particular point in the physical phase space in eq.~\eqref{eq:def:phy:ps}  with the setup \texttt{Precision=100}.  The numerical outputs for $\mathcal{I}_1$-$\mathcal{I}_7$ are subsequently compared against our solutions in eqs.~(\ref{eq:def:de:wt1}-\ref{eq:def:de:wt2}) and also those at higher weights, from which we solve for $\vec{f}_{i}(\xi_0; \varepsilon)$ up to $\mathcal{O}(\varepsilon^4)$. With the aid of the \texttt{PSLQ} algorithm \cite{PSLQ1,PSLQ2}, we  fit  $\vec{f}_{i}(\xi_0; \varepsilon)$ in terms of the Riemann zeta functions $\zeta_k$ and transcendental constants $\ln^k(2)$. Ultimately,  the inverse transformation is applied onto $\vec{f}_{i}(\xi; \varepsilon)$ to evaluate the master integrals $\mathcal{I}_1$-$\mathcal{I}_9$.
We have checked that, as for a few randomly selected points in the regular domain $m_X^2>0$, the expressions for $\mathcal{I}_1$-$\mathcal{I}_9$  in this paper are able to reproduce at least the first $\sim90$ significant digits in the outputs from \texttt{AMFlow} and \texttt{FiniteFlow}.

Hereafter, we will refer to these master integral expressions derived in the regular domain as
 \begin{align}   
 \label{eq:def:MI:reg:CRR}
\mathcal{I}_i\Bigg|_{m_X^2>0}=\mathcal{I}_i^{\mathrm{reg}}(x_1,x_2,z;\varepsilon)\,.
\end{align}

\subsubsection*{Resummation in the three-prong limit $m_X^2\to0$}

In the following,    the asymptotic limit of the  differential equations  will be exploited to derive the higher order $\varepsilon$-terms for  the master integrals $\mathcal{I}_i$.  At this moment, it merits recalling that the kinematic variables in eqs.~(\ref{eq:def:para:123}-\ref{eq:def:para:5}) all admit the relation
\begin{align}    
\label{eq:def;mx:yyb}
m_X^2=y\bar{y}\,.
\end{align}
Within the physical domain of eq.~\eqref{eq:def:phy:ps}, while the variable $\bar{y}$ always stays away from zero, $y$  could vanish as $m_X^2$ decreases.   Therefore,    the asymptotic  expansion of $\mathcal{I}_i$ in the three-prong limit $m_X^2\to0$ is equivalent to that neighbouring $y=0$.  

To study the leading contribution of $\mathcal{I}_i$ in the vicinity of $y=0$, we expand the differential equation of eq.~\eqref{eq:def:de} in the variable $y$, and only retain its  singular term for each topology. It yields that,
\begin{align}    
\label{eq:def:A1:y0}
\lim_{y\to0}\mathcal{A}^{[1]}_{y}(\alpha,y,\bar y) 
=&
\left[
\begin{array}{c c}
-\frac{1}{y}& 0 \\ [0.2cm]
+\frac{8}{y} & 0
\end{array}
\right]
+\mathcal{O}(y^0),
\\
\label{eq:def:A2:y0}
\lim_{y\to0}\mathcal{A}^{[2]}_{y}(\alpha,y,\bar y) 
=& 
\left[
\begin{array}{c c}
-\frac{1}{y}& 0 \\ [0.2cm]
0& -\frac{2}{y} 
\end{array}
\right]
+\mathcal{O}(y^0),
\\
\label{eq:def:A3:y0}
\lim_{y\to0}\mathcal{A}^{[3]}_{y}(\alpha,y,\bar y) 
 =&
\left[
\begin{array}{c c c}
-\frac{1}{y}& 0 &0 \\ [0.2cm]
0 & -\frac{1}{y} &0 \\ [0.2cm]
-\frac{2}{y}&   \frac{2}{y} &0
\end{array}
\right]
+\mathcal{O}(y^0),
\\
\label{eq:def:A5:y0}
\lim_{y\to0}\mathcal{A}^{[5]}_{y}(\alpha,y,\bar y) 
=&
\left[
\begin{array}{c c c}
-\frac{1}{y}& 0 &0 \\ [0.2cm]
0 & -\frac{1}{y} &0 \\ [0.2cm]
 \frac{2}{y}&   \frac{2}{y} &0
\end{array}
\right]
+\mathcal{O}(y^0).
\end{align}
Herein, the differential equations with respect to $y$ are reduced to a  diagonal form as to the first few blocks, from which we can solve for the leading term in $\vec{f}_{i}(\xi; \varepsilon)$, 
 \begin{align}  
 \label{eq:def:y0:app1}
\lim_{y\to0} \vec{f}^{(j)}_{i}(\xi; \varepsilon)= c_i^{(j)}(\bar{y},\alpha;\varepsilon) y^{- n \varepsilon}+\mathcal{O}(y)\,.
\end{align}
Therein,  $ {f}^{(j)}_{i}$ denotes the $j$-th component of the solution $\vec{f}_{i}$ in the $i$-th topology. The indices $i$  and $j$ run over  
\begin{align} 
\label{eq:fij:diag}
 (i,j)\in\{(1,1),(2,1),(2,2),(3,1),(3,2),(5,1),(5,2)\}.
\end{align}
 The integer $n$ participating into the exponent on the r.h.s of eq.~\eqref{eq:def:y0:app1} is determined by the expression of $\mathcal{A}^{[i]}_{y}$ in eqs.~(\ref{eq:def:A1:y0}-\ref{eq:def:A5:y0}). 
$c_i^{(j)}$ represent functions of the kinematic variables $\bar{y}$ and $\alpha$. 
To calibrate them, we expand the r.h.s of eq.~\eqref{eq:def:y0:app1} in $\varepsilon$ in the first place, and then compare each power of $\varepsilon$ with its counterpart of eq.~\eqref{eq:def:MI:reg:CRR} in the limit $y\to0$. 
Over the course, the function \texttt{ToFibrationBasis} from \texttt{PolyLogTools} \cite{Duhr:2019tlz} is utilized to sequence the arguments in GPL, such that $y$ always appears as the last argument. The reshuffled GPLs are subsequently processed by the built-in function \texttt{ExpandPolyLogs} \cite{Duhr:2019tlz} to extract the leading contribution in the limit $y\to0$. 

To calculate components beyond eq.~\eqref{eq:fij:diag}, we plug the results of eq.~\eqref{eq:def:y0:app1} back into the differential equation in eq.~\eqref{eq:def:de}, and then take the integrals over $y$ on the both sides. It follows that, 
 \begin{align}  
 \label{eq:def:asyexp:f1^2}
\lim_{y\to0} {f}_{1}^{(2)}(\xi; \varepsilon)=&-8c_1^{(1)}(\bar{y},\alpha;\varepsilon) y^{-\varepsilon}  +\mathcal{O}(y )\,,\\
 \label{eq:def:asyexp:f3^3}
\lim_{y\to0}  {f}_{3}^{(3)}(\xi; \varepsilon)=&2c_3^{(1)}(\bar{y},\alpha;\varepsilon) y^{-\varepsilon}-2c_3^{(2)}(\bar{y},\alpha;\varepsilon) y^{-\varepsilon}  +\mathcal{O}(y )\,,\\
 \label{eq:def:asyexp:f5^3}
\lim_{y\to0}  {f}_{5}^{(3)}(\xi; \varepsilon)=&-2c_5^{(1)}(\bar{y},\alpha;\varepsilon) y^{-\varepsilon}-2c_5^{(2)}(\bar{y},\alpha;\varepsilon) y^{-\varepsilon}  +\mathcal{O}(y )\,,
\end{align}
where the coefficient functions $c_i^{(j)}$ have been all solved in eq.~\eqref{eq:def:y0:app1}.  During the calculation, we adhere to the initial condition
 \begin{align}  
 \label{eq:def:ini:condi:asyexp}
 {f}_{1}^{(2)}(\bar y,y=0,\alpha; \varepsilon)= {f}_{3}^{(3)}(\bar y,y=0,\alpha;\varepsilon)={f}_{5}^{(3)}(\bar y,y=0,\alpha; \varepsilon)=0\,,
 \end{align}
 where $\bar y$ and $\alpha$ are taken from the physical domain illustrated in eq.~\eqref{eq:def:phy:ps}. Eq.~\eqref{eq:def:ini:condi:asyexp} can be a priori inferred from the radial integrals in eqs.~(\ref{eq:def:MI1}-\ref{eq:def:MI6}), 
 \begin{align}   
 \label{eq:def:aly:scale}
& \int d^dk_c \,\delta_+(k_c^2) \delta_+( (q_{\gamma}-\tilde{k}_i-\tilde{k}_j-k_c)^2)=\int \frac{d^{d-1}\vec{k}_c}{2E_c}\,\frac{d^{d-1}\vec{k}_d}{2E_d}\,\delta^d(q_{\gamma}-\tilde{k}_i-\tilde{k}_j-k_c-k_d)\nonumber\\
\propto&\, \int d\widetilde{\Omega}_c\, (m_X^2)^{-\varepsilon}\,,
 \end{align}
where $ \widetilde{\Omega}_c$ stands for the solid angle for the parton $c$.  In the last step,    the rest reference frame of the particles $c$ and $d$ is chosen to integrate out the delta function together with the radial distance. 
Thinking of the equivalence between  $m_X\to0$ and  $y\to0$ limits,   the canonical bases $\vec{f}_i$ of the master integrals $\mathcal{I}_k$ always yield vanishing results at the point $y=0$ as a result of the dimensional regularization scheme. 

In practice, we also tried to assume non-trivial boundary conditions beforehand for eq.~\eqref{eq:def:ini:condi:asyexp}, and then matched onto the regular solutions from eq.~\eqref{eq:def:MI:reg:CRR}, i.e.,
 \begin{align}  
 \label{eq:def:ini:condi:asyexp2:a}
& {f}_{1}^{(2)}(\bar y,y=0,\alpha; \varepsilon)= {c}_{1}^{(2)}(\bar y,\alpha; \varepsilon)\,,\quad 
 {f}_{3}^{(3)}(\bar y,y=0,\alpha;\varepsilon)= {c}_{3}^{(3)}(\bar y,\alpha; \varepsilon)\,,\\
  \label{eq:def:ini:condi:asyexp2:b}
 &{f}_{5}^{(3)}(\bar y,y=0,\alpha; \varepsilon)= {c}_{5}^{(3)}(\bar y,\alpha; \varepsilon)\,.
 \end{align}
However, we find that, at least up to $\mathcal{O}(\varepsilon^4)$, all the $c_{i}^{(j)}$ terms in eqs.~(\ref{eq:def:ini:condi:asyexp2:a}-\ref{eq:def:ini:condi:asyexp2:b}) have dropped  out. 

At last, we transform the solutions of differential equations from their canonical space to the master integral one, arriving at the resummation-improved expressions for eqs.~(\ref{eq:def:MI1}-\ref{eq:def:MI6}) in the three-prong regime $m_X^2\to0$,
\begin{align}   
  \label{eq:def:MI:asyexp:CRR}
\mathcal{I}_i\Bigg|_{m_X^2\to0}=\mathcal{I}_i^{\mathrm{res}}(x_1,x_2,z;\varepsilon)\,(m_X^2)^{-n_i\varepsilon}\,.
\end{align}
At variance with eq.~\eqref{eq:def:MI:reg:CRR}, the expressions here possess extra factor$\sim (m^2_X)^{-n_i\varepsilon}(n_i\neq0)$,  which will regularize all the singular behaviour $\propto m_X^{-2}$ from the IBP coefficients $d_{ij}^{[k]}$ in eq.~\eqref{eq:def:IBP:coeffs}.

\subsubsection{Combination and overlapping subtraction}\label{subsubsec:combination_and_subtraction}
With the master integrals in eq.~\eqref{eq:def:MI:reg:CRR} and eq.~\eqref{eq:def:MI:asyexp:CRR} at hand, we are ready to calculate the    coefficients in eqs.~(\ref{eq:def:Cqqb:RR}-\ref{eq:def:Cqq:RR}) now.  At this point, particular attention should be paid to the fact that the regular contribution $m_X^2>0$ has been accommodated in both expressions.  Hence, combining eq.~\eqref{eq:def:MI:reg:CRR} and eq.~\eqref{eq:def:MI:asyexp:CRR} necessitates consistent subtractions of the overlapping contribution  in between.  To this end, we organize the ingredients in the following way, 
 \begin{align}   \label{eq:def:combine:RR}
\mathcal{C}_{ij;q}^{\mathrm{RR}}\,=&\,\mathcal{C}_{ij;q,\mathrm{reg}}^{\mathrm{RR}}\,+\,\mathcal{C}_{ij;q,\mathrm{asy}}^{\mathrm{RR}}\,-\,\mathcal{C}_{ij;q,\mathrm{sub}}^{\mathrm{RR}}\,,
\end{align}
where $\mathcal{C}_{ij;q,\mathrm{reg}}^{\mathrm{RR}}$ represents the regular contribution from $m_X^2>0$. $\mathcal{C}_{ij;q,\mathrm{asy}}^{\mathrm{RR}}$ captures the resummation-improved  asymptotic behaviour in the limit $m_X^2\to0$. Their  overlapping contributions are excluded by $\mathcal{C}_{ij;q,\mathrm{sub}}^{\mathrm{RR}}$ in the third term. The expression of them read,
 \begin{align} 
   \label{eq:def:combine:RR:reg}
\mathcal{C}_{ij;q,\mathrm{reg}}^{\mathrm{RR}}=&\theta(m_X^2)\,e^{-\varepsilon L_h}\,\sum_{k=1}^9d^{[k]}_{ij;q}(x_1,x_2,z;\varepsilon)\,\mathcal{I}_{k}^{\mathrm{reg}}(x_1,x_2,z;\varepsilon) \,,\\
  \label{eq:def:combine:RR:asy}
\mathcal{C}_{ij;q,\mathrm{asy}}^{\mathrm{RR}}=&\theta(m_X^2)\,e^{-\varepsilon L_h}\,\sum_{k=1}^9d^{[k],(-1)}_{ij;q}(x_1,\tilde{x}_2, z ;\varepsilon)\,\mathcal{I}_k^{\mathrm{res}}(x_1,\tilde{x}_2,z;\varepsilon) \bigg\{-\frac{\delta[m_X^2]}{n_k\varepsilon} +\left[\frac{1}{m_X^2}\right]_*\nonumber\\
&-n_k\varepsilon\left[\frac{\ln(m_X^2)}{m_X^2}\right]_*+\dots\bigg\}\,,\\
  \label{eq:def:combine:RR:sub}
\mathcal{C}_{ij;q,\mathrm{sub}}^{\mathrm{RR}}=&\theta(m_X^2)\,e^{-\varepsilon L_h}\,\sum_{k=1}^9d^{[k],(-1)}_{ij;q}(x_1,\tilde{x}_2,z;\varepsilon)\,\mathcal{I}_k^{\mathrm{res}}(x_1,\tilde{x}_2,z;\varepsilon)\bigg\{\frac{1}{m_X^2}\nonumber\\
&-  n_k\varepsilon\,\frac{ \ln(m_X^2)}{m_X^2} +\dots\bigg\}\,.
\end{align}
Therein,  $\theta(m_X^2)$ is  to impose the kinematic constraint of eq.~\eqref{eq:def:phy:ps}. $d^{[k]}_{ij;q}$  encodes the IBP coefficients in front of the $k$th-master integral as in eqs.~\eqref{eq:def:IBP:coeffs}. 
 $d^{[k],(-1)}_{ij;q}$   arises from the expansion of $d^{[k]}_{ij;q}$ around $x_2=\tilde{x}_2$ in $x_2$, more specifically, 
 \begin{align}   
 \label{eq:def:dkij:exp}
\lim_{x_2\to\tilde{x}_2}d^{[k]}_{ij;q}(x_1,x_2,z;\varepsilon ) =\sum_{l}\left({m_X^2}\right)^l \,d^{[k],(l)}_{ij;q}(x_1,\tilde{x}_2,z;\varepsilon)\,.
\end{align}
Here we define 
 \begin{align}   
 \label{eq:def:tildex2}
\tilde{x}_2\equiv \frac{1-x_1}{1-zx_1}
\end{align}
 and also utilised the relation below to carry out the expansion on the r.h.s of eq.~\eqref{eq:def:dkij:exp},
 \begin{align}   {x_2\to\tilde{x}_2-\frac{m_X^2}{1-zx_1}}\,.\end{align}
In writing eq.~\eqref{eq:def:combine:RR:asy}, we have expanded $(m_X^2)^{-n_i\varepsilon}$ in $\varepsilon$ for each singular coefficient, which ends up with star distributions, akin to those proposed in \cite{DeFazio:1999ptt,Bosch:2004th}, 
 \begin{align}   
 \label{eq:def:starD}
&\int^{\tau}_0 dm_X^2 f(m_X^2) \left[\frac{\ln^k(m_X^2)}{m_X^2}\right]_*\equiv\int_0^{\tau} \frac{dm_X^2}{m_X^2} \Big[f(m_X^2)-f(0)\Big]{\ln^k(m_X^2)}+ \frac{\ln^{k+1}(\tau)}{k+1}f(0)\,.
\end{align}
Herein $f(m_X^2)$ refers to an arbitrary test function of $m_X^2$ that is infinitely differentiable over the interval $m_X^2\in[0,\tau]$.

\subsection{Real-virtual correction}
\label{sec:def:NLO:RV}
\begin{figure}[tb]
    \centering
    \includegraphics[width=\linewidth]{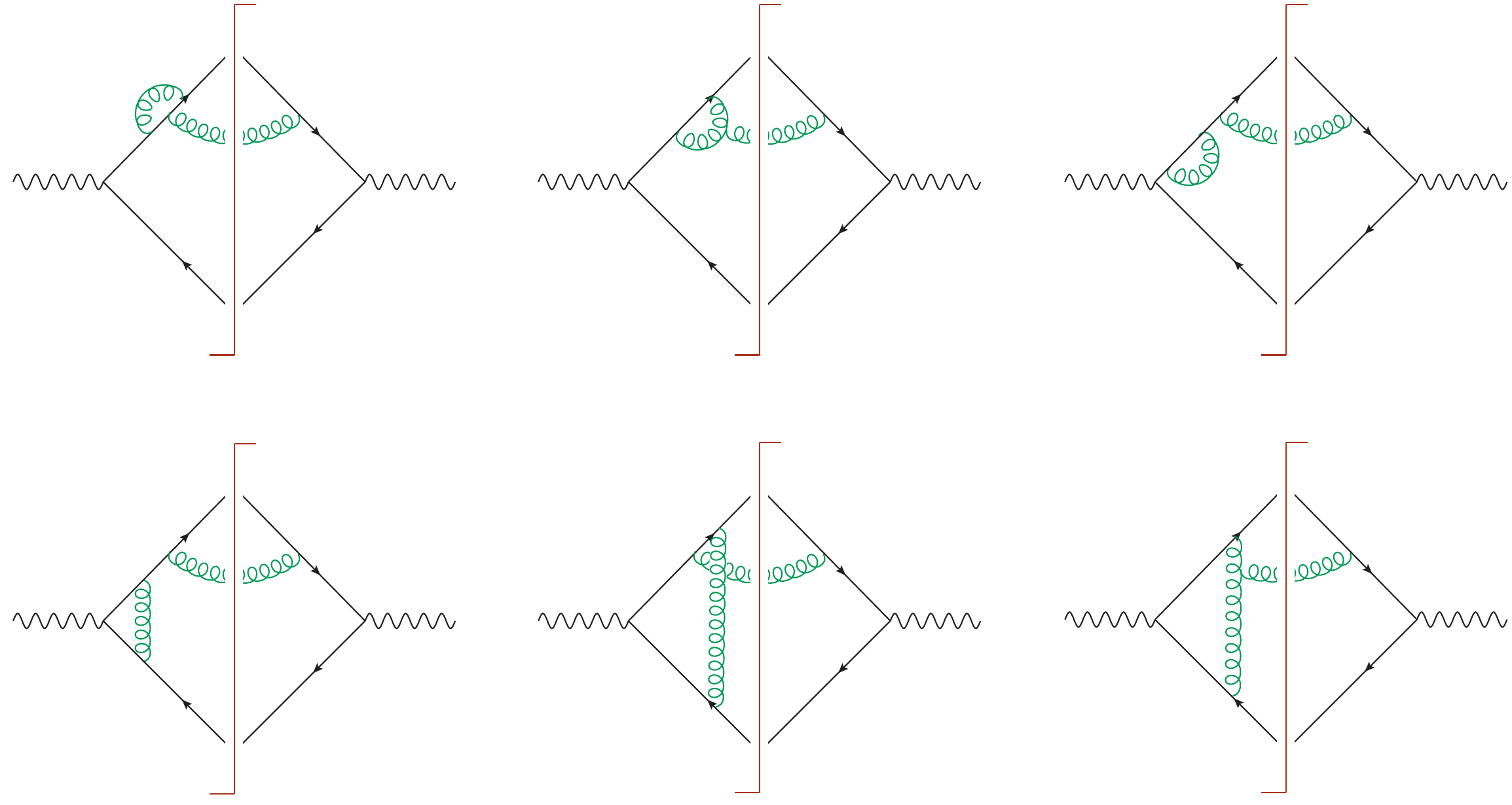}
    \caption{Typical Feynman diagrams for the real-virtual correction.}
    \label{fig:feynman_RV}
\end{figure}

The coefficient $\mathcal{C}_{ij;q}^{\mathrm{RV}}$ governing the real-virtual contribution entails the one loop amplitudes for the partonic process $\gamma^*\to q\bar{q}g$, namely, 
\begin{align} \label{eq:def:amp2:RV}   
&\left(\frac{\alpha_s}{4\pi}\right)^2\mathcal{H}^{(2)}_{\mathrm{RV}}(x_q,x_{\bar{q}};\varepsilon;\mu)
\equiv\sum_{\mathrm{pol,col}} 2\,\mathbf{Re}\Big\{\mathcal{M}^{\dagger}_{{\mathrm{B}}}(\gamma^*\to q\bar{q}g)\mathcal{M}_{{\mathrm{V}}}(\gamma^*\to q\bar{q}g)\Big\}\,  ,
\end{align}
where $x_q$ and $x_{\bar q}$ are introduced as in eq.~\eqref{eq:def:xq:xqb}. $\mathcal{M}_{{\mathrm{B}}}$ and $\mathcal{M}_{{\mathrm{V}}}$ signal the tree-level and one-loop level amplitudes, respectively. Their  Feynman diagrams are in part exhibited in fig.~\ref{fig:feynman_RV}. To calculate $\mathcal{H}^{(2)}_{\mathrm{RV}}$, the \texttt{Mathematica} packages \texttt{FeynArts}~\cite{Kublbeck:1990xc}  and~\texttt{FeynCalc}~\cite{Shtabovenko:2023idz,Shtabovenko:2020gxv,Shtabovenko:2016sxi,Mertig:1990an} are employed to generate the amplitudes, evaluate the Dirac traces and perform the tensor decomposition of one-loop Feynman integrals into the Passarino-Veltman scalar integrals \cite{Passarino:1978jh}. The analytic expressions for those scalar integrals are provided by \texttt{Package}-\texttt{X} \cite{Patel:2015tea}, which has been interfaced to~\texttt{FeynCalc} through~\texttt{FeynHelpers}~\cite{Shtabovenko:2016whf}.

In this work, we adhere to the $\overline{\mathrm{MS}}$ scheme \cite{Bardeen:1978yd} to renormalise the UV divergence from the strong coupling,
\begin{align}
\label{eq:def:Zas}
\alpha_s^{b}\,\to\, \alpha_s(\mu^2)\,\Bigg[1- \frac{\alpha_s(\mu^2)}{4\pi\varepsilon} \left(\frac{11}{3}C_A-\frac{2}{3}N_f\right)+\mathcal{O}(\alpha_s^2)\Bigg]\,,
\end{align}
where  $C_A=N_C$ stands for the color factor in QCD.  Plugging the renormalised strong couplings above into the LO squared amplitudes in eq.~\eqref{eq:def:amp2:LO}, we obtain the counter term for the UV divergence in the real-virtual contribution,
\begin{align} \label{eq:def:amp2:RV:CT}   
& \mathcal{H}^{(2)}_{\mathrm{CT}}(x_q,x_{\bar{q}};\varepsilon;\mu)
=- \frac{1}{ \varepsilon} \left(\frac{11}{3}C_A-\frac{2}{3}N_f\right)\mathcal{H}^{(1)}(x_q,x_{\bar{q}};\varepsilon;\mu)\,  .
\end{align} 
We have checked the sum of eqs.~\eqref{eq:def:amp2:RV} and \eqref{eq:def:amp2:RV:CT}  indeed replicates the result of~\cite{Ellis:1980wv} upon synchronizing the choice of the renormalization scale.

Eventually, we plug eqs.~\eqref{eq:def:amp2:RV} and \eqref{eq:def:amp2:RV:CT}  back into eq.~\eqref{eq:def:Cij} and then integrate out the momenta of the identified partons via Dirac delta functions. It follows that
\begin{align}    
\label{eq:def:sig:qqb:RV}
 \hat{\sigma}_q \, \mathcal{C}^{ \mathrm{RV}}_{q\bar{q};q}(x_1,x_2;\varepsilon;\mu)&=   \frac{(2\pi) \delta\left(m_X^2\right)}{Q^2}\, \kappa(z;\varepsilon)\,\left( x_1x_2\right)^{1-2\varepsilon}\,\Big[\mathcal{H}^{(2)}_{\mathrm{RV}}(x_1,x_2;\varepsilon;\mu)\,\nonumber\\
  &+\mathcal{H}^{(2)}_{\mathrm{CT}}(x_1,x_2;\varepsilon;\mu)\Big]  \,,
  \\
  \label{eq:def:sig:qg:RV}
\hat{\sigma}_q \,\mathcal{C}^{\mathrm{RV}}_{qg;q}(x_1,x_2;\varepsilon;\mu)&=\hat{\sigma}_q \, \mathcal{C}^{\mathrm{RV}}_{g\bar{q};q}(x_2,x_1;\varepsilon;\mu)\nonumber\\
&=   \frac{(2\pi)\delta\left(m_X^2\right)}{Q^2}\, \kappa(z;\varepsilon)\,\left( x_1x_2\right)^{1-2\varepsilon} \,\Big[\mathcal{H}^{(2)}_{\mathrm{RV}}(x_1,2-x_1-x_2;\varepsilon;\mu)  \nonumber\\
&+\mathcal{H}^{(2)}_{\mathrm{CT}}(x_1,2-x_1-x_2;\varepsilon;\mu) 
\Big]\,.
  \end{align}
\label{eq:def:sig:qbg:RV}
The results for $ \mathcal{C}^{ \mathrm{RV}}_{\bar{q}q;q}$, $ \mathcal{C}^{ \mathrm{RV}}_{gq;q}$, and $ \mathcal{C}^{ \mathrm{RV}}_{\bar{q}g;q}$ can be derived by 
swapping the momentum fractions $x_1\leftrightarrow x_2$ in eqs.~(\ref{eq:def:sig:qqb:RV}-\ref{eq:def:sig:qbg:RV}) as appropriate.

\subsection{Fragmentation function renormalization}
\label{sec:def:NLO:renFFs}

The fragmentation functions in eq.~\eqref{eq:def:fac} are subject to the renormalization up to $\mathcal{O}(\alpha_s)$, 
\begin{align}
\label{eq:def:ff:renor}
D^{b}_{H/i}\left(x\right)=\frac{\alpha_s(\mu)}{4\pi\varepsilon}\, \sum_k \int^1_{x} \frac{\mathrm{d} y}{y}D_{H/k}\left(\frac{x }{y },\mu\right) \, P^{\mathrm{T},(0)}_{ki}(y )+\mathcal{O}(\alpha_s^2)\,,
\end{align}
where $D^{b}_{H/i}$ on the l.h.s stands for the bare fragmentation function.  The renormalized one is denoted by $D_{H/k}$ on the r.h.s with the explicit dependence on the collinear factorization scale $\mu$. 
$P^{\mathrm{T},(0)}_{ij}$ represents the LO time-like splitting function \cite{Altarelli:1977zs}, which can be generally expressed as 
\begin{align}
\label{eq:def:sp:dec}
P^{\mathrm{T},(0)}_{ij}(x)=P^{\mathrm{T},(0)}_{ij,\mathrm{reg}}(x)+P^{\mathrm{T},(0)}_{ij,\delta}\,\delta\left(1-x\right)+P^{\mathrm{T},(0)}_{ij,+}\,\left[\frac{1}{1-x}\right]_+\,.
\end{align}
Herein, the plus distribution is defined as 
\begin{align}
\int^{1}_{\tau}\mathrm{d}x f(x)\left[\frac{1}{1-x}\right]_+\equiv
\int^{1}_{\tau}\mathrm{d}x  \frac{f(x)-f(1)}{1-x} +f(1)\ln(1-\tau)\,,
\end{align}
where $f(x)$ denotes a test function of $x$ that is analytic over the domain $[\tau,1]$.

In Eq.~\eqref{eq:def:sp:dec}, $P^{\mathrm{T},(0)}_{ij,\mathrm{reg}}(x)$ encodes the regular contribution for the splitting function at LO, while $P^{\mathrm{T},(0)}_{ij,\delta}$ and $P^{\mathrm{T},(0)}_{ij,+}$ are the constant coefficients in front of the delta function and plus distribution, respectively. We have summarized the expression of $P^{\mathrm{T},(0)}_{ij}$ for each partonic channel in Sec.~\ref{eq:sec:app:sp}.

Combining eq.~\eqref{eq:def:ff:renor} with the LO    coefficients $\mathcal{C}_{ij;q}^{(1)}$ in eq.~\eqref{eq:def:sig:All:LO} leads to the counter terms for the NLO    coefficients, i.e., $\mathcal{C}_{ij;q}^{\mathrm{CT}}$. According to the contributing sectors in the splitting function of eq.~\eqref{eq:def:sp:dec}, $\mathcal{C}_{ij;q}^{\mathrm{CT}}$ can be categorised into three groups plus their counterparts subjecting to $H_1\leftrightarrow H_2$ substitutions, 
\begin{align}
\label{eq:def:cT:all}
\mathcal{C}_{ij;q}^{\mathrm{CT}}\equiv\mathcal{C}_{ij;q,\mathrm{reg}}^{\mathrm{CT}}+\mathcal{C}_{ij;q,\delta}^{\mathrm{CT}}+\mathcal{C}_{ij;q,*}^{\mathrm{CT}}+(H_1\leftrightarrow H_2)\,.
\end{align}
The computation of $\mathcal{C}_{ij;q,\mathrm{reg}}^{\mathrm{CT}}$ entails the convolution of $P^{\mathrm{T},(0)}_{ij,\mathrm{reg}}$ with $\mathcal{C}_{ij;q}^{(1)}$, more explicitly, 
\begin{align}
\label{eq:def:cT:reg1}
&\frac{\alpha_s(\mu)}{4\pi}
\sum_{i,j}\int_{\tau_1}^1\frac{\mathrm{d}x_1}{x_1} D_{H_1/i}\left(\frac{\tau_1}{x_1},\mu\right)\,
  \int_{\tau_2}^1\frac{\mathrm{d}x_2}{x_2} D_{H_2/j}\left(\frac{\tau_2}{x_2},\mu\right)\,
 \mathcal{C}_{ij;q}^{\mathrm{CT}}\left(x_1,x_2,z;\varepsilon\right)
 \nonumber\\
= &
 \sum_{i,k}\int_{\tau_1}^1\frac{\mathrm{d}x_1}{x_1} D_{H_1/i}\left(\frac{\tau_1}{x_1},\mu\right)\,
  \int_{\tau_2}^1\frac{\mathrm{d}y_2}{y_2} D^b_{H_2/k}\left(\frac{\tau_2}{y_2}\right)\,
 \mathcal{C}_{ik;q}^{(1)}\left(x_1,y_2,z;
\varepsilon\right) \,.
\end{align}
At this moment,  only $P^{\mathrm{T},(0)}_{ij,\mathrm{reg}}$ is taken into account when renormalizing the fragmentation functions. 
 Thus, plugging eq.~\eqref{eq:def:ff:renor} into eq.~\eqref{eq:def:cT:reg1}, it evaluates to,
 \begin{align}
 \label{eq:def:cT:reg2}
 & \sum_k \int_{\tau_2}^1\frac{\mathrm{d}y_2}{y_2} D^b_{H_2/k}\left(\frac{\tau_2}{y_2}\right)\,
 \mathcal{C}_{ik;q}^{(1)}\left(x_1,y_2,z;\varepsilon\right)\nonumber\\
 =&\frac{\alpha_s(\mu)}{4\pi\varepsilon}\sum_{k,j}\int_{\tau_2}^1\frac{\mathrm{d}y_2}{y_2} \,
 \mathcal{C}_{ik;q,\delta}^{(1)}\left(x_1,y_2,z;\varepsilon\right)\,
 \delta\left(m_X^2\right)\int^1_{\frac{\tau_2}{y_2}} \frac{\mathrm{d} x_2}{x_2}D_{H_2/j}\left(\frac{\tau_2 }{y_2x_2 },\mu\right) \, P^{\mathrm{T},(0)}_{jk,\mathrm{reg}}(x_2 )\nonumber\\
 =&\frac{\alpha_s(\mu)}{4\pi\varepsilon}\sum_{k,j}\int^1_{\tau_2}\frac{\mathrm{d}x_2}{x_2} D_{H_2/j}\left(\frac{\tau_2 }{x_2 },\mu\right)P^{\mathrm{T},(0)}_{jk,\mathrm{reg}}\left(\frac{x_2}{\tilde{x}_2}\right)
 \frac{\mathcal{C}_{ik;q,\delta}^{(1)}\left(x_1,\tilde{x}_2,z;\varepsilon\right)}{1-x_1}\,\theta\!\left(m_X^2\right)\,.
\end{align}
In the last step, we have completed the integral over $y_2$ with the aid of $\delta(m_X^2)$, and then re-parametrized the remaining integration variable as 
$$x_2\to \dfrac{x_2}{\tilde{x}_2}\,.$$
The definition of $\tilde{x}_2$ has been presented in eq.~\eqref{eq:def:tildex2}.  
The resulting expression in eq.~\eqref{eq:def:cT:reg2} now matches that on the l.h.s of eq.~\eqref{eq:def:cT:reg1}. Substituting  eq.~\eqref{eq:def:cT:reg2} into eq.~\eqref{eq:def:cT:reg1} and  factoring out the fragmentation functions with the common mother partons,  $\mathcal{C}_{ij;q,\mathrm{reg}}^{\mathrm{CT}}$ can be extracted, i.e., 
   \begin{align}
   \label{eq:res:cT:reg}
\mathcal{C}_{ij;q,\mathrm{reg}}^{\mathrm{CT}}=\frac{1}{\varepsilon}\, \frac{\theta\!\left(m_X^2\right)}{1-x_1}\,\sum_k\,
\mathcal{C}_{ik;q,\delta}^{(1)}\left(x_1,\tilde{x}_2,z;\varepsilon\right)\,P^{\mathrm{T},(0)}_{jk,\mathrm{reg}}\left(\frac{x_2}{\tilde{x}_2}\right)\,.
  \end{align}
   
   Repeating the procedures in eqs.~(\ref{eq:def:cT:reg1}-\ref{eq:def:cT:reg2}) with the other two constituents in eq.~\eqref{eq:def:sp:dec} enables us to calculate $\mathcal{C}_{ij;q,\delta}^{\mathrm{CT}}$ and $\mathcal{C}_{ij;q,*}^{\mathrm{CT}}$,   
    \begin{align}
      \label{eq:res:cT:del}
\mathcal{C}_{ij;q,\delta}^{\mathrm{CT}}&=\frac{1}{\varepsilon} \,\sum_k\,\mathcal{C}_{ik;q,\delta}^{(1)}\left(x_1,\tilde{x}_2,z;\varepsilon\right)\,P^{\mathrm{T},(0)}_{jk,\delta}\delta(m_X^2)\,,\\
      \label{eq:res:cT:plus}
\mathcal{C}_{ij;q,*}^{\mathrm{CT}}&=\frac{\theta(m_X^2)}{\varepsilon} \,\sum_k\,\mathcal{C}_{ik;q,\delta}^{(1)}\left(x_1,\tilde{x}_2,z;\varepsilon\right)\,P^{\mathrm{T},(0)}_{jk,+}\,\left\{\left[\frac{1}{m_X^2}\right]_*-\ln(1-x_1)\delta(m_X^2)\right\}\,.
  \end{align}  
   The result in eq.~\eqref{eq:res:cT:del} is an immediate conclusion of eq.~\eqref{eq:def:cT:reg2} once $P^{\mathrm{T},(0)}_{jk,\mathrm{reg}}$ is replaced with the second term in eq.~\eqref{eq:def:sp:dec}.  To derive eq.~\eqref{eq:res:cT:plus}, the following identity has been exploited to transform the plus distribution in the third term of eq.~\eqref{eq:def:sp:dec}  to the star type as in eq.~\eqref{eq:def:starD}, 
     \begin{align}
     \label{eq:id:plus:star}
\left[\frac{1}{1-x_2'}\right]_+\Bigg|_{x_2'\to\frac{x_2}{\tilde{x}_2}}=(1-x_1)\left\{\left[\frac{1}{m_X^2}\right]_*-\ln(1-x_1)\delta(m_X^2)\right\}\,.
  \end{align}  
  The validity of eq.~\eqref{eq:id:plus:star} can be confirmed by multiplying both sides by a smooth test function of $x_2$ and integrating the resulting expression with respect to $x_2$ over the kinematic interval
$$x_2\in\left[\tau_2, \frac{1-x_1}{1-zx_1}\right].$$ 
Here the  boundaries are determined by the physical domain in eq.~\eqref{eq:def:phy:ps}.
   
Eventually, the combination of eqs.~(\ref{eq:res:cT:reg}-\ref{eq:res:cT:plus}) with eq.~\eqref{eq:def:cT:all} leads to the counter term of the    coefficients  at NLO, i.e., $\mathcal{C}_{ij;q}^{\mathrm{CT}}$. 
We have checked that for all ensuing partonic channels, our $\mathcal{C}_{ij;q}^{\mathrm{CT}}$ exactly cancel the poles in  the sum of $\mathcal{C}^{ \mathrm{RV}}_{ij;q}$ in eqs.~(\ref{eq:def:sig:qqb:RV}-\ref{eq:def:sig:qbg:RV}) and  $\mathcal{C}^{ \mathrm{RR}}_{ij;q}$ in eqs.~(\ref{eq:def:Cqqb:RR}-\ref{eq:def:Cqqp:RR}). This serves a non-trivial validation for the collinear factorization formula in eq.~\eqref{eq:def:fac} and also our    coefficients on the level of $\mathcal{O}(\alpha_s^2)$.

\section{The finite Wilson coefficients for dihadron production}
\label{sec:def:finite:coeffs}

In this section, we present the expression of the NLO coefficient $\mathcal{C}^{(2)}_{ij;q}$ in eq.~\eqref{eq:def:coeff:nlo:finite}, evaluated by summing the double real emission in eq.~\eqref{eq:def:combine:RR}, the real-virtual contributions in eqs.~(\ref{eq:def:sig:qqb:RV}-\ref{eq:def:sig:qg:RV}), and the counter terms in eq.~\eqref{eq:def:cT:all}. To facilitate the discussion, we  categorize  $\mathcal{C}^{(2)}_{ij;q}$ according to its analytic structure near $m_X^2=0$, 
\begin{align}
\label{eq:finite_wilson}
\mathcal{C}_{ij;q}^{(2)}(x_1,x_2,z;\mu)\, =\,& \theta(m_X^2)\,R_{ij;q}(x_1,x_2,z;\mu)\,+\,  V_{ij;q}(x_1,\tilde{x}_2;\mu)\delta(m_X^2) \nonumber\\
&\, +\theta(m_X^2)\sum_{k=0}^1\,U^{[k]}_{ij;q}(x_1,\tilde{x}_2;\mu)\,\Bigg(\left[\frac{\ln^k(m_X^2)}{m_X^2}\right]_* -\frac{\ln^k(m_X^2)}{m_X^2}\Bigg) \, ,
\end{align}
where $R_{ij;q}$ encapsulates the DE solutions in eq.~\eqref{eq:def:MI:reg:CRR} derived in the domain $m_X^2>0$, together with the regular term $\mathcal{C}_{ij;q,\mathrm{reg}}^{\mathrm{CT}}$ of eq.~\eqref{eq:res:cT:reg} from the FF renormalization. 
$V_{ij;q}$ and $U_{ij;q}$ consist of the renormalization improved DE solution in eq.~\eqref{eq:def:MI:asyexp:CRR},  the real-virtual contributions in eqs.~(\ref{eq:def:sig:qqb:RV}-\ref{eq:def:sig:qg:RV}), and the distribution terms $\mathcal{C}_{ij;q,\delta}^{\mathrm{CT}}$ and $\mathcal{C}_{ij;q,*}^{\mathrm{CT}}$ in eqs.~(\ref{eq:res:cT:del}-\ref{eq:res:cT:plus}). The definition of $m_X^2$ has been delivered in eq.~\eqref{eq:def:phy:ps}.  
In writing $V_{ij;q}$ and $U_{ij;q}$, the argument $\tilde{x}_2$ is entailed as in eq.~\eqref{eq:def:tildex2} due to the asymptotic expansion in eq.~\eqref{eq:def:dkij:exp}.

In the following, we focus on the partonic processes with the identified partons in the six sectors $$\{ij\}\in\{q\bar{q},qq^\prime,q^\prime\bar{q}^\prime,qg,qq,gg\}.$$ The Wilson coefficients for the other combinations of identified partons can be obtained by the relationships in eqs.~(\ref{eq:def:Cgqb}-\ref{eq:def:Cqbqbp}).
For notational simplicity, the subscript ``$q$'' in $R_{ij;q}$, $V_{ij;q}$, and $U_{ij;q}$, indicating the flavor of the quark directly coupled to the photon, will be omitted below.

\subsection{The \texorpdfstring{$m_X^2>0$}{mxsq>0} part}

\begin{figure}[tb]
    \centering
    \includegraphics[width=0.6\linewidth]{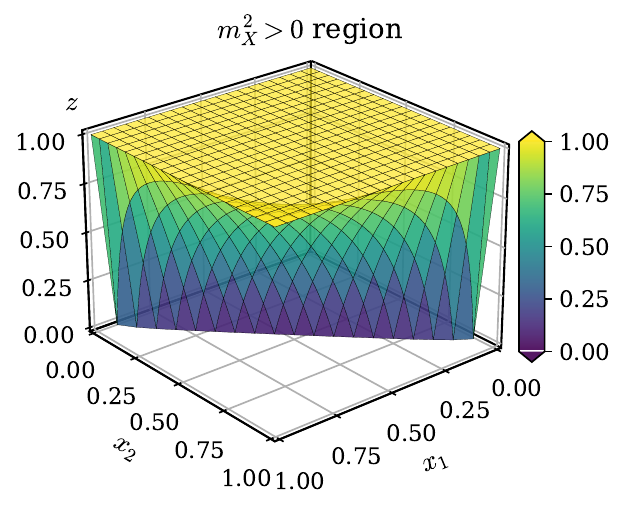}
    \caption{Kinematically allowed region in $\{x_1, x_2, z\}$ space for $m_{X}^2 > 0$, with the color bar representing the values of $z$.}
    \label{fig:region_mxp}
\end{figure}
 $R_{ij}$ governs the contribution from  the region $m_X^2>0$,  as illustrated in fig.~\ref{fig:region_mxp} from the $\{x_1,\, x_2,\, z\}$ coordinate space.  Up to NLO, 
its expression can be recast in terms of transcend\\entality-one functions, more explicitly, 
\begin{align}
\label{eq:def:Rijq:expr}
R_{ij}(x_1,x_2,z;\mu)=\sum_{k=0}^{14}  r^{(k)}_{ij}(x_1,x_2,z;\mu)\,\ell^{(k)}\,,%
\end{align}
where the bases $\ell^{(k)}$ with $k\in[1,14]$ contain 
\begin{align}
&\ell^{(0)}=1,\,\qquad\ell^{(1)}=\ln(2),\,\qquad\ell^{(3)}=\ln(1-x_1),\,\qquad\ell^{(4)}=\ln(1-x_2),\,\nonumber\\
&\ell^{(5)}=\ln(x_1),\,\qquad\ell^{(6)}=\ln(x_2),\,\qquad\ell^{(7)}=\ln(1-x_1z),\,\qquad\ell^{(8)}=\ln(1-x_2z),\,\nonumber\\
&\ell^{(9)}=\ln\Big[x_1(1-z)+x_2(1-x_1z)\Big],\,\qquad\ell^{(10)}=\ln\Big[x_2(1-z)+x_1(1-x_2z)\Big],\,\nonumber\\
&\ell^{(11)}=\ln(x_1+x_2-\beta),\,\qquad\ell^{(12)}=\ln(x_1+x_2+\beta),\,\qquad\ell^{(13)}=\ln(m_X^2),\,\nonumber\\
&\ell^{(14)}=\ln(1-z)\,.
\end{align}
In eq.~\eqref{eq:def:Rijq:expr}, the coefficient $ r^{(k)}_{ij}$ is defined as 
\begin{align}
    r^{(k)}_{ij}(x_1,x_2,z;\mu)= \mathcal{N}^{(k)}_{ij}(x_1,x_2,z;\mu)  \prod_{(m,n)\in\mathcal{S}^{(k)}_{ij}}  \left(\frac{1}{\mathcal{D}^{(m)}} \right)^{n} \,.
\end{align}
Here $\mathcal{N}^{(k)}_{ij}$ is made up of the polynomials in the kinematic variables $\{x_1,x_2,z\}$ and also the logarithm $L_h=\ln(Q^2/\mu^2)$.  $\mathcal{D}^{(m)}$ stands for the denominator resulting from IBP reduction or the Feynman propagator of the squared amplitudes. In this work, $R_{ij}$   invokes $11$ sorts of denominators in total, 
\begin{align}
\label{eq:def:dens:Rijq}
& \mathcal{D}^{(1)}=x_1,\,\qquad\mathcal{D}^{(2)}=x_2,\,\qquad\mathcal{D}^{(3)}=z,\,\qquad\mathcal{D}^{(4)}=x_1+x_2,\, \nonumber\\
& \mathcal{D}^{(5)}=1-x_1,\,\qquad\mathcal{D}^{(6)}=1-x_2,\,\qquad\mathcal{D}^{(7)}=1-z,\,\nonumber\\
& \mathcal{D}^{(8)}=1-x_1z,\,\qquad\mathcal{D}^{(9)}=1-x_2z,\,\qquad\mathcal{D}^{(10)}=m_X^2,\,\qquad\mathcal{D}^{(11)}=\beta.
\end{align}
These denominators are organized via  $\mathcal{S}^{(k)}_{ij}$, which is a set of $(m,n)$ with $m\in[1,11]$ and $n\in[0,5]$. Herein, the first index ``$m$'' specifies the type of denominator in eq.~\eqref{eq:def:dens:Rijq}, with the latter ``$n$'' indicating  the power.

In deriving eq.~\eqref{eq:def:Rijq:expr}, the \texttt{MultivariateApart}~\cite{Heller:2021qkz} and  \texttt{Singular\_pfd}~\cite{Boehm:2020ijp} packages, which implement the partial fraction algorithms for multiple variables, are utilized in this work.  The resulting  expressions for $R_{ij}$ are provided in the ancillary files of the \textit{arXiv} submission.

From our results, it is interesting to note that not all of the coefficients $r^{(k)}_{ij}$ are independent. For all partonic sectors under consideration, the coefficients accompanying $\ell^{(11)}$ and $\ell^{(12)}$ are related through
\begin{equation}
r^{(11)}_{ij}\Big|_{\beta \to -\beta} = r^{(12)}_{ij}.
\end{equation}
Moreover, as to the sectors $\{ij\} \in \{q\bar{q},\, q'\bar{q}',\, gg,\, qq\}$, the partonic coefficients $\mathcal{C}_{ij}^{(2)}=R_{ij}$ exhibit symmetry under the exchange $x_1 \leftrightarrow x_2$, which leads to the following relations among the coefficients:
\begin{equation}
r^{(3)}_{ij} = r^{(4)}_{ij}, \quad r^{(5)}_{ij} = r^{(6)}_{ij}, \quad r^{(7)}_{ij} = r^{(8)}_{ij}, \quad r^{(9)}_{ij} = r^{(10)}_{ij}.
\end{equation}
In the case of the $qg$ sector, we find that $r^{(10)}_{qg} = 0$, implying that $\ell^{(10)}$ does not contribute to its functional basis. For the $qq'$ sector, the function space of $R_{qq'}$ is restricted to the set
\begin{equation}
\{\ell^{(1)},\, \ell^{(2)},\, \ell^{(3)} - \ell^{(6)} - \ell^{(7)},\, \ell^{(11)},\, \ell^{(12)},\, \ell^{(13)}\}.
\end{equation}
On the $q'\bar{q}'$ sector, the Wilson coefficient depends solely on the two logarithmic functions $\ell^{(11)}$ and $\ell^{(12)}$, and takes the form
\begin{equation}
\mathcal{C}^{(2)}_{q'\bar{q}'} = R_{q'\bar{q}'} = r^{(11)}_{q'\bar{q}'} \,\left[\ell^{(11)} - \ell^{(12)}\right] + r^{(0)}_{q'\bar{q}'},
\end{equation}
with the constraint $r^{(11)}_{q'\bar{q}'} + r^{(12)}_{q'\bar{q}'} = 0$.

While the Wilson coefficients $R_{ij}$ remain finite within the physical phase space defined in eq.~\eqref{eq:def:phy:ps}, they can exhibit volatile asymptotic behavior in certain kinematic limits, particularly as $\{x_1,\, x_2,\, z\} \to \{0,1\}$, owning to the soft and collinear nature of QCD interactions. In what follows, we deliver a detailed analysis of these asymptotic regimes. We expect that this discussion will facilitate future studies on QCD factorization and resummation.

\subsubsection{Asymptotic behaviour of  \texorpdfstring{$R_{ij}$}{Rij} in the limits \texorpdfstring{$x_\varrho\to\{0,1\}$}{xto0or1}}

\begin{table}[tb]
    \renewcommand{\arraystretch}{1.2} 
    \caption{Leading contributions of $R_{ij}$ in the kinematic limits of $x_{1}$ and $x_{2}$ with $z\in(0,1)$. ``K.L.'' denotes kinematic limit. Here, $w_{\varrho} = 1 - x_{\varrho}$ with $\varrho =1,2$, and $h^{(l)}_{\varrho\varsigma} = (x_\varrho)^k x_\varsigma^{-l-k}$ with $\{\varrho, \varsigma\} = 1,2$ and $k = -1,0,1,2$. We further define $s^{(l)}_{\varrho\varsigma} = (x_\varrho)^k (1-x_\varsigma)^{-l-k}$ and $\tilde{s}^{(l)}_{\varrho\varsigma} = (x_\varrho)^{-l-k} (1-x_\varsigma)^k$, with $k = -1,0,1,2$ for the $qg$ and $gg$ sectors, and $k =0,1,2$ for the others.}
    \label{tab:div_fix_z}
    \centering
    \resizebox{\textwidth}{!}{
    \begin{tabular}{|c|c|c|c|c|c|c|} 
        \hline
        \diagbox{K.L.}{Leading terms}{Sectors} & $q\bar{q}$                                     & $qq'$                       & $q'\bar{q}'$                                   & $qg$                                        & $gg$                                       & $qq$ \\
        \hline
        $x_1\to 0$                               & $\ln x_1$                                      & $\ln x_1$                   & $\ln x_1$                                      & $\ln x_1$                                   & $\frac{\ln x_1}{x_1}$                      & $\ln x_1$  \\
        \hline
        $x_2\to 0$                               & $\ln x_2$                                      & $\ln x_2$                   & $\ln x_2$                                      & $\frac{\ln x_2}{x_2}$                       & $\frac{\ln x_2}{x_2}$                      & $\ln x_2$  \\
        \hline
        $x_1\to 0$                               & \multirow{2}{*}{$h^{(2)}_{\varrho\varsigma}\ln x_{1,2}$}     &\multirow{2}{*}{$\ln x_\varrho$}   & \multirow{2}{*}{$h^{(2)}_{\varrho\varsigma}\ln x_{1,2}$}     & \multirow{2}{*}{$\frac{\ln x_\varrho}{x_\varsigma}$}      & \multirow{2}{*}{$h^{(2)}_{\varrho\varsigma}\ln x_{1,2}$} & \multirow{2}{*}{$\ln x_\varrho$}  \\
        $x_2 \to 0$                              &                                                &                             &                                                &                                             &                                            &                \\
        \hline
        $x_1\to 1$                               & $s^{(2)}_{21}\ln w_1$                          & $s^{(2)}_{21}\ln w_1$       & \multirow{2}{*}{$\ln x_2$}                     & $s(\tilde{s})^{(2)}_{21}\ln w_1$            & $s(\tilde{s})^{(1)}_{21}\ln w_1$           & $s^{(2)}_{21}\ln w_1$\\
        $x_2\to 0$                               & $s^{(2)}_{21}\ln x_2$                          & $s^{(2)}_{21}\ln x_2$       &                                                & $s(\tilde{s})^{(2)}_{21}\ln x_2$            & $s(\tilde{s})^{(1)}_{21}\ln x_2$           & $s^{(2)}_{21}\ln x_2$\\                                     
        \hline
        $x_1\to 0$                               & $s^{(2)}_{12}\ln w_2$                          &$\ln w_2$                    & \multirow{2}{*}{$\ln x_1$}                     & $s(\tilde{s})^{(1)}_{12}\ln w_2$            & $s(\tilde{s})^{(1)}_{12}\ln w_2$           & $s^{(2)}_{12}\ln w_2$\\ 
        $x_2\to 1$                               & $s^{(2)}_{12}\ln x_1$                          &$\ln x_1$                    &                                                & $s(\tilde{s})^{(1)}_{12}\ln x_1$            & $s(\tilde{s})^{(1)}_{12}\ln x_1$           & $s^{(2)}_{12}\ln x_1$\\                                                              
        \hline
    \end{tabular}
    }
\end{table} 

\begin{figure}[tb]
    \centering
    \includegraphics[width=\linewidth]{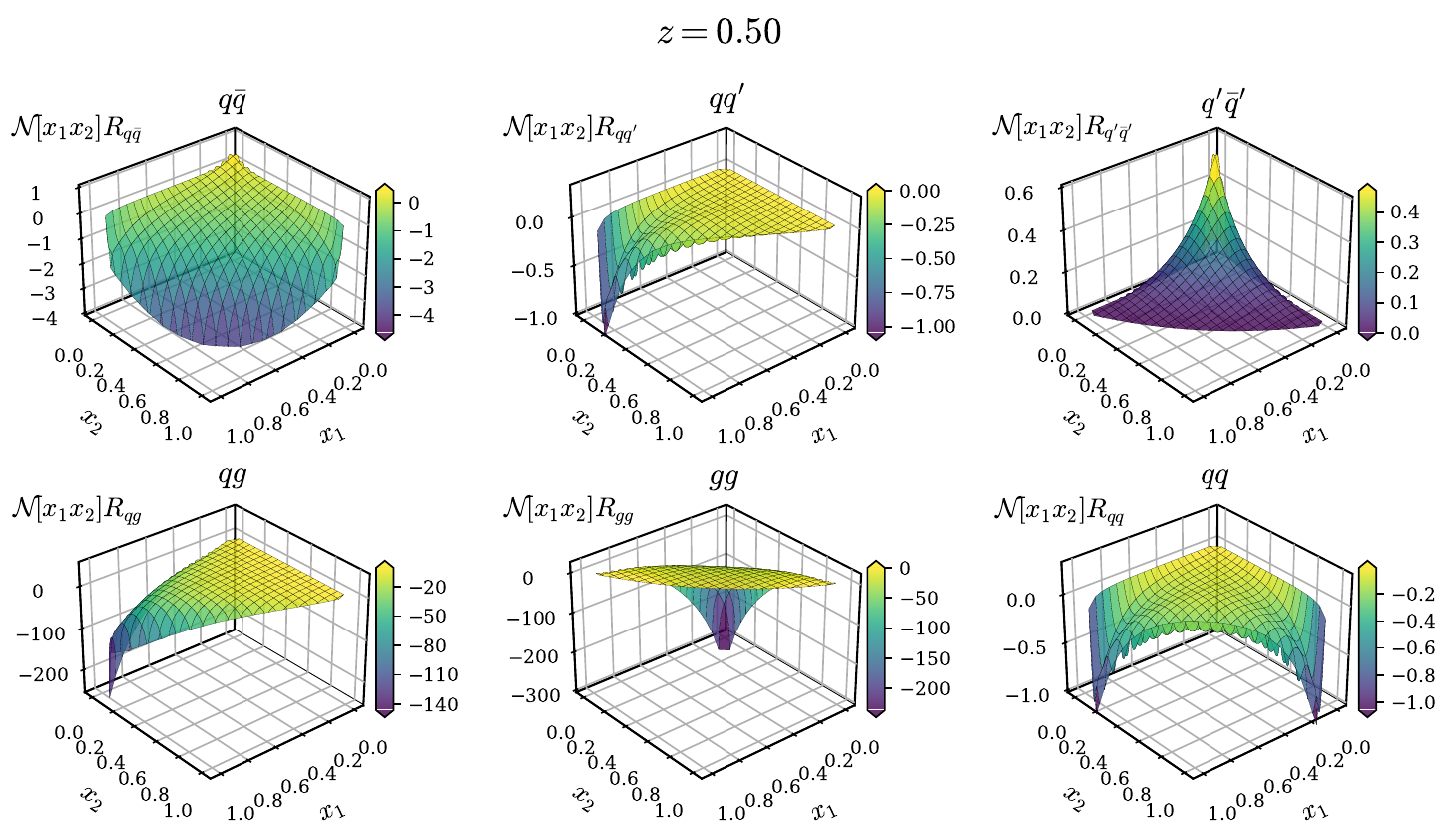}
    \caption{The Wilson coefficients for $m_X^2>0$ at $z=1/2$ as functions of $x_1,\,x_2$ in the $q\bar{q}$ (top-left), $qq^{\prime}$ (top-middle), $q^{\prime}\bar{q}^{\prime}$(top-right), $qg$ (bottom-left), $gg$ (bottom-middle), and $qq$ (bottom-right) sectors. The prefactor $\mathcal{N}[x_1x_2]=N_C x_1x_2m_X^2/(4\pi^2)$ is multiplied to mitigate the volatile asymptotic behaviors of the Wilson coefficients. The values of the $\mathcal{N}[x_1x_2]$ weighted Wilson coefficients are encoded in color.}
    \label{fig:sigmahat_xxmX_z1o2}
\end{figure}

\begin{figure}[tb]
    \centering
    \includegraphics[width=\linewidth]{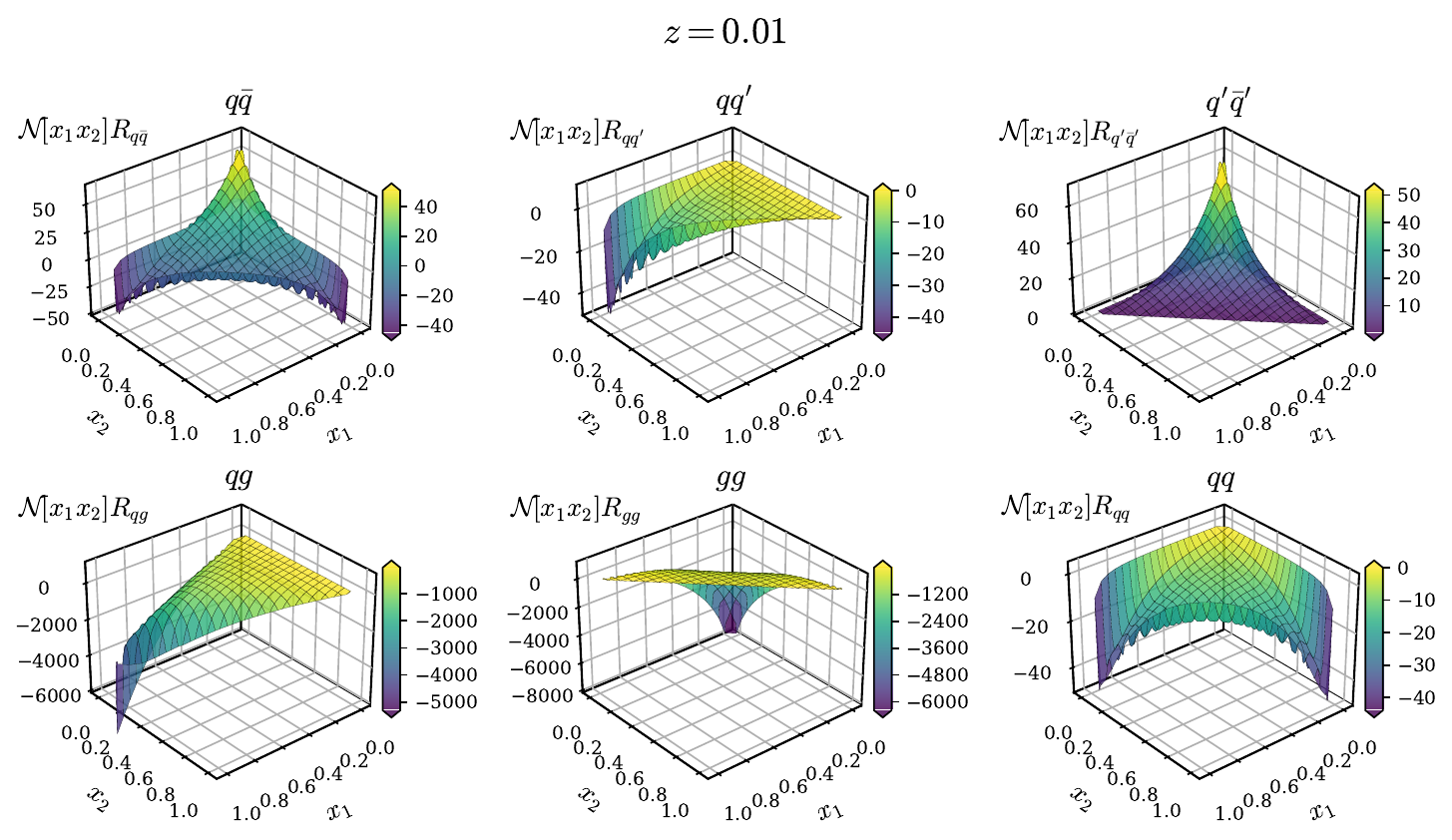}
    \caption{The Wilson coefficients for $m_X^2>0$ at $z=1/100$ as functions of $x_1,\,x_2$. The notations are same as fig.~\ref{fig:sigmahat_xxmX_z1o2}.}
    \label{fig:sigmahat_xxmX_z1o100}
\end{figure}

\begin{figure}[tb] 
    \centering
    \includegraphics[width=\linewidth]{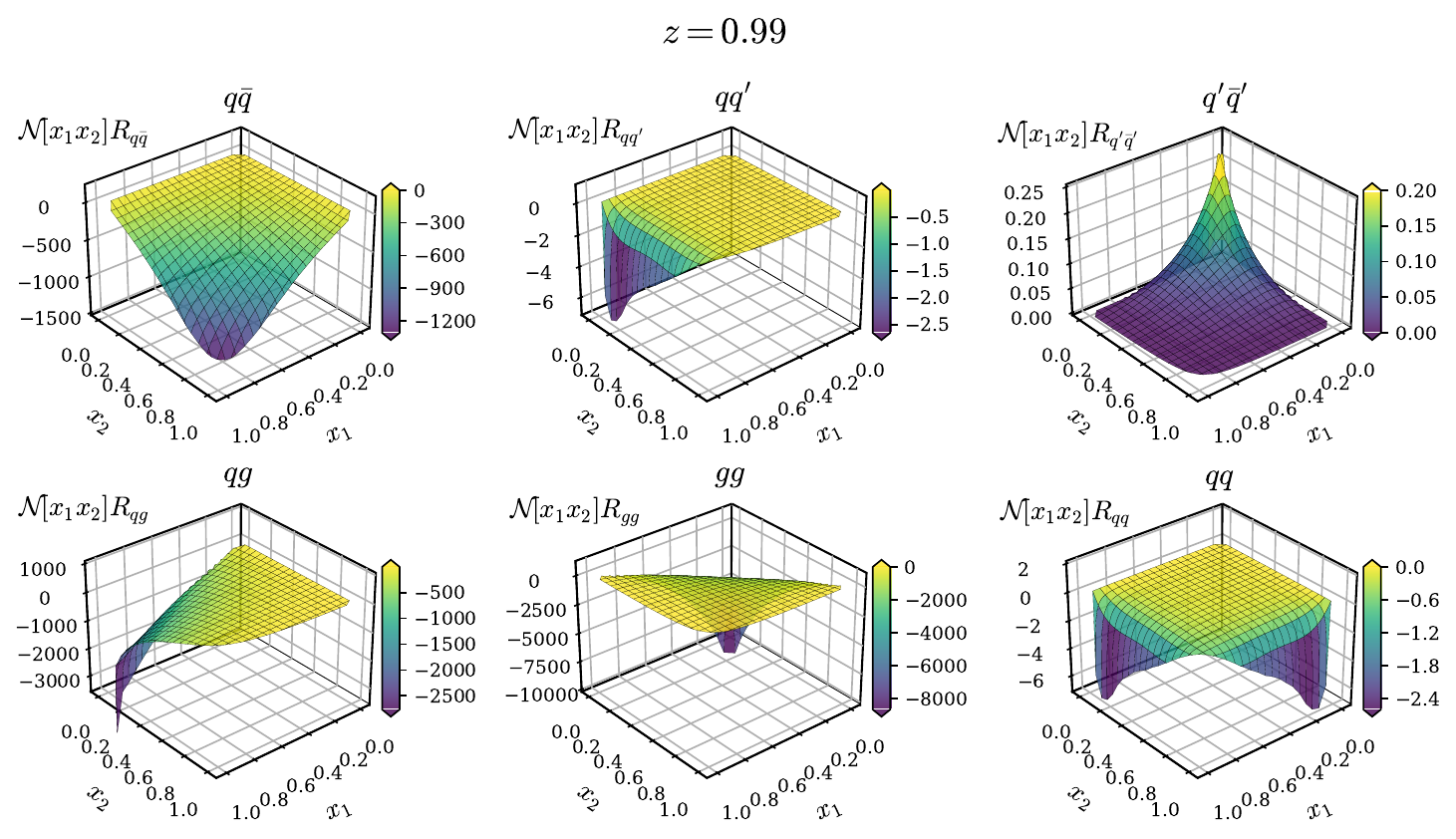}
    \caption{The Wilson coefficients for $m_X^2>0$ at $z=99/100$ as functions of $x_1,\,x_2$. The notations are same as fig.~\ref{fig:sigmahat_xxmX_z1o2}.}
    \label{fig:sigmahat_xxmX_z99o100}
\end{figure}

For a fixed $z \in (0,1)$ and under the constraint $m_X^2 > 0$, we emphasize the singular behavior of the Wilson coefficient $R_{ij}$ arising from the following kinematic limits:
\begin{align}
  &x_1 \to 0, \qquad x_2 \to 0, \qquad x_1 \to 0\;\&\;x_2 \to 0, \nonumber \\
   &x_1 \to 1 \;\&\; x_2 \to 0,\qquad x_1 \to 0 \;\&\; x_2 \to 1.
\end{align}
Here, $x_1 \to 0$ denotes the limit in which $x_1$ approaches zero while $x_2$ remains finite and separated from $1$. An analogous interpretation applies to $x_2 \to 0$ as well.

The regular parts $R_{ij}$ of the NLO Wilson coefficients for all six sectors are displayed as functions of $\{x_1,\, x_2\}$ for representative values of $z = 1/2$, $1/100$, and $99/100$ in figs.~\ref{fig:sigmahat_xxmX_z1o2}, \ref{fig:sigmahat_xxmX_z1o100}, and \ref{fig:sigmahat_xxmX_z99o100}, respectively. To mitigate the volatile asymptotic behaviors as $\{x_1,\, x_2,\, m_X^2\} \to 0$, the results are presented with an additional multiplicative factor $$\mathcal{N}[x_1x_2] \equiv \frac{N_C x_1 x_2 m_X^2}{4\pi^2}.$$ The numerical evaluations are performed using the parameter choices $N_C = 3$, $N_f = 5$, and $\mu = Q$.  The analytic expressions for the leading singular terms in each sector are summarized in table~\ref{tab:div_fix_z}. 

$$~$$

\noindent
\textbf{Region}~$\mathbf {x_1 \to 0:}$ For \(x_1 \to 0\), the leading term in $R_{gg}$ scales \(\sim\ln x_1 / x_1\), whereas those in the other sectors behave like \(R_{ij} \sim \ln x_1\). This stronger singular behavior in the \(gg\) sector arises from the configuration where one of the identified gluons becomes soft. 

\noindent
\textbf{Region}~$\mathbf {x_2 \to 0:}$~
For the same reason, the Wilson coefficients in the \(qg\) and \(gg\) sectors observe enhanced leading contributions \(R_{qg,gg} \sim \ln x_2 / x_2\) as \(x_2 \to 0\), while the other four sectors retain only logarithmic scaling, \(R_{ij} \sim \ln x_2\). 

\noindent
\textbf{Region}~$\mathbf {x_1 \to0\;\&\; x_2 \to 0:}$~When both momentum fractions vanish simultaneously, i.e.,  \(x_1 \to0\;\&\; x_2 \to 0\), the identified \(q\bar{q}\), \(q'\bar{q}'\), and \(gg\) pairs can originate from the splitting of a soft gluon. Consequently, their corresponding Wilson coefficients exhibit stronger leading terms in this limit, scaling as
\begin{equation}
R_{q\bar{q},\,q'\bar{q}',\,gg} \sim  \frac{x_\varrho^k}{x_\varsigma^{k+2}} \ln x_{1,2},\quad\mathrm{with}~\{\varrho,\varsigma\}=1,2~\mathrm{and}~k = -1,0,1,2.
\end{equation}
The \(R_{qg}\) follows the same scaling behavior as in the limit \(x_2 \to 0\),
\begin{equation}
R_{qg} \sim  \frac{\ln x_\varrho}{x_\varsigma},\; \quad\mathrm{with}~\{\varrho,\varsigma\}=1,2\,,
\end{equation}
which again reflects the contribution of a soft gluon. In contrast, the Wilson coefficients \(R_{qq}\) and \(R_{qq'}\) remain governed by milder logarithmic contributions, \(R_{qq,qq'} \sim \{\ln x_\varrho,\, \varrho=1,2\}\), since the two identified partons in these sectors are not associated with the splitting of a single soft gluon.  

\noindent
\textbf{Region}~$\mathbf { x_1 (x_2) \to 1 \;\&\; x_2 (x_1) \to 0 :}$ 
Due to the interplay between the three-prong limit of \(m_{X}^2  \to 0\) and the region \(x_1 (x_2) \to 1 \;\&\; x_2 (x_1) \to 0\),  the analysis here is involved.
For the double-real correction at \(\mathcal{O}(\alpha_s^2)\), \(m^2_X\) is proportional to the invariant mass of the  unmeasured particles,
\begin{align}
 m_X^2 = m_{34}^2 = \frac{(k_c + k_d)^2}{Q^2} = x_3 x_4 z_{34},
\end{align}
where \(x_3 = 2 k_c \cdot q_{\gamma} / Q^2\) and \(x_4 = 2 k_d \cdot q_{\gamma} / Q^2\) are the energy fractions of the unidentified partons, and \(z_{34} = (1 - \cos \theta_{34}) / 2\), with \(\theta_{34}\) being the open angle between unresolved partons.  
The energy-momentum conservation imposes the constraint $ x_1 + x_2 + x_3 + x_4 = 2$. 

In the regime where \( m_X^2 \to 0 \) but \( \{x_1,\, x_2\} \not\to \{0,\,1\} \), the dominant contributions arise from the limits \( x_3 \to 0 \), \( x_4 \to 0 \), and \( z_{34} \to 0 \). Among all the sectors, the Wilson coefficients in the \( q\bar{q} \) and \( qg \) sectors exhibit enhanced singular behavior,\begin{align}
    R_{q\bar{q}, qg} \sim \frac{\ln m_X^2}{m_X^2}.
\end{align}
In the \( q\bar{q} \) sector, the leading contributions originate from configurations in which the unresolved parton pairs \( q'\bar{q}' \), \( gg \), or \( q\bar{q} \), produced via gluon splitting, become collinear (\( z_{34} \to 0 \)), as in the \( \gamma^*\to q\bar{q}q'\bar{q}' \), \( \gamma^*\to q\bar{q}gg \), and \( \gamma^*\to q\bar{q}q\bar{q} \) channels, respectively. Additional singularities appear in the \( \gamma^*\to q\bar{q}gg \) channel when one of the unresolved gluons becomes soft (\( x_3(x_4) \to 0 \)).
Similarly, in the \( qg \) sector, leading singularities emerge from configurations where the unresolved \( \bar{q}g \) pair from \( \bar{q} \to \bar{q}g \) splitting becomes collinear, or when the gluon becomes soft in the \( \gamma^*\to qg\bar{q}g \) channel.
The remaining four sectors do not exhibit such enhanced singular behavior. In these cases, the Wilson coefficients scale more mildly as
\begin{align}
    R_{ij} \sim \ln m_X^2,
\end{align}
without the \( 1/m_X^2 \) enhancement.

 In the limit \( x_1 \to 1 \) and \( x_2 \to 0 \), the Wilson coefficients in the \( q\bar{q} \), \( qq' \), and \( qq \) sectors endure enhanced singularities, 
\begin{align}
    R_{q\bar{q}, qq', qq} \sim \left\{ \frac{x_2^k}{(1 - x_1)^{k+2}} \ln x_2,\, \frac{x_2^k}{(1 - x_1)^{k+2}} \ln (1 - x_1) \right\}, \quad k = 0, 1, 2.
\end{align}
Beyond the leading singularities associated with the unresolved final-state radiation in the limit \( m_X^2 \to 0 \), the \( q\bar{q} \) sector also receives significant contributions when the identified antiquark (\( \bar{q} \)) and an unresolved quark are produced from the splitting of a soft gluon. This occurs in the \( \gamma^*\to q\bar{q}q\bar{q} \) process in the simultaneous soft limit \( x_2 \to 0 \) and \( x_3 \to 0 \).
Similarly, in the \( qq^{(')} \) sectors, enhanced contributions arise when a soft gluon splits into the identified \( q^{(')} \) and an unresolved \( \bar{q}^{(')} \), corresponding to the configuration \( x_2 \to 0 \) together with \( x_3 \to 0 \) or \( x_4 \to 0 \) in the \( qq^{(')}\bar{q}\bar{q}^{(')} \) final state.

In the limit \( x_1 \to 1 \) and \( x_2 \to 0 \), the Wilson coefficient in the \( qg \) sector exhibits stronger  singular behavior compared to the \( q\bar{q} \), \( qq' \), and \( qq \) sectors. Specifically, it scales as
\begin{align}
    R_{qg} \sim \left\{ 
    \frac{x_2^k}{(1 - x_1)^{k+2}} \ln x_2,\,
    \frac{x_2^k}{(1 - x_1)^{k+2}} \ln(1 - x_1),\,
    \frac{(1 - x_1)^k}{x_2^{k+2}} \ln x_2,\,
    \frac{(1 - x_1)^k}{x_2^{k+2}} \ln(1 - x_1) 
    \right\}\,,
    \end{align}
    where  $ k = -1, 0, 1, 2$.
In addition to the leading contributions arising from the unresolved emissions in the \( m_X^2 \to 0 \) limit, further enhancement occurs when both the identified and unidentified gluons in the \(\gamma^*\to qg\bar{q}g \) channel become soft (\( x_2 \to 0 \) and \( x_4 \to 0 \)). This configuration corresponds to the splitting of a soft gluon into two softer gluons, for instance, \( g \to gg \).

The Wilson coefficient \( R_{gg} \) in the \( gg \) sector features weaker singular behavior:
\begin{align}
    R_{gg} \sim
    & \Bigg\{ 
    \frac{x_2^k}{(1 - x_1)^{k+1}} \ln x_2,\,
    \frac{x_2^k}{(1 - x_1)^{k+1}} \ln(1 - x_1),\,
    \frac{(1 - x_1)^k}{x_2^{k+1}} \ln x_2,\,\\
    &
    \frac{(1 - x_1)^k}{x_2^{k+1}} \ln(1 - x_1) 
    \Bigg\} .
\end{align}
  where  $ k = -1, 0, 1, 2$.
This milder behavior comes from the fact that the \( gg \) sector includes only a single identified soft gluon in the \( \gamma^*\to ggq\bar{q} \) channel.

For the \( q'\bar{q}' \) sector, where the identified quark pair originates from the splitting of a hard gluon in the \( q'\bar{q}'q\bar{q} \) final state, the Wilson coefficient exhibits only logarithmic dependence,
\begin{align}
    R_{q'\bar{q}'} \sim \ln x_2.
\end{align}

Thinking of the Wilson coefficients in the opposite limit, \( x_1 \to 0 \) and \( x_2 \to 1 \), the leading singularities from the \( q\bar{q} \), \( q'\bar{q}' \), \( gg \), and \( qq \) sectors can be derived from those in the \( x_1 \to 1 \) and \( x_2 \to 0 \) limit via  the exchange \( x_1 \leftrightarrow x_2 \).  Concerning the \( qg \) sector, the leading term observes weaker enhancement than that in the $x_1\to 1\;\&\; x_2\to 0$ limit,
\begin{align}
    R_{qg} \sim \left\{ 
    \frac{x_2^k}{(1 - x_1)^{k+1}} \ln x_2,\,
    \frac{x_2^k}{(1 - x_1)^{k+1}} \ln(1 - x_1),\,
    \frac{(1 - x_1)^k}{x_2^{k+1}} \ln x_2 
    \right\},  
\end{align}
where  \(k = -1, 0, 1, 2\), 
due to the fact that  only the unresolved gluon can be soft.  Regarding  the \( qq' \) sector, the Wilson coefficient   displays at most logarithmic behavior in this limit, 
\begin{align}
    R_{qq'} \sim \left\{ \ln(1 - x_2), \quad \ln x_1\right\},
\end{align}
as the \( qq' \) pair in the \( qq'\bar{q}\bar{q}' \) channel originates from a hard gluon.

 \subsubsection{Asymptotic behaviour of  \texorpdfstring{$R_{ij}$}{Rij} in the limits \texorpdfstring{$z \to\{0,1\}$}{zto0or1}} 

\begin{table}[tb]
    \renewcommand{\arraystretch}{1.2}
\caption {   Leading contributions of $R_{ij}$ in the kinematic limits of $x_{\varrho}$ and $z$. ``K.L.'' denotes kinematic limit. 
Here, $w_{\varrho} = 1 - x_{\varrho}$ with $\varrho \in \{1,2\}$, and  $w_z=1-z$. We further define $o^{(l)}_{\varrho}=(1-x_\varrho)^{-l-k}(1-z)^{k}$, and $\tilde{o}^{(l)}_{\varrho}=(1-z)^{-l-k}(1-x_\varrho)^{k}$, $k=-1,0,1,2$. 
    The $o(\tilde{o})$ refers to the Wilson coefficient containing both kinds of singularities. The ``$\mathcal{O}(1)$'' in the table indicates the Wilson coefficients are finite at the corresponding phase-space point.}
    \label{tab:div_fix_x}
    \centering
    \resizebox{\textwidth}{!}{
    \begin{tabular}{|c|c|c|c|c|c|c|}
        \hline
        \diagbox{K.L.}{Leading terms}{Sectors} & $q\bar{q}$                               & $qq'$                             & $q'\bar{q}'$                      & $qg$                                         & $gg$                              & $qq$                 \\
        
        \hline
        $z\to 0$                               & $\frac{\ln z}{z}$                        & $\frac{\ln z}{z}$                 & $\frac{\ln z}{z}$                 & $\frac{\ln z}{z}$                            & $\frac{\ln z}{z}$                 & $\frac{\ln z}{z}$    \\
        
        \hline
        $z\to 1$                               & $\frac{\ln w_z}{w_z}$                    &   $\mathcal{O}(1)$                 &    $\mathcal{O}(1)$                 & $\frac{\ln w_z}{w_z}$                        & $\frac{\ln w_z}{w_z}$             & $\ln w_z$        \\

        \hline 
        $x_1\to 0$                             & $\frac{\ln x_1}{z}$                      & $\frac{\ln x_1}{z}$               & $\frac{\ln x_1}{z}$               & $\frac{\ln x_1}{z}$                          & $\frac{\ln x_1}{x_1z}$            & $\frac{\ln x_1}{z}$  \\
        $z \to 0$                              & $\frac{\ln z}{z}$                        & $\frac{\ln z}{z}$                 & $\frac{\ln z}{z}$                 & $\frac{\ln z}{z}$                            & $\frac{\ln z}{x_1z}$              & $\frac{\ln z}{z}$    \\
        
        \hline
        $x_1\to 0$                             & $\frac{\ln x_1}{w_z}$                    & \multirow{2}{*}{$\ln x_1$}        & \multirow{2}{*}{$\ln x_1$}        & $\frac{\ln x_1}{w_z}$                        & $\frac{\ln x_1}{x_1 w_z}$         & $\ln x_1$            \\
        $z\to 1$                               & $\frac{\ln w_z}{w_z}$                    &                                   &                                   & $\frac{\ln w_z}{w_z}$                        & $\frac{\ln w_z}{x_1 w_z}$         & $\ln w_z$         \\
                                                                                                                                                                                                              
        \hline
        $x_1\to 1$                             & $o(\tilde{o})^{(2)}_{1}\ln w_1$          & $o(\tilde{o})^{(2)}_{1}\ln w_1$   & \multirow{2}{*}{$\mathcal{O}(1)$}   & $o(\tilde{o})^{(2)}_{1}\ln w_1$              & $o(\tilde{o})^{(1)}_{1}\ln w_1$   & $o(\tilde{o})^{(2)}_{1}\ln w_1$\\ 
        $z\to 1$                               & $o(\tilde{o})^{(2)}_{1}\ln w_z$          & $o(\tilde{o})^{(2)}_{1}\ln w_z$   &                                   & $o(\tilde{o})^{(2)}_{1}\ln w_z$              & $o(\tilde{o})^{(1)}_{1}\ln w_z$   & $o(\tilde{o})^{(2)}_{1}\ln w_z$\\
                                                                                                                                                                                                             
        \hline

        $x_2\to 0$                             & $\frac{\ln x_2}{z}$                      & $\frac{\ln x_2}{z}$               & $\frac{\ln x_2}{z}$               & $\frac{\ln x_2}{x_2 z}$                      & $\frac{\ln x_2}{x_2 z}$           & $\frac{\ln x_2}{z}$  \\
        $z \to 0$                              & $\frac{\ln z}{z}$                        & $\frac{\ln z}{z}$                 & $\frac{\ln z}{z}$                 & $\frac{\ln z}{x_2 z}$                        & $\frac{\ln z}{x_2 z}$             & $\frac{\ln z}{z}$    \\
        
        \hline
        $x_2\to 0$                             & $\frac{\ln x_2}{w_z}$                    & \multirow{2}{*}{$\ln x_2$}        & \multirow{2}{*}{$\ln x_2$}        & $\frac{\ln x_2}{x_2 w_z}$                    & $\frac{\ln x_2}{x_2 w_z}$         & $\ln x_2$            \\
        $z\to 1$                               & $\frac{\ln w_z}{w_z}$                    &                                   &                                   & $\frac{\ln w_z}{x_2 w_z}$                    & $\frac{\ln w_z}{x_2 w_z}$         & $\ln w_z$         \\
                                                                                                                                                                                                              
        \hline
        $x_2\to 1$                             & $o(\tilde{o})^{(2)}_{2}\ln w_2$          & $\ln w_2$                         & \multirow{2}{*}{$\mathcal{O}(1)$}   & $o(\tilde{o})^{(1)}_{2}\ln w_2$              & $o(\tilde{o})^{(1)}_{2}\ln w_2$   & $o(\tilde{o})^{(2)}_{2}\ln w_2$\\ 
        $z\to 1$                               & $o(\tilde{o})^{(2)}_{2}\ln w_z$          & $\ln w_z$                         &                                   & $o(\tilde{o})^{(1)}_{2}\ln w_z$              & $o(\tilde{o})^{(1)}_{2}\ln w_z$   & $o(\tilde{o})^{(2)}_{2}\ln w_z$\\
                                                                                                                                                                                                             
        \hline
    \end{tabular}
    }
\end{table}

\begin{figure}[tb]
\centering
\subfigure[]{\includegraphics[width=0.495\linewidth]{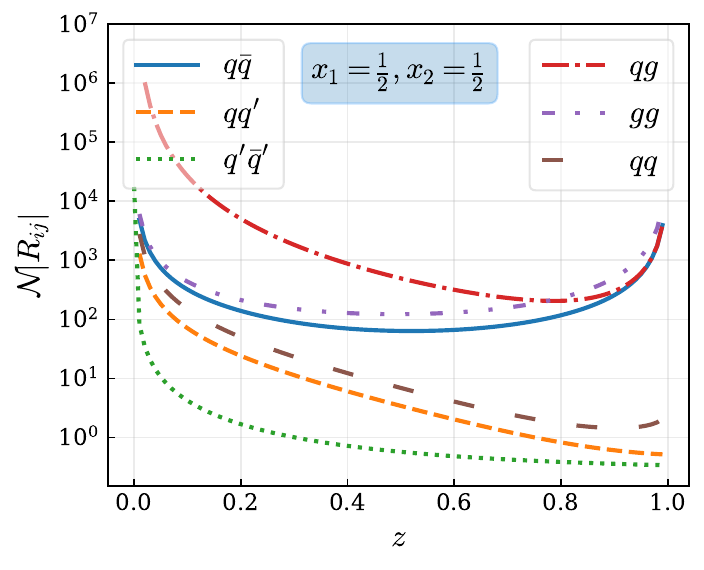}}
\subfigure[]{\includegraphics[width=0.495\linewidth]{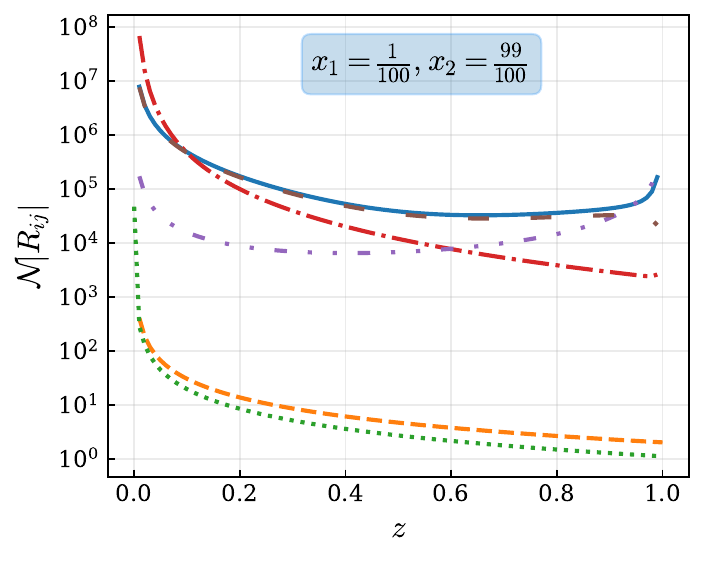}}
\subfigure[]{\includegraphics[width=0.495\linewidth]{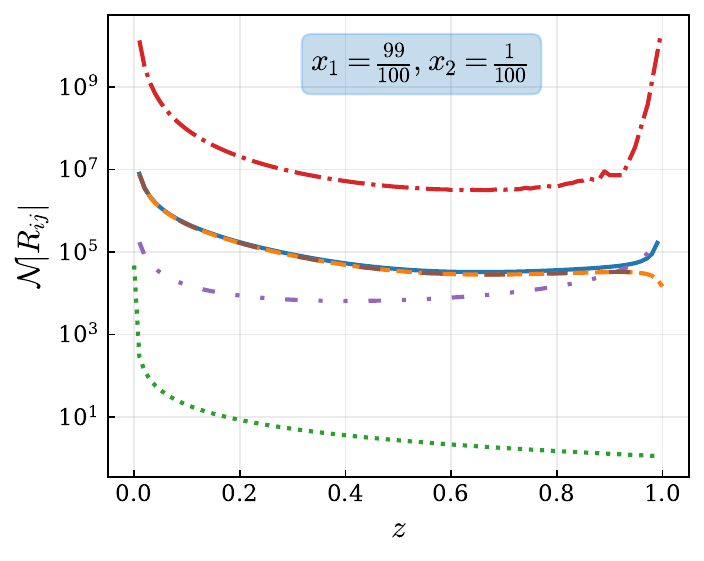}}
\subfigure[]{\includegraphics[width=0.495\linewidth]{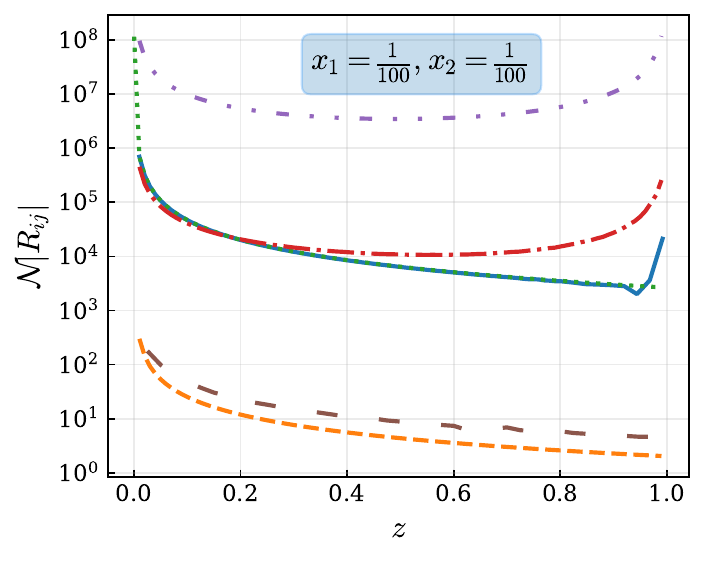}}
\caption{The magnitudes of  Wilson coefficients in the six sectors with the prefactor $\mathcal{N}=N_C/(4\pi^2)$ at $(x_1,x_2)=(1/2,1/2)$ (top-left), $(x_1,x_2)=(1/100,99/100)$ (top-right),
$(x_1,x_2)=(99/100,1/100)$ (bottom-left), $(x_1,x_2)=(1/100,1/100)$ (bottom-right) as functions of $z$.}
    \label{fig:sigmahat_z}
\end{figure}

\begin{figure}[htb]
    \centering
    \includegraphics[width=\linewidth]{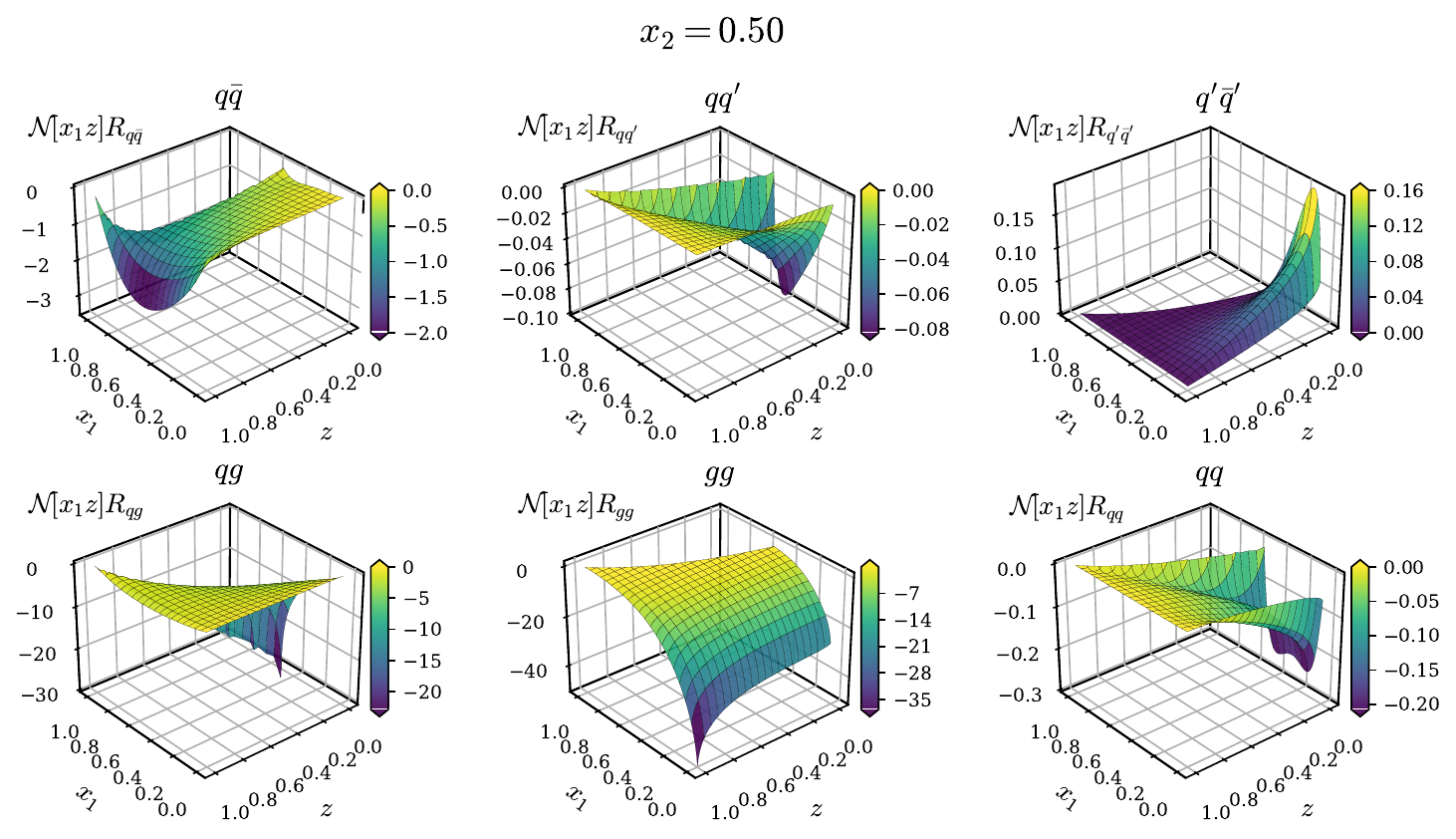}
    \caption{The Wilson coefficients for $m_X^2>0$ at $x_2=1/2$ as functions of $x_1,\,z$ in the $q\bar{q}$ (top-left), $qq^{\prime}$ (top-middle), $q^{\prime}\bar{q}^{\prime}$(top-right), $qg$ (bottom-left), $gg$ (bottom-middle), and $qq$ (bottom-right) sectors. The prefactor $\mathcal{N}[x_1z]=N_C x_1(1-x_1)z(1-z)m_X^2/(4\pi^2)$ is multiplied to mitigate the volatile asymptotic behaviors  of the Wilson coefficients. The values of the $\mathcal{N}[x_1z]$ weighted Wilson coefficients are encoded in color.}
    \label{fig:sigmahat_x1z_1o2}
\end{figure}

\begin{figure}[ht]
    \centering
    \includegraphics[width=\linewidth]{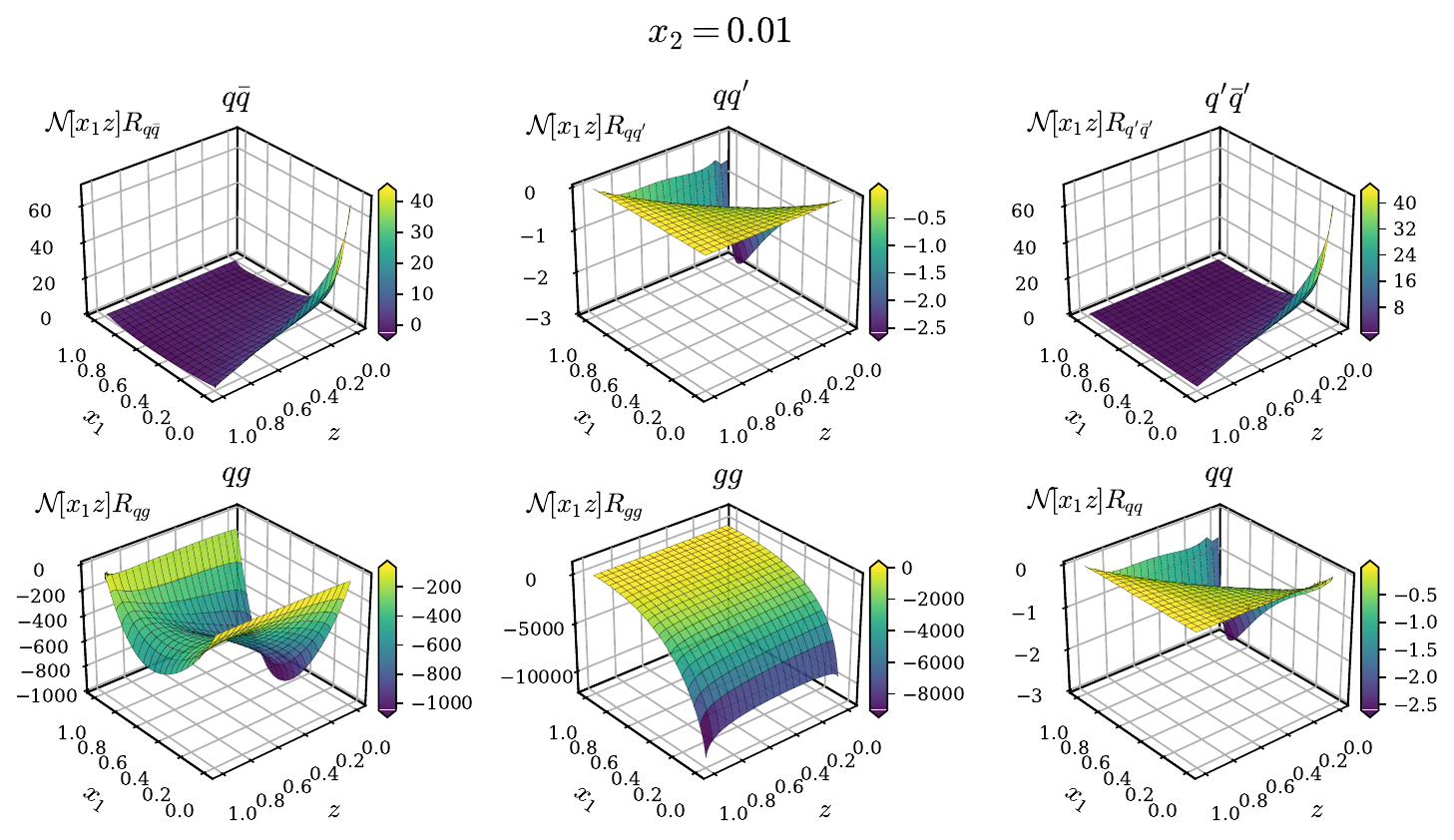}
    \caption{The Wilson coefficients for $m_X^2>0$ at $x_2=1/100$ as functions of $x_1,\,z$. The notations are same as fig.~\ref{fig:sigmahat_x1z_1o2}.}
    \label{fig:sigmahat_x1z_1o100}
\end{figure}

\begin{figure}[htb]
    \centering
    \includegraphics[width=\linewidth]{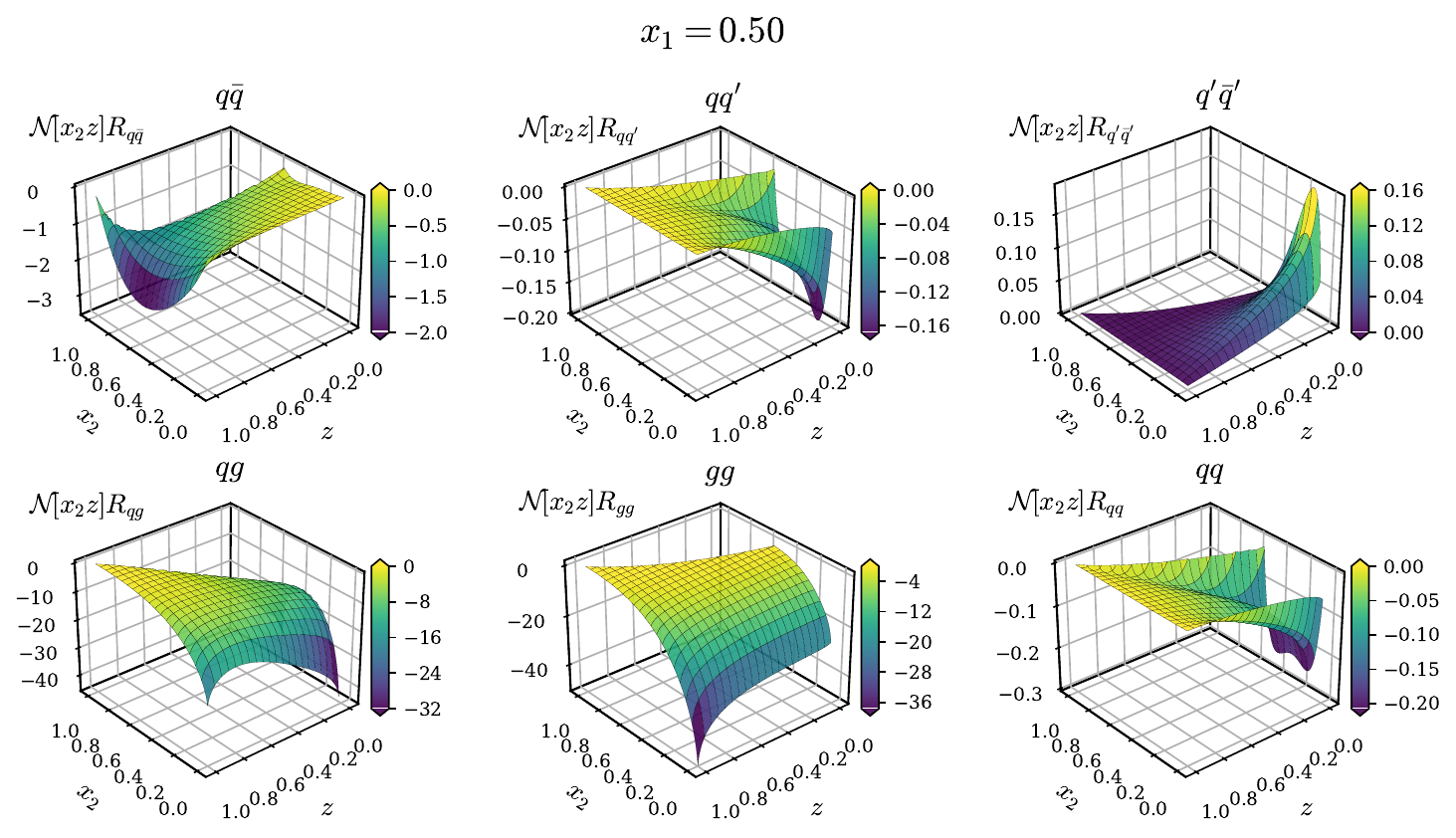}
    \caption{The Wilson coefficients for $m_X^2>0$ at $x_1=1/2$ as functions of $x_2,\,z$ in the $q\bar{q}$ (top-left), $qq^{\prime}$ (top-middle), $q^{\prime}\bar{q}^{\prime}$(top-right), $qg$ (bottom-left), $gg$ (bottom-middle), and $qq$ (bottom-right) sectors. The prefactor $\mathcal{N}[x_2z]=N_C x_2(1-x_2)z(1-z)m_X^2/(4\pi^2)$ is multiplied to mitigate the volatile asymptotic behaviors of the Wilson coefficients. The values of the $\mathcal{N}[x_2z]$ weighted Wilson coefficients are encoded in color.}
    \label{fig:sigmahat_x2z_1o2}
\end{figure}

\begin{figure}[htb]
    \centering
    \includegraphics[width=\linewidth]{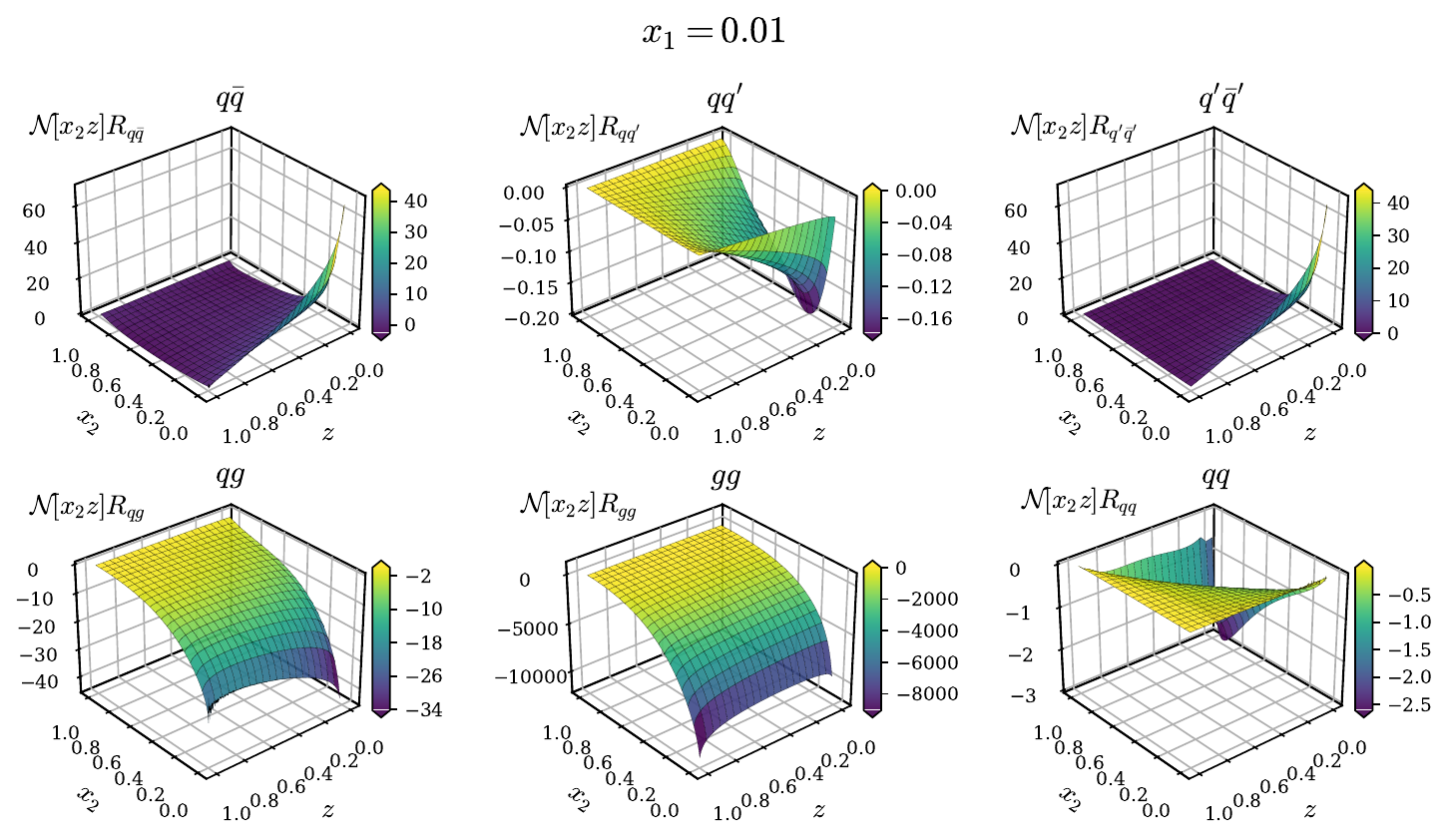}
    \caption{The Wilson coefficients for $m_X^2>0$ at $x_1=1/100$ as functions of $x_2,\,z$. The notations are same as fig.~\ref{fig:sigmahat_x2z_1o2}.}
    \label{fig:sigmahat_x2z_1o100}
\end{figure}

For fixed values of \( x_\varrho\in \{0,1\} \) with \( \varrho= 1,2 \), and under the constraint \( m_X^2 > 0 \), the Wilson coefficients  present  enhanced contributions in  kinematic limits: 
\begin{equation}
\begin{aligned}
  &z \to 0, \qquad z\to 1, \qquad x_\varrho \to 0\;\&\;z \to 0, \\
  &x_\varrho \to 1\;\&\;z \to 0,\qquad x_\varrho \to 0\;\&\;z \to 1,
\end{aligned}
\end{equation}
where the limits \( z \to \{0,\,1\} \) are defined with \( x_1 \) and \( x_2 \) held in the interior of the unit interval, i.e., \( x_{1,2} \not\to \{0,\,1\} \). Similarly, the composite limits such as \( x_1 \to 0 \) with \( z \to 0 \) are taken with \( x_2 \) fixed to a regular value in \( (0,1) \), and analogous interpretations apply to the other cases. It is important to note that when \( x_\varrho \to 1 \), it is necessary that \( z \to 1 \) to ensure that \( m_X^2 > 0 \) being satisfied.

To illustrate the behavior of the Wilson coefficients in these regimes, we consider \( N_C = 3 \), \( N_f = 5 \), and \( \mu = Q \). The absolute value \( \mathcal{N} |R_{ij}| \) of the regular parts of the Wilson coefficients across all six partonic sectors is displayed in fig.~\ref{fig:sigmahat_z} as a function of \( z \), for representative points \( (x_1, x_2) = \{ (1/2, 1/2), (1/100, 99/100), (99/100, 1/100), (1/100, 1/100) \} \). Here, a normalization factor $$ \mathcal{N} = \frac{N_C}{4\pi^2} $$ is applied.
Furthermore, the behavior of \( R_{ij} \) is also shown in figs.~\ref{fig:sigmahat_x1z_1o2} and \ref{fig:sigmahat_x1z_1o100} as functions of \( \{ x_1, z \} \), with \( x_2 \in \{ 1/2, 1/100 \} \) fixed, and in figs.~\ref{fig:sigmahat_x2z_1o2} and \ref{fig:sigmahat_x2z_1o100} as functions of \( \{ x_2, z \} \), with \( x_1 \in \{ 1/2, 1/100 \} \) fixed.
Figures corresponding to \( x_{1,2} \to 1 \) are omitted due to the limited available phase space in these regions.

In order to moderate the volatility in the limits \( \{ x_1, z, m_X^2 \} \to 0 \) and \( \{ x_1, z \} \to 1 \), an additional damping factor
\begin{align}
    \mathcal{N}[x_1 z] = \frac{N_C\, x_1 (1 - x_1) z (1 - z) m_X^2}{4\pi^2}
\end{align}
is applied to the Wilson coefficients in the \( x_2 \)-fixed scans. An analogous factor,
\begin{align}
    \mathcal{N}[x_2 z] = \frac{N_C\, x_2 (1 - x_2) z (1 - z) m_X^2}{4\pi^2},
\end{align}
is used when \( x_1 \) is held fixed. 

\noindent
\textbf{Region}~$\mathbf { z\to 0 :}$   In the limit \( z \to 0 \), the regular parts \( R_{ij} \) of the Wilson coefficients exhibit collinear singularities of the form \( R_{ij} \sim \ln z / z \) across all sectors. In the \(q\bar{q}\), \(q'\bar{q}'\) and \(gg\) sectors, these singularities arise predominantly from collinear gluon splittings such as \( g \to q\bar{q} \), \( g \to q'\bar{q}' \), and \( g \to gg \), in the \( \gamma^*\to q\bar{q}q\bar{q} \), \( \gamma^*\to  q'\bar{q}' q\bar{q} \), and \( \gamma^*\to ggq\bar{q} \) channels, respectively. In the \( qg \), \( qq' \), and \( qq \) sectors, the dominant contributions stem from collinear quark splittings. Specifically, the identified \( qg \) pairs in the \( \gamma^*\to  qg\bar{q}g \) channel can originate from the process \( q \to qg \), while the \( qq^{(')} \) configurations can arise via successive splittings \( q \to qg \to qq^{(')}\bar{q}^{(')} \) in the \(\gamma^*\to qq^{(')}\bar{q}\bar{q}^{(')}\) channel.

\noindent\textbf{Region}~$\mathbf { z \to 1 :}$ As \( z \to 1 \), the identified partons approach a back-to-back configuration. In this limit, \( R_{q\bar{q}} \), \( R_{qg} \), and \( R_{gg} \) exhibit leading singular behavior of the form \( \ln(1-z)/(1-z) \). These contributions can be attributed to the collinear splitting of a (anti)quark into an identified parton and an unresolved parton. In the \( q\bar{q} \) sector, additional enhancements may result from successive splittings such as \( q(\bar{q}) \to q(\bar{q})g \to \{ q(\bar{q}) q'\bar{q}',\, q(\bar{q})gg,\, q(\bar{q}) q\bar{q} \} \), where the second branching products are unresolved. In \( R_{qg} \), the enhancement originates from \( q \to qg \), with the gluon unidentified. Similarly, \( R_{gg} \) receives contributions from \( q \to qg \), with the gluon identified. The \( qq \) sector exhibits only logarithmic enhancements \( R_{qq} \sim \ln(1-z) \), while \( R_{qq'}, R_{q'\bar{q}'} \) remain regular as \( z \to 1 \) with \( x_{1,\,2} \not \to \{0,\,1\} \).

\noindent\textbf{Region}~$\mathbf { x_1 (x_2) \to 0 \;\&\; z \to 0 :}$ In the combined limit \( x_1 \to 0 \) and \( z \to 0 \), the \( gg \) sector exhibits enhanced singular behavior of the form \( R_{gg} \sim \{ \ln x_1 / (x_1 z),\, \ln z / (x_1 z) \} \), driven by soft and collinear identified gluons. In other sectors, the dominant singularities remain collinear, with \( R_{ij} \sim \{ \ln x_1 / z,\, \ln z / z \} \). Analogously, for \( x_2 \to 0 \) and \( z \to 0 \), the \( qg \) and \( gg \) sectors display similar enhanced behavior, \( R_{qg,gg} \sim \{ \ln x_2 / (x_2 z),\, \ln z / (x_2 z) \} \), while the remaining sectors continue to exhibit standard collinear scaling.

\noindent\textbf{Region}~$\mathbf { x_1 (x_2) \to 0 \;\&\; z \to 1 :}$ For \( x_1 \to 0 \) and \( z \to 1 \), the structure of the singular contributions mirrors that in the \( x_1 \to 0 \) and  \( z \to 0 \) case. The \( gg \) sector develops enhanced contributions of the form \( R_{gg} \sim \{ \ln x_1 / [x_1 (1-z)],\, \ln(1-z) / [x_1 (1-z)] \} \), due to overlapping soft and collinear limits. The \( q\bar{q} \) and \( qg \) sectors exhibit leading singular behavior \( R \sim \{ \ln x_1 / (1-z),\, \ln(1-z)/(1-z) \} \). The \( qq' \), \( q'\bar{q}' \), and \( qq \) sectors display only logarithmic dependence: \( R_{ij} \sim\{ \ln x_1  , \, \ln(1-z)\} \). The limit \( x_2 \to 0 \) and \( z \to 1 \) yields analogous behavior, with the most enhanced contributions again occurring in the \( qg \) and \( gg \) sectors.

\noindent\textbf{Region}~$\mathbf { x_1 (x_2) \to 1 \;\&\; z \to 1 :}$In the limit \( x_1 \to 1 \) and \( z \to 1 \), the singular structure becomes more involved due to the vanishing invariant mass \( m_X^2 = m_{34}^2 = 1 - x_1 - x_2 + x_1 x_2 z \to 0 \), entailing configurations where \( x_3 \to 0 \), \( x_4 \to 0 \), and \( z_{34} \to 0 \). In this regime, the \( q\bar{q} \), \( qq' \), \( qg \), and \( qq \) sectors exhibit leading singular behavior scaling as:
\begin{align}
    R_{ij} \sim &\bigg\{ \frac{(1-x_1)^k}{(1-z)^{k+2}} \ln(1-x_1),\, \frac{(1-z)^k}{(1-x_1)^{k+2}} \ln(1-x_1),\, \frac{(1-x_1)^k}{(1-z)^{k+2}} \ln(1-z),\,\nonumber\\
    & \frac{(1-z)^k}{(1-x_1)^{k+2}} \ln(1-z) \bigg\}, \quad k = -1, 0, 1, 2.
\end{align}
In the \( q\bar{q} \) and \( qg \) sectors, in addition to standard collinear enhancements as \( z \to 1 \), there are further singular contributions from collinear configurations among the unresolved partons (\( z_{34} \to 0 \)) and soft gluon emissions. The \( qq^{(')} \) sectors can receive enhanced contributions from successive splittings such as \( \bar{q} \to \bar{q}g \to \bar{q} q^{(')}\bar{q}^{(')} \). In the \( gg \) sector, the dominant singularities again arise from collinear limits near \( z \to 1 \), with scaling:
\begin{align}
    R_{gg} \sim& \bigg\{ \frac{(1-x_1)^k}{(1-z)^{k+1}} \ln(1-x_1),\, \frac{(1-z)^k}{(1-x_1)^{k+1}} \ln(1-x_1),\, \frac{(1-x_1)^k}{(1-z)^{k+1}} \ln(1-z),\, \nonumber\\
    &\frac{(1-z)^k}{(1-x_1)^{k+1}} \ln(1-z) \bigg\},\,k=-1,0,1,2\,.
\end{align}
By contrast, the \( q'\bar{q}' \) sector remains free of leading singular behavior in this limit.

For \( x_2 \to 1 \) and \( z \to 1 \), the leading behavior of the \( q\bar{q} \), \( q'\bar{q}' \), \( gg \), and \( qq \) sectors can be obtained by the substitution \( x_1 \leftrightarrow x_2 \), due to the symmetry of the Wilson coefficients under this exchange. In the asymmetric sectors, the \(R_{qq'} \) and \( R_{qg} \) exhibit milder singular structure. Specifically, \( R_{qq'} \sim \{ \ln(1-z),\, \ln(1-x_2) \} \), while \( R_{qg} \) displays collinear scaling of the form:
\begin{align}
    R_{qg} \sim& \bigg\{ \frac{(1-x_1)^k}{(1-z)^{k+1}} \ln(1-x_1),\, \frac{(1-z)^k}{(1-x_1)^{k+1}} \ln(1-x_1),\, \frac{(1-x_1)^k}{(1-z)^{k+1}} \ln(1-z),\,\nonumber\\
    & \frac{(1-z)^k}{(1-x_1)^{k+1}} \ln(1-z) \bigg\}\,,~~~\, k=-1,0,1,2
\end{align}
A summary of the leading singular behavior of the Wilson coefficients \( R_{ij} \) at various singular phase-space boundaries for fixed \( x_\varrho \in (0,1) \) is provided in table~\ref{tab:div_fix_x}.

\subsection{The \texorpdfstring{$m_X^2\to 0$}{mxsq->0} part}

The coefficients $V_{ij}$ and $U^{[k]}_{ij}$ of the delta and star distributions are only non zero for the $q\bar{q}$ and $qg$ sectors, i.e., 
\begin{align}
V_{ij}=U^{[k]}_{ij}=0,\,\qquad\{ij\}\in\{qq,qq^\prime,q^\prime\bar{q}^\prime,gg\}.\, 
\end{align}
As shown in the last section, the Wilson coefficients in the $q\bar{q}$ and $qg$ sectors have singular behavior
$R_{q\bar{q},qg}\sim \ln m_X^2/m_X^2$, and these divergences are isolated into the star distributions in eq.~\eqref{eq:finite_wilson}, which makes the coefficients $V_{ij}$ and $U_{ij}^{[k]},\,k=0,\,1$ non-vanishing in these two sectors. To shorten the expressions, 
we give the exact form of $V_{ij}$ and $U_{ij}^{[k]}$ as functions of $\{x_1,\, z\}$ in terms of the arguments 
$\{x_1,\,\tilde{x}_2=(1-x_1)/(1-x_1z)\}$ in the following. 
The arguments $\{\tilde{x}_1=(1-x_2)/(1-x_2z),\, x_2\}$ can also be used to express these coefficients as functions of $\{x_2,\, z\}$.

The coefficients of the $[\ln m_X^2/m_X^2]_*$ are rational functions, 
\begin{align}
   U_{q\bar{q}}^{[1]}(x_1,\tilde{x}_2)&= \frac{4 N_A \left(N_A-1\right) x_1 \tilde{x}_2 \left(x_1^2+\tilde{x}_2^2\right) }{N_C^2 w_1 \tilde{w}_2 },\\
   U_{qg}^{[1]}(x_1,\tilde{x}_2)&=\frac{2 N_A \left(5 N_A+4\right) x_1 \tilde{x}_2 \left(x_1^2+x_{X}^2\right) }{N_C^2 w_1 w_{X} },
\end{align}
where $w_\varrho=1-x_\varrho\, (\varrho=\{1,X\})$, $\tilde{w}_2=1-\tilde{x}_2$, $x_{X}=2-x_1-\tilde{x}_2$,
and the coefficients of $[1/m_X^2]_*$ contain logarithmic functions;
\begin{align}
    U_{q\bar{q}}^{[0]}(x_1,\tilde{x}_2)=&\frac{ N_A x_1 \tilde{x}_2 \left(x_1^2+\tilde{x}_2^2\right)  \left[ N_C (2 N_f-11 N_C)+12 \ell_*^{[0]}\right]}{3 N_C^2 w_1 \tilde{w}_2 }\nonumber\\
    &+L_h\frac{4 N_A^2  x_1 \tilde{x}_2 \left(x_1^2+\tilde{x}_2^2\right) }{N_C^2 w_1 \tilde{w}_2 },\\
    U_{qg}^{[0]}(x_1,\tilde{x}_2)=& -\frac{N_A x_1 \tilde{x}_2 \left(x_1^2+x_{X}^2\right)  \left(3N_A+8 N_C^2 \ell_*^{[0]}\right)}{2 N_C^2 w_1 w_{X} }\nonumber\\
    &+L_h \frac{2 N_A \left(3 N_A+2\right)  x_1 \tilde{x}_2 \left(x_1^2+x_{X}^2\right) }{N_C^2 w_1 w_{X}},
\end{align}
where $L_h=\ln (Q^2/\mu^2)$ and $\ell_*^{[0]}=\ln w_1+ \ln \tilde{w}_2- \ln w_X$.

\begin{figure}[tb]
    \centering
    \includegraphics[width=\linewidth]{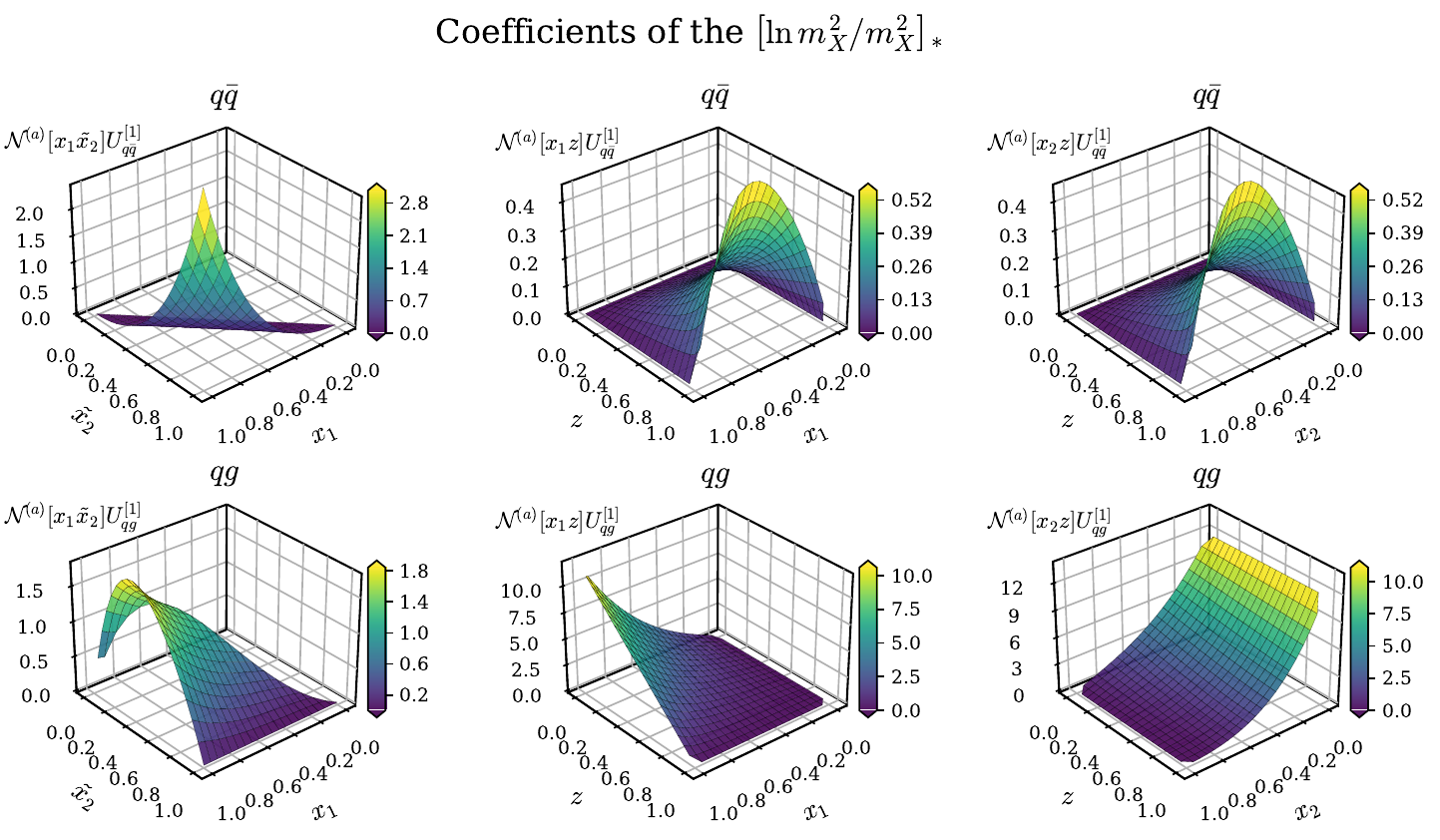}
    \caption{Coefficients $U_{ij}^{[1]}$ of the $\left[\ln m_X^2/m_X^2\right]_*$ in the $q\bar{q}$ (first row) and $qg$ (second row) sectors as functions of $\{x_1,\,\tilde{x}_2\}$ (first column), $\{x_1,\, z\}$ (second column) and $\{x_2,\,z\}$ (third column). In the first column, the $\tilde{x}_2$ is treated as an independent variable. The prefactors $\mathcal{N}^{(a)}[x_1\tilde{x}_2]=N_C(1-x_1)(1-\tilde{x}_2)(x_1+\tilde{x}_2-1)/(4\pi^2)$, 
    $\mathcal{N}^{(a)}[x_1z]=N_C(1-x_1)x_1(1-z)z/(4\pi^2)$ and $\mathcal{N}^{(a)}[x_2z]=N_C(1-x_2)x_2(1-z)z/(4\pi^2)$ in the first, second, and third column are multiplied to mitigate the strong asymptotic singular behaviors of the coefficients. The values of $U_{ij}^{[1]}$ weighted by the corresponding prefactors are encoded in color.}
    \label{fig:sigmahat_Coeffdlog_plus}
\end{figure}

\begin{figure}[tb]
    \centering
    \includegraphics[width=\linewidth]{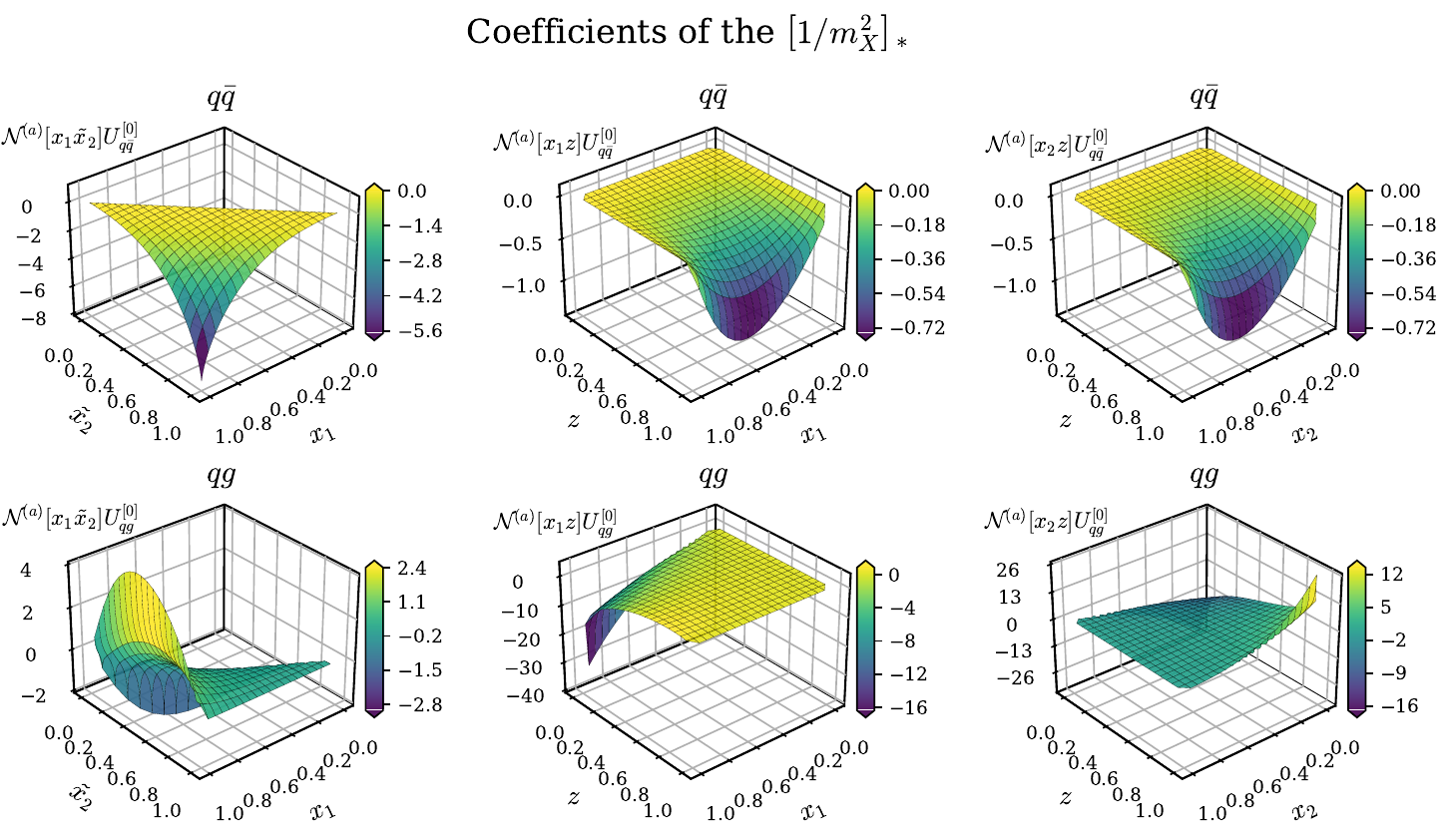}
    \caption{Coefficients $U_{ij}^{[0]}$ of the $\left[1/m_X^2\right]_*$ in the $q\bar{q}$ (first row) and $qg$ (second row) sectors as functions of different arguments. The notations and prefactors are same as those in fig.~\ref{fig:sigmahat_Coeffdlog_plus}. The values of $U_{ij}^{[0]}$ weighted by the corresponding prefactors are encoded in color. }
    \label{fig:sigmahat_Coeffd_plus}
\end{figure}

The coefficients $V_{ij}$ of $\delta(m_X^2)$ include both double-real and real-virtual contributions, 
and contain functions to transcendentality-two. $V_{ij}$ can be expressed as 
\begin{align}
    V_{ij}(x_1,\tilde{x}_2;\mu)=\sum_{k=0}^{3}(r_{\delta})^{(k)}_{ij}(x_1,\tilde{x}_2;\mu)\ell_{\delta}^{(k)}
    +\sum_{k=1}^{8}(pr_{\delta})^{(k)}_{ij}(x_1,\tilde{x}_2;\mu)(\ell\ell)_{\delta}^{(k)},
\end{align}
where $\ell_{\delta}^{(0)}=1$, and $\{\ell_{\delta}^{(k)},\, k=1,2,3\}$ are the weight-one basis
\begin{align}
    \ell_{\delta}^{(1)}=\ln w_1, \qquad \ell_{\delta}^{(2)}=\ln \tilde{w}_2, \qquad \ell_{\delta}^{(3)}=\ln w_X,
\end{align} 
$\{(\ell\ell)_{\delta}^{(k)}, k=1,\cdots,8\}$ are the weight-two basis,
\begin{align}
    (\ell\ell)_{\delta}^{(1)}&=\ln x_1 \ln w_1+\mathrm{Li}_2 (w_1),\quad (\ell\ell)_{\delta}^{(2)}=\ln \tilde{x}_2 \ln \tilde{w}_2+\mathrm{Li}_2 (\tilde{w}_2),\nonumber\\
    (\ell\ell)_{\delta}^{(3)}&=\ln x_X \ln w_X+\mathrm{Li}_2 (w_X),\quad (\ell\ell)_{\delta}^{(4)}=\ln^2 (w_1),\quad (\ell\ell)_{\delta}^{(5)}=\ln^2 (\tilde{w}_2)\nonumber\\
    (\ell\ell)_{\delta}^{(6)}&=\ln w_1 \ln \tilde{w}_2,\quad (\ell\ell)_{\delta}^{(7)}=\ln w_1\ln w_X ,\quad (\ell\ell)_{\delta}^{(8)}=\ln \tilde{w}_2\ln w_X,
\end{align}
and the coefficients of these bases $\{(r_{\delta})^{(k)}_{ij},\, k=0,1,2,3\}$ and $\{(pr_{\delta})^{(k)}_{ij},\, k=1,\cdots,8\}$ are rational functions.
It is worthwhile to mention that the polylogarithmic functions in the weight-two basis only come from the real-virtual contributions. The 
weight-two functions in the double-real contributions can be simplified to the product of two logarithmic functions through the 
identity
\begin{align}
    \mathrm{Li}_2(x)+\mathrm{Li}_2(1-x)+\ln (x)\ln (1-x)=\zeta_2,
\end{align}
and the shuffle algebra of the GPL functions 
\begin{align}
    G(a,b,x)+G(b,a,x)=G(a,x)G(b,x).
\end{align}
\begin{figure}[tb]
    \centering
    \includegraphics[width=\linewidth]{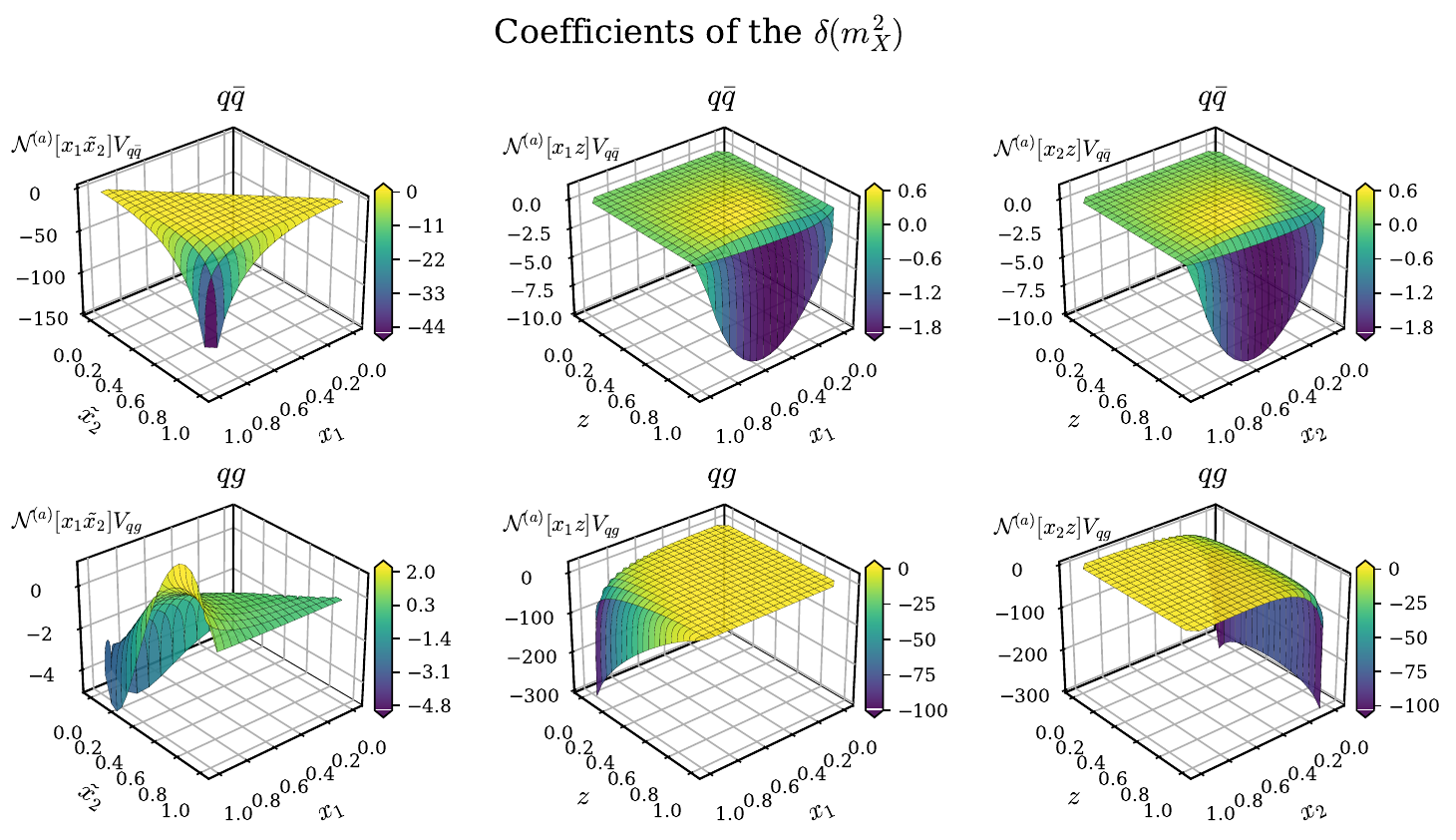}
    \caption{Coefficients of the $\delta(m_X^2)$ in the $q\bar{q}$ (first row) and $qg$ (second row) sectors as functions of different arguments. The notations and prefactors are same as those in fig.~\ref{fig:sigmahat_Coeffdlog_plus}. The values of $V_{ij}$ weighted by the corresponding prefactors are encoded in color.}
    \label{fig:sigmahat_Coeffd_delta}
\end{figure}

In the $q\bar{q}$ sector, the weight-two basis elements $(\ell\ell)_{\delta}^{(4)}$, $(\ell\ell)_{\delta}^{(5)}$ and $(\ell\ell)_{\delta}^{(6)}$
combine into a single term $(\ell\ell)_{\delta}^{(4)}+(\ell\ell)_{\delta}^{(5)}+2(\ell\ell)_{\delta}^{(6)}=(\ln w_1+\ln \tilde{w}_2)^2$.
Since the Wilson coefficient is symmetrical under the $x_1\leftrightarrow \tilde{x}_2$ swap ($w_1\leftrightarrow \tilde{w}_2$ at the same time),
there are only 4 independent coefficients for the weight-two basis, and two independent coefficients for the 
weight-one basis.
The rational coefficients in the $q\bar{q}$ sector are 
\begin{align}
    (r_{\delta})^{(0)}_{q\bar{q}} =& \frac{ N_A^2 x_X  \left[\left(4+9 \pi ^2\right) x_1 \tilde{x}_2 \left(x_1^2+\tilde{x}_2^2\right)-9 w_X^2 \left(x_1+\tilde{x}_2\right)\right]}{9 N_C^2 w_1 \tilde{w}_2 x_X } \nonumber \\
    & -\frac{ N_A x_1 \tilde{x}_2 \left[\left(x_1^2+\tilde{x}_2^2\right) \left(10  N_C N_f x_X+67 \left(x_1+\tilde{x}_2\right)-152\right)\right]}{9N_C^2 w_1 \tilde{w}_2 x_X } \nonumber \\
    & +\frac{ N_A x_1 \tilde{x}_2 \left[ \pi ^2 x_X \left(2 w_X^2+x_1^2+\tilde{x}_2^2\right)-12 w_X\right]}{3 N_C^2 w_1 \tilde{w}_2 x_X } \nonumber \\
    & -L_h \frac{ N_A x_1 \tilde{x}_2 \left(x_1^2+\tilde{x}_2^2\right)   \left[2 N_C \left(N_C-N_f\right)+9\right]}{3  N_C^2 w_1 \tilde{w}_2},
\end{align}

\begin{align}
    (r_{\delta})^{(1)}_{q\bar{q}}=(r_{\delta})^{(2)}_{q\bar{q}}|_{x_1\leftrightarrow \tilde{x}_2}=&\frac{  N_A \tilde{x}_2 \left[N_A \left(3 x_1 \tilde{x}_2+w_X\right) +2 w_1 x_1\right]}{N_C^2 x_1 }\nonumber\\
    &-L_h\frac{2 N_A^2 x_1 \tilde{x}_2 \left(x_1^2+\tilde{x}_2^2\right) }{ N_C^2 w_1 \tilde{w}_2 },
\end{align}

\begin{align}
   (r_{\delta})^{(3)}_{q\bar{q}}= -\frac{4 N_A x_1 \tilde{x}_2 w_X \left(x_X+1\right) }{N_C^2 x_X^2 },
\end{align}

\begin{align}
    (pr_{\delta})^{(1)}_{q\bar{q}}=(pr_{\delta})^{(2)}_{q\bar{q}}|_{x_1\leftrightarrow \tilde{x}_2}
    =\frac{2 N_A x_1 \tilde{x}_2  \left[N_A(x_1^2+\tilde{x}_2^2) -w_X^2+\tilde{x}_2^2\right]}{ N_C^2 w_1 \tilde{w}_2},
\end{align}

\begin{align}
    (pr_{\delta})^{(3)}_{q\bar{q}}=
    -\frac{2 N_A x_1 \tilde{x}_2 \left(x_1^2+\tilde{x}_2^2+2w_X^2\right) }{N_C^2 w_1 \tilde{w}_2 },
\end{align}

\begin{align}
    (pr_{\delta})^{(4)}_{q\bar{q}}=(pr_{\delta})^{(5)}_{q\bar{q}}
    =\frac{1}{2}(pr_{\delta})^{(6)}_{q\bar{q}}=-\frac{ N_A \left(N_A+2\right) x_1 \tilde{x}_2 \left(x_1^2+\tilde{x}_2^2\right) }{ N_C^2 w_1 \tilde{w}_2 },
\end{align}

\begin{align}
    (pr_{\delta})^{(7)}_{q\bar{q}}=(pr_{\delta})^{(8)}_{q\bar{q}}|_{x_1\leftrightarrow \tilde{x}_2}
    =\frac{2 N_A x_1 \tilde{x}_2 \left(2 x_1^2+\tilde{x}_2^2+w_X^2\right) }{N_C^2 w_1 \tilde{w}_2 }.
\end{align}

In the $qg$ sector, the weight-two basis elements $(\ell\ell)_{\delta}^{(3)}$ and $(\ell\ell)_{\delta}^{(8)}$
combine into a single term $(\ell\ell)_{\delta}^{(3)}-(\ell\ell)_{\delta}^{(8)}=\mathrm{Li}_2(w_X)+\ln w_X(\ln x_X-\ln \tilde{w}_2)$,
the coefficient $(pr_{\delta})^{(4)}_{qg}=0$, and the coefficients of $(\ell\ell)_{\delta}^{(5)}$ and $(\ell\ell)_{\delta}^{(7)}$
are proportional to each other. Therefore, there are 5 independent coefficients of the weight-two basis.
The rational coefficients in the $qg$ sector are 

\begin{align}
    (r_{\delta})^{(0)}_{qg} =&\frac{N_A^2 \tilde{x}_2  \left[\left(8 \pi ^2-21\right) x_1 x_X \left(x_1^2+x_X^2\right)-6 \tilde{w}_2^2 \left(x_1+x_X\right)\right]}{6 N_C^2 w_1 w_X x_X } \nonumber \\
    &+\frac{ N_A x_1 x_X \left[\pi ^2 \tilde{x}_2 \left(2 \tilde{w}_2^2+3 \left(x_1^2+x_X^2\right)\right)+6 \left(w_1^2+w_X^2\right)\right]}{3 N_C^2 w_1 w_X x_X } \nonumber \\
    &+L_h\frac{3 N_A^2 x_1 \tilde{x}_2 \left(x_1^2+x_X^2\right)  }{2 N_C^2 w_1 w_X },
\end{align}

\begin{align}
    (r_{\delta})^{(1)}_{qg}=&\frac{  N_A \tilde{x}_2 \left[N_A \left(3 x_1 x_X+\tilde{w}_2\right) +2 w_1 x_1\right]}{N_C^2 x_1 }\nonumber\\
    &-L_h\frac{4 N_A x_1 \tilde{x}_2 \left(x_1^2+x_X^2\right) }{w_1 w_X },
\end{align}

\begin{align}
    (r_{\delta})^{(2)}_{qg}=&\frac{4 N_A x_1 \left(\tilde{x}_2^2-1\right) }{N_C^2 \tilde{x}_2 }-L_h\frac{2 N_A^2 x_1 \tilde{x}_2 \left(x_1^2+x_X^2\right)  }{N_C^2 w_1 w_X },
\end{align}

\begin{align}
   (r_{\delta})^{(3)}_{qg}= \frac{ N_A x_1 \tilde{x}_2  \left[N_A \left(\tilde{w}_2+3 x_1 x_X\right)+2 w_X x_X\right]}{N_C^2 x_X^2 }
\end{align}

\begin{align}
    (pr_{\delta})^{(1)}_{qg}=\frac{2 N_A x_1 \tilde{x}_2  \left[N_A \left(x_1^2+x_X^2\right) -\tilde{w}_2^2+x_X^2\right]}{N_C^2 w_1 w_X },
\end{align}

\begin{align}
    (pr_{\delta})^{(2)}_{qg}=-\frac{2 N_A x_1 \tilde{x}_2  \left(2 \tilde{w}_2^2+x_1^2+x_X^2\right)}{N_C^2 w_1 w_X },
\end{align}

\begin{align}
    (pr_{\delta})^{(3)}_{qg}=-(pr_{\delta})^{(8)}_{qg}=\frac{2 N_A x_1 \tilde{x}_2  \left[N_A \left(x_1^2+x_X^2\right) -\tilde{w}_2^2+x_1^2\right]}{N_C^2 w_1 w_X },
\end{align}

\begin{align}
    (pr_{\delta})^{(5)}_{qg}=-\frac{N_C^2+1}{4 N_C^2}(pr_{\delta})^{(7)}_{qg}
    =\frac{ N_A \left(N_A+2\right) x_1 \tilde{x}_2 \left(x_1^2+x_X^2\right) }{N_C^2 w_1 w_X },
\end{align}

\begin{align}
    (pr_{\delta})^{(6)}_{qg}=\frac{2 N_A x_1 \tilde{x}_2  \left[N_A \left(x_1^2+x_X^2\right) +\tilde{w}_2^2+2 x_1^2+x_X^2\right]}{N_C^2 w_1 w_X }.
\end{align}

From the above expressions, one can see that the coefficients $U_{ij}^{[k]},\, V_{ij}$ of the distributions can be divergent at 
$\{x_1,\,\tilde{x}_2\}\to\{ 0,\,1\}$ as functions of $\{x_1,\,\tilde{x}_2\}$, at $x_\varrho\to \{0,\,1\}$, $z\to 1$, and $z\to 0 \;\&\; x_\varrho\to \{0,\,1\}$ as functions of 
$\{x_\varrho,\,z\},\, \varrho=1,2$. 
The divergent behavior can  be more prominent at $\{x_\varrho,\,z \}\to 1,\, \varrho=1,2$ because of the $w_1$, $\tilde{w}_2$, $w_X$ and $x_X$ in the denominators. 
For the illustration of these coefficients, the $U_{ij}^{[1]}$, $U_{ij}^{[0]}$, and $V_{ij}$ as functions of different arguments 
$\{x_1,\,\tilde{x}_2\}$ ($\tilde{x}_2$ is regarded as an independent variable), $\{x_1,\,z\}$ and $\{x_2,\,z\}$ are shown in figs.~\ref{fig:sigmahat_Coeffdlog_plus}, \ref{fig:sigmahat_Coeffd_plus}, 
and \ref{fig:sigmahat_Coeffd_delta}, respectively, where $N_C=3,\,N_f=5$, and $\mu=Q$ are chosen. 
The factors 
\begin{align}
    \mathcal{N}^{(a)}[x_1\tilde{x}_2]&=\frac{N_C(1-x_1)(1-\tilde{x}_2)(x_1+\tilde{x}_2-1)}{4\pi^2},
\end{align}
\begin{align}
    \mathcal{N}^{(a)}[x_1z]&=\frac{N_C(1-x_1)x_1(1-z)z}{4\pi^2},
\end{align}
and 
\begin{align}
    \mathcal{N}^{(a)}[x_2z]=\frac{N_C(1-x_2)x_2(1-z)z}{4\pi^2}
\end{align}
are multiplied to the functions with arguments 
$\{x_1,\,\tilde{x}_2\}$, $\{x_1,\,z\}$ and $\{x_2,\,z\}$ respectively to mitigate their strong asymptotic singular behaviors. 
One can see that the functions of $\{x_1,\,\tilde{x}_2\}$ are defined on the region $\{(x_1,\tilde{x}_2)|0<x_1<1,\,0<\tilde{x}_2<1,\,0<x_1+\tilde{x}_2-1<1\}$,
and the functions of $\{x_\varrho,\,z\}$ are defined on the region $\{(x_\varrho,\,z)|0<x_\varrho<1,\,0<z<1,\,\varrho=1,\,2\}$.
When we perform the convolution of the Wilson coefficients and the FFs in eq.~\eqref{eq:def:fac} to get the physical dihadron angular correlation functions, 
the arguments of $U_{ij}^{[k]},\, V_{ij}$ should be $\{x_1,\,z\}$ or $\{x_2,\,z\}$ for the consistency of the definition regions of the 
$m_X^2>0$ and $m_X^2\to 0$ parts of the Wilson coefficients. 

The figs.~\ref{fig:sigmahat_Coeffdlog_plus}, \ref{fig:sigmahat_Coeffd_plus}, and \ref{fig:sigmahat_Coeffd_delta} show that as functions of $\{x_1,\,\tilde{x}_2\}$, 
the coefficients of the delta function and star distributions are mainly divergent at $x_1\to 1\;\& \;\tilde{x}_2\to 1$ in the $q\bar{q}$ sector, 
and at $x_1 \to 1$ in the $qg$ sector; as functions of $\{x_1,\,z\}$, 
the coefficients are mainly divergent at $z\to 1$ in the $q\bar{q}$ sector and at $x_1\to 1,\,z\to 0$ in the $qg$ sector; 
and as functions of $\{x_2,\,z\}$, the coefficients are mainly divergent at $z \to 1$ in the $q\bar
q$ sector, and at $x_2\to 0$ in the $qg$ sector.

\section{Conclusion}
\label{sec:def:recap}

This paper calculates the analytic expressions of the partonic coefficients $\mathcal{C}_{ij;q}(x_1,x_2,z)$ in the triply differential spectrum $\mathrm{d}^3\sigma/(\mathrm{d}\tau_1\mathrm{d}\tau_2\mathrm{d}z)$ of the process $e^+e^- \to H_1 H_2 + X$  at NLO, i.e., $\mathcal{O}(\alpha_s^2)$. Here $x_1(\tau_1)$ and $x_2(\tau_2)$ stand for the momentum fractions of the incident partons (hadrons). $z$ characterizes the opening angle between the detected hadrons.  

At this order, the QCD corrections consist of two components: the real-virtual contributions from one-loop amplitudes for photon decay into three partons, and the double-real contributions induced by tree-level amplitudes for photon decay into four partons. For the former, we use the \texttt{Mathematica} packages \texttt{FeynArts} and \texttt{FeynCalc} to generate amplitudes and perform tensor reduction, and extract the one-loop scalar integrals analytically via \texttt{Package-X} through \texttt{FeynHelpers}.

 The computation of the double real corrections is at the core of this work. To this end, the IBP identities are utilized in the first place to recast the phase space integrals in terms of 6 independent master integrals. To facilitate the subsequent application of the DE approach, we reparametrize the master integrals with the kinematic variables  $\xi\in\{\alpha,y,\bar y\}$, as defined in eqs.~(\ref{eq:def:para:123}-\ref{eq:def:para:5}), in place of $l\in\{x_1,x_2,z\}$, such that the letters of alphabets from the resulting   DEs are free of any quadratic terms in the canonical bases, which  enables us to recursively access the solution of DEs in terms of GPLs. In this work, we figure out the DE solutions up to $\texttt{weight}=5$, as preparation for a future N$^2$LO analysis.

In the three-prong limit where $m_X^2 \to 0$, divergences can arise from the IBP coefficients. To this end,  we resum  the higher order contributions in the dimensional regulator $\varepsilon$ for each master integral based on the asymptotic performance of their DEs. It is observed that  in the limit  $m^2_X\to0$, the leading terms from the resummation-improved master integrals   all possess extra factor $(m^2_X)^{-k\varepsilon}$, with $k$ being integers,  which are capable of  harnessing any emerging divergences from the IBP coefficients, thanks to the dimensional regularization scheme.

After consistently combining the virtual and real corrections with the counter terms in the fragmentation  function renormalization,  we verify that the pole terms indeed cancel out in each partonic channels, echoing the collinear factorization formalism.  
To facilitate the future implementation of our analytic results  in event generators, we further transform the underlying transcendental functions in  $\mathcal{C}_{ij;q}(x_1,x_2,z)$ from GPLs to the classical (poly)logarithms.  
The complete analytic expressions for $\mathcal{C}_{ij;q}(x_1,x_2,z)$ are provided in the ancillary files of the \textit{arXiv} submission.

\acknowledgments

W.-L.J. acknowledges the support of the Natural Sciences and Engineering Research Council of Canada (NSERC).
Z.-Y.S. is supported by President's Fund for Undergraduate Research Training Program at Peking University.
T.-Z.Y. is supported by the European Research Council (ERC) under the European Union's Horizon 2020 research and innovation programme grant agreement 101019620 (ERC Advanced Grant TOPUP) and Guangdong Major Project of Basic and Applied Basic Research~(No. 2020B0301030008). Z.-H.Z. and H.X.Z. are supported by the National Natural Science Foundation of China under contract No. 12425505 and The Fundamental Research Funds for the Central Universities, Peking University.

\appendix

\section{Splitting functions}\label{eq:sec:app:sp}
LO time like splitting functions involved in  eq.~\eqref{eq:def:sp:dec} read \cite{Altarelli:1977zs},
\begin{align}
P^{\mathrm{T},(0)}_{qq}(z) &= 2 C_F \left[ p_{qq}(z) +\frac{3}{2} \delta (1-z) \right] ,\qquad P^{\mathrm{T},(0)}_{gq}(z) = 2 C_F \, p_{gq}(z) \, ,   \\
P^{\mathrm{T},(0)}_{gg}(z) &= 4 C_A \, p_{gg}(z) + \delta(1-z) \left( \frac{11}{3}  C_A - \frac{2}{3}   N_f \right) ,\qquad P^{\mathrm{T},(0)}_{qg}(z) =   \, p_{qg}(z) \, ,
\end{align}
where
\begin{align}
p_{qq}(z) &= 2\left[\frac{1}{1-z}\right]_+-z-1   , \qquad p_{gq}(z) = \frac{1+(1-z)^2}{z}  , \\
p_{gg}(z) &= \left[\frac{1}{1-z}\right]_+-z^2+z+\frac{1}{z}-2 \, , \qquad p_{qg}(z) = z^2+ (1-z)^2 \, .
\end{align}

\bibliographystyle{JHEP}
\bibliography{paper_dihadron}

\providecommand{\href}[2]{#2}\begingroup\raggedright\begin{thebibliography}{100}

\bibitem{Collins:1989gx}
J.C.~Collins, D.E.~Soper and G.F.~Sterman, \emph{{Factorization of Hard Processes in QCD}}, \href{https://doi.org/10.1142/9789814503266_0001}{\emph{Adv. Ser. Direct. High Energy Phys.} {\bfseries 5} (1989) 1} [\href{https://arxiv.org/abs/hep-ph/0409313}{{\ttfamily hep-ph/0409313}}].

\bibitem{Collins:1981ta}
J.C.~Collins and G.F.~Sterman, \emph{{Soft Partons in {QCD}}}, \href{https://doi.org/10.1016/0550-3213(81)90370-9}{\emph{Nucl. Phys. B} {\bfseries 185} (1981) 172}.

\bibitem{Amati:1978by}
D.~Amati, R.~Petronzio and G.~Veneziano, \emph{{Relating Hard QCD Processes Through Universality of Mass Singularities. 2.}}, \href{https://doi.org/10.1016/0550-3213(78)90430-3}{\emph{Nucl. Phys. B} {\bfseries 146} (1978) 29}.

\bibitem{Ellis:1978ty}
R.K.~Ellis, H.~Georgi, M.~Machacek, H.D.~Politzer and G.G.~Ross, \emph{{Perturbation Theory and the Parton Model in QCD}}, \href{https://doi.org/10.1016/0550-3213(79)90105-6}{\emph{Nucl. Phys. B} {\bfseries 152} (1979) 285}.

\bibitem{Libby:1978qf}
S.B.~Libby and G.F.~Sterman, \emph{{Jet and Lepton Pair Production in High-Energy Lepton-Hadron and Hadron-Hadron Scattering}}, \href{https://doi.org/10.1103/PhysRevD.18.3252}{\emph{Phys. Rev. D} {\bfseries 18} (1978) 3252}.

\bibitem{Almasy:2011eq}
A.A.~Almasy, S.~Moch and A.~Vogt, \emph{{On the Next-to-Next-to-Leading Order Evolution of Flavour-Singlet Fragmentation Functions}}, \href{https://doi.org/10.1016/j.nuclphysb.2011.08.028}{\emph{Nucl. Phys. B} {\bfseries 854} (2012) 133} [\href{https://arxiv.org/abs/1107.2263}{{\ttfamily 1107.2263}}].

\bibitem{Rijken:1996vr}
P.J.~Rijken and W.L.~van Neerven, \emph{{O (alpha-s**2) contributions to the longitudinal fragmentation function in e+ e- annihilation}}, \href{https://doi.org/10.1016/0370-2693(96)00898-2}{\emph{Phys. Lett. B} {\bfseries 386} (1996) 422} [\href{https://arxiv.org/abs/hep-ph/9604436}{{\ttfamily hep-ph/9604436}}].

\bibitem{Rijken:1996npa}
P.J.~Rijken and W.L.~van Neerven, \emph{{O (alpha-s**2) contributions to the asymmetric fragmentation function in e+ e- annihilation}}, \href{https://doi.org/10.1016/S0370-2693(96)01529-8}{\emph{Phys. Lett. B} {\bfseries 392} (1997) 207} [\href{https://arxiv.org/abs/hep-ph/9609379}{{\ttfamily hep-ph/9609379}}].

\bibitem{Rijken:1996ns}
P.J.~Rijken and W.L.~van Neerven, \emph{{Higher order QCD corrections to the transverse and longitudinal fragmentation functions in electron - positron annihilation}}, \href{https://doi.org/10.1016/S0550-3213(96)00669-4}{\emph{Nucl. Phys. B} {\bfseries 487} (1997) 233} [\href{https://arxiv.org/abs/hep-ph/9609377}{{\ttfamily hep-ph/9609377}}].

\bibitem{Mitov:2006wy}
A.~Mitov and S.-O.~Moch, \emph{{QCD Corrections to Semi-Inclusive Hadron Production in Electron-Positron Annihilation at Two Loops}}, \href{https://doi.org/10.1016/j.nuclphysb.2006.05.018}{\emph{Nucl. Phys. B} {\bfseries 751} (2006) 18} [\href{https://arxiv.org/abs/hep-ph/0604160}{{\ttfamily hep-ph/0604160}}].

\bibitem{Xu:2024rbt}
Z.~Xu and H.X.~Zhu, \emph{{Threshold Resummation for Semi-Inclusive Single-Hadron Production with Effective Field Theory}},  \href{https://arxiv.org/abs/2411.11595}{{\ttfamily 2411.11595}}.

\bibitem{He:2025hin}
C.-Q.~He, H.~Xing, T.-Z.~Yang and H.X.~Zhu, \emph{{Single-inclusive hadron production in electron-positron annihilation at next-to-next-to-next-to-leading order in QCD}},  \href{https://arxiv.org/abs/2503.20441}{{\ttfamily 2503.20441}}.

\bibitem{Goyal:2023zdi}
S.~Goyal, S.-O.~Moch, V.~Pathak, N.~Rana and V.~Ravindran, \emph{{Next-to-Next-to-Leading Order QCD Corrections to Semi-Inclusive Deep-Inelastic Scattering}}, \href{https://doi.org/10.1103/PhysRevLett.132.251902}{\emph{Phys. Rev. Lett.} {\bfseries 132} (2024) 251902} [\href{https://arxiv.org/abs/2312.17711}{{\ttfamily 2312.17711}}].

\bibitem{Bonino:2024qbh}
L.~Bonino, T.~Gehrmann and G.~Stagnitto, \emph{{Semi-Inclusive Deep-Inelastic Scattering at Next-to-Next-to-Leading Order in QCD}}, \href{https://doi.org/10.1103/PhysRevLett.132.251901}{\emph{Phys. Rev. Lett.} {\bfseries 132} (2024) 251901} [\href{https://arxiv.org/abs/2401.16281}{{\ttfamily 2401.16281}}].

\bibitem{Liu:2023fsq}
C.~Liu, X.~Shen, B.~Zhou and J.~Gao, \emph{{Automated calculation of jet fragmentation at NLO in QCD}}, \href{https://doi.org/10.1007/JHEP09(2023)108}{\emph{JHEP} {\bfseries 09} (2023) 108} [\href{https://arxiv.org/abs/2305.14620}{{\ttfamily 2305.14620}}].

\bibitem{Zhou:2024cyk}
B.~Zhou and J.~Gao, \emph{{The impact of data from future lepton colliders on light hadrons fragmentation functions}}, \href{https://doi.org/10.1007/JHEP02(2025)003}{\emph{JHEP} {\bfseries 02} (2025) 003} [\href{https://arxiv.org/abs/2407.10059}{{\ttfamily 2407.10059}}].

\bibitem{Bertone:2017tyb}
{\scshape NNPDF} collaboration, \emph{{A determination of the fragmentation functions of pions, kaons, and protons with faithful uncertainties}}, \href{https://doi.org/10.1140/epjc/s10052-017-5088-y}{\emph{Eur. Phys. J. C} {\bfseries 77} (2017) 516} [\href{https://arxiv.org/abs/1706.07049}{{\ttfamily 1706.07049}}].

\bibitem{Soleymaninia:2018uiv}
M.~Soleymaninia, M.~Goharipour and H.~Khanpour, \emph{{First QCD analysis of charged hadron fragmentation functions and their uncertainties at next-to-next-to-leading order}}, \href{https://doi.org/10.1103/PhysRevD.98.074002}{\emph{Phys. Rev. D} {\bfseries 98} (2018) 074002} [\href{https://arxiv.org/abs/1805.04847}{{\ttfamily 1805.04847}}].

\bibitem{Borsa:2022vvp}
I.~Borsa, R.~Sassot, D.~de~Florian, M.~Stratmann and W.~Vogelsang, \emph{{Towards a Global QCD Analysis of Fragmentation Functions at Next-to-Next-to-Leading Order Accuracy}}, \href{https://doi.org/10.1103/PhysRevLett.129.012002}{\emph{Phys. Rev. Lett.} {\bfseries 129} (2022) 012002} [\href{https://arxiv.org/abs/2202.05060}{{\ttfamily 2202.05060}}].

\bibitem{AbdulKhalek:2022laj}
{\scshape MAP (Multi-dimensional Analyses of Partonic distributions)} collaboration, \emph{{Pion and kaon fragmentation functions at next-to-next-to-leading order}}, \href{https://doi.org/10.1016/j.physletb.2022.137456}{\emph{Phys. Lett. B} {\bfseries 834} (2022) 137456} [\href{https://arxiv.org/abs/2204.10331}{{\ttfamily 2204.10331}}].

\bibitem{Gao:2025hlm}
J.~Gao, X.~Shen, H.~Xing, Y.~Zhao and B.~Zhou, \emph{{Fragmentation functions of charged hadrons at next-to-next-to-leading order and constraints on proton PDFs}},  \href{https://arxiv.org/abs/2502.17837}{{\ttfamily 2502.17837}}.

\bibitem{Gao:2025bko}
J.~Gao, C.~Liu, M.~Li, X.~Shen, H.~Xing, Y.~Zhao et~al., \emph{{Global analysis of fragmentation functions to light neutral hadrons}},  \href{https://arxiv.org/abs/2503.21311}{{\ttfamily 2503.21311}}.

\bibitem{Collins:1985kw}
J.C.~Collins and D.E.~Soper, \emph{{Transverse Momentum in $e^+ e^- \to$ a + $B$ + X}}, {\emph{Acta Phys. Polon. B} {\bfseries 16} (1985) 1047}.

\bibitem{Collins:1981uk}
J.C.~Collins and D.E.~Soper, \emph{{Back-To-Back Jets in QCD}}, \href{https://doi.org/10.1016/0550-3213(81)90339-4}{\emph{Nucl. Phys. B} {\bfseries 193} (1981) 381}.

\bibitem{Collins:1981va}
J.C.~Collins and D.E.~Soper, \emph{{Back-To-Back Jets: Fourier Transform from B to K-Transverse}}, \href{https://doi.org/10.1016/0550-3213(82)90453-9}{\emph{Nucl. Phys. B} {\bfseries 197} (1982) 446}.

\bibitem{Collins:1984kg}
J.C.~Collins, D.E.~Soper and G.F.~Sterman, \emph{{Transverse Momentum Distribution in Drell-Yan Pair and W and Z Boson Production}}, \href{https://doi.org/10.1016/0550-3213(85)90479-1}{\emph{Nucl. Phys. B} {\bfseries 250} (1985) 199}.

\bibitem{Catani:2000vq}
S.~Catani, D.~de~Florian and M.~Grazzini, \emph{{Universality of nonleading logarithmic contributions in transverse momentum distributions}}, \href{https://doi.org/10.1016/S0550-3213(00)00617-9}{\emph{Nucl. Phys. B} {\bfseries 596} (2001) 299} [\href{https://arxiv.org/abs/hep-ph/0008184}{{\ttfamily hep-ph/0008184}}].

\bibitem{Bozzi:2005wk}
G.~Bozzi, S.~Catani, D.~de~Florian and M.~Grazzini, \emph{{Transverse-momentum resummation and the spectrum of the Higgs boson at the LHC}}, \href{https://doi.org/10.1016/j.nuclphysb.2005.12.022}{\emph{Nucl. Phys. B} {\bfseries 737} (2006) 73} [\href{https://arxiv.org/abs/hep-ph/0508068}{{\ttfamily hep-ph/0508068}}].

\bibitem{Bozzi:2007pn}
G.~Bozzi, S.~Catani, D.~de~Florian and M.~Grazzini, \emph{{Higgs boson production at the LHC: Transverse-momentum resummation and rapidity dependence}}, \href{https://doi.org/10.1016/j.nuclphysb.2007.09.034}{\emph{Nucl. Phys. B} {\bfseries 791} (2008) 1} [\href{https://arxiv.org/abs/0705.3887}{{\ttfamily 0705.3887}}].

\bibitem{Collins:2011zzd}
J.~Collins, \emph{{Foundations of Perturbative QCD}}, vol.~32, Cambridge University Press (2011), \href{https://doi.org/10.1017/9781009401845}{10.1017/9781009401845}.

\bibitem{Ebert:2016gcn}
M.A.~Ebert and F.J.~Tackmann, \emph{{Resummation of Transverse Momentum Distributions in Distribution Space}}, \href{https://doi.org/10.1007/JHEP02(2017)110}{\emph{JHEP} {\bfseries 02} (2017) 110} [\href{https://arxiv.org/abs/1611.08610}{{\ttfamily 1611.08610}}].

\bibitem{Monni:2016ktx}
P.F.~Monni, E.~Re and P.~Torrielli, \emph{{Higgs Transverse-Momentum Resummation in Direct Space}}, \href{https://doi.org/10.1103/PhysRevLett.116.242001}{\emph{Phys. Rev. Lett.} {\bfseries 116} (2016) 242001} [\href{https://arxiv.org/abs/1604.02191}{{\ttfamily 1604.02191}}].

\bibitem{Bizon:2017rah}
W.~Bizon, P.F.~Monni, E.~Re, L.~Rottoli and P.~Torrielli, \emph{{Momentum-space resummation for transverse observables and the Higgs p$_{\perp}$ at N$^{3}$LL+NNLO}}, \href{https://doi.org/10.1007/JHEP02(2018)108}{\emph{JHEP} {\bfseries 02} (2018) 108} [\href{https://arxiv.org/abs/1705.09127}{{\ttfamily 1705.09127}}].

\bibitem{Bizon:2019zgf}
W.~Bizon, A.~Gehrmann-De~Ridder, T.~Gehrmann, N.~Glover, A.~Huss, P.F.~Monni et~al., \emph{{The transverse momentum spectrum of weak gauge bosons at N ${}^3$ LL + NNLO}}, \href{https://doi.org/10.1140/epjc/s10052-019-7324-0}{\emph{Eur. Phys. J. C} {\bfseries 79} (2019) 868} [\href{https://arxiv.org/abs/1905.05171}{{\ttfamily 1905.05171}}].

\bibitem{Bizon:2018foh}
W.~Bizo\'n, X.~Chen, A.~Gehrmann-De~Ridder, T.~Gehrmann, N.~Glover, A.~Huss et~al., \emph{{Fiducial distributions in Higgs and Drell-Yan production at N$^{3}$LL+NNLO}}, \href{https://doi.org/10.1007/JHEP12(2018)132}{\emph{JHEP} {\bfseries 12} (2018) 132} [\href{https://arxiv.org/abs/1805.05916}{{\ttfamily 1805.05916}}].

\bibitem{Becher:2010tm}
T.~Becher and M.~Neubert, \emph{{Drell-Yan Production at Small $q_T$, Transverse Parton Distributions and the Collinear Anomaly}}, \href{https://doi.org/10.1140/epjc/s10052-011-1665-7}{\emph{Eur. Phys. J. C} {\bfseries 71} (2011) 1665} [\href{https://arxiv.org/abs/1007.4005}{{\ttfamily 1007.4005}}].

\bibitem{GarciaEchevarria:2011rb}
M.G.~Echevarria, A.~Idilbi and I.~Scimemi, \emph{{Factorization Theorem For Drell-Yan At Low $q_T$ And Transverse Momentum Distributions On-The-Light-Cone}}, \href{https://doi.org/10.1007/JHEP07(2012)002}{\emph{JHEP} {\bfseries 07} (2012) 002} [\href{https://arxiv.org/abs/1111.4996}{{\ttfamily 1111.4996}}].

\bibitem{Becher:2011dz}
T.~Becher and G.~Bell, \emph{{Analytic Regularization in Soft-Collinear Effective Theory}}, \href{https://doi.org/10.1016/j.physletb.2012.05.016}{\emph{Phys. Lett. B} {\bfseries 713} (2012) 41} [\href{https://arxiv.org/abs/1112.3907}{{\ttfamily 1112.3907}}].

\bibitem{Chiu:2011qc}
J.-y.~Chiu, A.~Jain, D.~Neill and I.Z.~Rothstein, \emph{{The Rapidity Renormalization Group}}, \href{https://doi.org/10.1103/PhysRevLett.108.151601}{\emph{Phys. Rev. Lett.} {\bfseries 108} (2012) 151601} [\href{https://arxiv.org/abs/1104.0881}{{\ttfamily 1104.0881}}].

\bibitem{Chiu:2012ir}
J.-Y.~Chiu, A.~Jain, D.~Neill and I.Z.~Rothstein, \emph{{A Formalism for the Systematic Treatment of Rapidity Logarithms in Quantum Field Theory}}, \href{https://doi.org/10.1007/JHEP05(2012)084}{\emph{JHEP} {\bfseries 05} (2012) 084} [\href{https://arxiv.org/abs/1202.0814}{{\ttfamily 1202.0814}}].

\bibitem{Li:2016axz}
Y.~Li, D.~Neill and H.X.~Zhu, \emph{{An exponential regulator for rapidity divergences}}, \href{https://doi.org/10.1016/j.nuclphysb.2020.115193}{\emph{Nucl. Phys. B} {\bfseries 960} (2020) 115193} [\href{https://arxiv.org/abs/1604.00392}{{\ttfamily 1604.00392}}].

\bibitem{Li:2016ctv}
Y.~Li and H.X.~Zhu, \emph{{Bootstrapping Rapidity Anomalous Dimensions for Transverse-Momentum Resummation}}, \href{https://doi.org/10.1103/PhysRevLett.118.022004}{\emph{Phys. Rev. Lett.} {\bfseries 118} (2017) 022004} [\href{https://arxiv.org/abs/1604.01404}{{\ttfamily 1604.01404}}].

\bibitem{Moult:2018jzp}
I.~Moult and H.X.~Zhu, \emph{{Simplicity from Recoil: The Three-Loop Soft Function and Factorization for the Energy-Energy Correlation}}, \href{https://doi.org/10.1007/JHEP08(2018)160}{\emph{JHEP} {\bfseries 08} (2018) 160} [\href{https://arxiv.org/abs/1801.02627}{{\ttfamily 1801.02627}}].

\bibitem{Moffat:2019pci}
E.~Moffat, T.C.~Rogers, N.~Sato and A.~Signori, \emph{{Collinear factorization in wide-angle hadron pair production in $e^+e^-$ annihilation}}, \href{https://doi.org/10.1103/PhysRevD.104.059904}{\emph{Phys. Rev. D} {\bfseries 100} (2019) 094014} [\href{https://arxiv.org/abs/1909.02951}{{\ttfamily 1909.02951}}].

\bibitem{Hautmann:2022xuc}
F.~Hautmann, M.~Hentschinski, L.~Keersmaekers, A.~Kusina, K.~Kutak and A.~Lelek, \emph{{A parton branching with transverse momentum dependent splitting functions}}, \href{https://doi.org/10.1016/j.physletb.2022.137276}{\emph{Phys. Lett. B} {\bfseries 833} (2022) 137276} [\href{https://arxiv.org/abs/2205.15873}{{\ttfamily 2205.15873}}].

\bibitem{Gao:2023ivm}
A.~Gao, H.T.~Li, I.~Moult and H.X.~Zhu, \emph{{The transverse energy-energy correlator at next-to-next-to-next-to-leading logarithm}}, \href{https://doi.org/10.1007/JHEP09(2024)072}{\emph{JHEP} {\bfseries 09} (2024) 072} [\href{https://arxiv.org/abs/2312.16408}{{\ttfamily 2312.16408}}].

\bibitem{Boussarie:2023izj}
R.~Boussarie et~al., \emph{{TMD Handbook}},  \href{https://arxiv.org/abs/2304.03302}{{\ttfamily 2304.03302}}.

\bibitem{Ji:2004wu}
X.-d.~Ji, J.-p.~Ma and F.~Yuan, \emph{{QCD factorization for semi-inclusive deep-inelastic scattering at low transverse momentum}}, \href{https://doi.org/10.1103/PhysRevD.71.034005}{\emph{Phys. Rev. D} {\bfseries 71} (2005) 034005} [\href{https://arxiv.org/abs/hep-ph/0404183}{{\ttfamily hep-ph/0404183}}].

\bibitem{Ji:2004xq}
X.-d.~Ji, J.-P.~Ma and F.~Yuan, \emph{{QCD factorization for spin-dependent cross sections in DIS and Drell-Yan processes at low transverse momentum}}, \href{https://doi.org/10.1016/j.physletb.2004.07.026}{\emph{Phys. Lett. B} {\bfseries 597} (2004) 299} [\href{https://arxiv.org/abs/hep-ph/0405085}{{\ttfamily hep-ph/0405085}}].

\bibitem{Bacchetta:2006tn}
A.~Bacchetta, M.~Diehl, K.~Goeke, A.~Metz, P.J.~Mulders and M.~Schlegel, \emph{{Semi-inclusive deep inelastic scattering at small transverse momentum}}, \href{https://doi.org/10.1088/1126-6708/2007/02/093}{\emph{JHEP} {\bfseries 02} (2007) 093} [\href{https://arxiv.org/abs/hep-ph/0611265}{{\ttfamily hep-ph/0611265}}].

\bibitem{Bacchetta:2017gcc}
A.~Bacchetta, F.~Delcarro, C.~Pisano, M.~Radici and A.~Signori, \emph{{Extraction of partonic transverse momentum distributions from semi-inclusive deep-inelastic scattering, Drell-Yan and Z-boson production}}, \href{https://doi.org/10.1007/JHEP06(2017)081}{\emph{JHEP} {\bfseries 06} (2017) 081} [\href{https://arxiv.org/abs/1703.10157}{{\ttfamily 1703.10157}}].

\bibitem{Kang:2015msa}
Z.-B.~Kang, A.~Prokudin, P.~Sun and F.~Yuan, \emph{{Extraction of Quark Transversity Distribution and Collins Fragmentation Functions with QCD Evolution}}, \href{https://doi.org/10.1103/PhysRevD.93.014009}{\emph{Phys. Rev. D} {\bfseries 93} (2016) 014009} [\href{https://arxiv.org/abs/1505.05589}{{\ttfamily 1505.05589}}].

\bibitem{Bastami:2018xqd}
S.~Bastami et~al., \emph{{Semi-Inclusive Deep Inelastic Scattering in Wandzura-Wilczek-type approximation}}, \href{https://doi.org/10.1007/JHEP06(2019)007}{\emph{JHEP} {\bfseries 06} (2019) 007} [\href{https://arxiv.org/abs/1807.10606}{{\ttfamily 1807.10606}}].

\bibitem{Boer:2011xd}
D.~Boer, L.~Gamberg, B.~Musch and A.~Prokudin, \emph{{Bessel-Weighted Asymmetries in Semi Inclusive Deep Inelastic Scattering}}, \href{https://doi.org/10.1007/JHEP10(2011)021}{\emph{JHEP} {\bfseries 10} (2011) 021} [\href{https://arxiv.org/abs/1107.5294}{{\ttfamily 1107.5294}}].

\bibitem{Collins:2004nx}
J.C.~Collins and A.~Metz, \emph{{Universality of soft and collinear factors in hard-scattering factorization}}, \href{https://doi.org/10.1103/PhysRevLett.93.252001}{\emph{Phys. Rev. Lett.} {\bfseries 93} (2004) 252001} [\href{https://arxiv.org/abs/hep-ph/0408249}{{\ttfamily hep-ph/0408249}}].

\bibitem{Xue:2020xba}
S.-C.~Xue, X.~Wang, D.-M.~Li and Z.~Lu, \emph{{The Collins asymmetry in electroproduction of Kaon at the electron ion colliders within TMD factorization}}, \href{https://doi.org/10.1140/epjc/s10052-020-8263-5}{\emph{Eur. Phys. J. C} {\bfseries 80} (2020) 685} [\href{https://arxiv.org/abs/2003.05679}{{\ttfamily 2003.05679}}].

\bibitem{Sun:2014dqm}
P.~Sun, J.~Isaacson, C.P.~Yuan and F.~Yuan, \emph{{Nonperturbative functions for SIDIS and Drell\textendash{}Yan processes}}, \href{https://doi.org/10.1142/S0217751X18410063}{\emph{Int. J. Mod. Phys. A} {\bfseries 33} (2018) 1841006} [\href{https://arxiv.org/abs/1406.3073}{{\ttfamily 1406.3073}}].

\bibitem{Li:2020bub}
H.T.~Li, I.~Vitev and Y.J.~Zhu, \emph{{Transverse-Energy-Energy Correlations in Deep Inelastic Scattering}}, \href{https://doi.org/10.1007/JHEP11(2020)051}{\emph{JHEP} {\bfseries 11} (2020) 051} [\href{https://arxiv.org/abs/2006.02437}{{\ttfamily 2006.02437}}].

\bibitem{Caucal:2023nci}
P.~Caucal, F.~Salazar, B.~Schenke, T.~Stebel and R.~Venugopalan, \emph{{Back-to-back inclusive dijets in DIS at small x: gluon Weizs\"acker-Williams distribution at NLO}}, \href{https://doi.org/10.1007/JHEP08(2023)062}{\emph{JHEP} {\bfseries 08} (2023) 062} [\href{https://arxiv.org/abs/2304.03304}{{\ttfamily 2304.03304}}].

\bibitem{Scimemi:2019cmh}
I.~Scimemi and A.~Vladimirov, \emph{{Non-perturbative structure of semi-inclusive deep-inelastic and Drell-Yan scattering at small transverse momentum}}, \href{https://doi.org/10.1007/JHEP06(2020)137}{\emph{JHEP} {\bfseries 06} (2020) 137} [\href{https://arxiv.org/abs/1912.06532}{{\ttfamily 1912.06532}}].

\bibitem{Bury:2021sue}
M.~Bury, A.~Prokudin and A.~Vladimirov, \emph{{Extraction of the Sivers function from SIDIS, Drell-Yan, and $W^\pm/Z$ boson production data with TMD evolution}}, \href{https://doi.org/10.1007/JHEP05(2021)151}{\emph{JHEP} {\bfseries 05} (2021) 151} [\href{https://arxiv.org/abs/2103.03270}{{\ttfamily 2103.03270}}].

\bibitem{Li:2021txc}
H.T.~Li, Y.~Makris and I.~Vitev, \emph{{Energy-energy correlators in Deep Inelastic Scattering}}, \href{https://doi.org/10.1103/PhysRevD.103.094005}{\emph{Phys. Rev. D} {\bfseries 103} (2021) 094005} [\href{https://arxiv.org/abs/2102.05669}{{\ttfamily 2102.05669}}].

\bibitem{Bhattacharya:2025bqa}
S.~Bhattacharya, Z.-B.~Kang, D.~Padilla and J.~Penttala, \emph{{Probing the Sivers Asymmetry with Transverse Energy-Energy Correlators in the Small-$x$ Regime}},  \href{https://arxiv.org/abs/2504.10475}{{\ttfamily 2504.10475}}.

\bibitem{Bozzi:2010xn}
G.~Bozzi, S.~Catani, G.~Ferrera, D.~de~Florian and M.~Grazzini, \emph{{Production of Drell-Yan lepton pairs in hadron collisions: Transverse-momentum resummation at next-to-next-to-leading logarithmic accuracy}}, \href{https://doi.org/10.1016/j.physletb.2010.12.024}{\emph{Phys. Lett. B} {\bfseries 696} (2011) 207} [\href{https://arxiv.org/abs/1007.2351}{{\ttfamily 1007.2351}}].

\bibitem{Becher:2011xn}
T.~Becher, M.~Neubert and D.~Wilhelm, \emph{{Electroweak Gauge-Boson Production at Small $q_T$: Infrared Safety from the Collinear Anomaly}}, \href{https://doi.org/10.1007/JHEP02(2012)124}{\emph{JHEP} {\bfseries 02} (2012) 124} [\href{https://arxiv.org/abs/1109.6027}{{\ttfamily 1109.6027}}].

\bibitem{Banfi:2011dx}
A.~Banfi, M.~Dasgupta and S.~Marzani, \emph{{QCD predictions for new variables to study dilepton transverse momenta at hadron colliders}}, \href{https://doi.org/10.1016/j.physletb.2011.05.028}{\emph{Phys. Lett. B} {\bfseries 701} (2011) 75} [\href{https://arxiv.org/abs/1102.3594}{{\ttfamily 1102.3594}}].

\bibitem{Banfi:2011dm}
A.~Banfi, M.~Dasgupta, S.~Marzani and L.~Tomlinson, \emph{{Probing the low transverse momentum domain of Z production with novel variables}}, \href{https://doi.org/10.1007/JHEP01(2012)044}{\emph{JHEP} {\bfseries 01} (2012) 044} [\href{https://arxiv.org/abs/1110.4009}{{\ttfamily 1110.4009}}].

\bibitem{Banfi:2012du}
A.~Banfi, M.~Dasgupta, S.~Marzani and L.~Tomlinson, \emph{{Predictions for Drell-Yan $\phi^*$ and $Q_T$ observables at the LHC}}, \href{https://doi.org/10.1016/j.physletb.2012.07.035}{\emph{Phys. Lett. B} {\bfseries 715} (2012) 152} [\href{https://arxiv.org/abs/1205.4760}{{\ttfamily 1205.4760}}].

\bibitem{Catani:2015vma}
S.~Catani, D.~de~Florian, G.~Ferrera and M.~Grazzini, \emph{{Vector boson production at hadron colliders: transverse-momentum resummation and leptonic decay}}, \href{https://doi.org/10.1007/JHEP12(2015)047}{\emph{JHEP} {\bfseries 12} (2015) 047} [\href{https://arxiv.org/abs/1507.06937}{{\ttfamily 1507.06937}}].

\bibitem{Scimemi:2017etj}
I.~Scimemi and A.~Vladimirov, \emph{{Analysis of vector boson production within TMD factorization}}, \href{https://doi.org/10.1140/epjc/s10052-018-5557-y}{\emph{Eur. Phys. J. C} {\bfseries 78} (2018) 89} [\href{https://arxiv.org/abs/1706.01473}{{\ttfamily 1706.01473}}].

\bibitem{Bacchetta:2019sam}
A.~Bacchetta, V.~Bertone, C.~Bissolotti, G.~Bozzi, F.~Delcarro, F.~Piacenza et~al., \emph{{Transverse-momentum-dependent parton distributions up to N$^{3}$LL from Drell-Yan data}}, \href{https://doi.org/10.1007/JHEP07(2020)117}{\emph{JHEP} {\bfseries 07} (2020) 117} [\href{https://arxiv.org/abs/1912.07550}{{\ttfamily 1912.07550}}].

\bibitem{Becher:2020ugp}
T.~Becher and T.~Neumann, \emph{{Fiducial $q_T$ resummation of color-singlet processes at N$^3$LL+NNLO}}, \href{https://doi.org/10.1007/JHEP03(2021)199}{\emph{JHEP} {\bfseries 03} (2021) 199} [\href{https://arxiv.org/abs/2009.11437}{{\ttfamily 2009.11437}}].

\bibitem{Ebert:2020dfc}
M.A.~Ebert, J.K.L.~Michel, I.W.~Stewart and F.J.~Tackmann, \emph{{Drell-Yan $q_{T}$ resummation of fiducial power corrections at N$^{3}$LL}}, \href{https://doi.org/10.1007/JHEP04(2021)102}{\emph{JHEP} {\bfseries 04} (2021) 102} [\href{https://arxiv.org/abs/2006.11382}{{\ttfamily 2006.11382}}].

\bibitem{Re:2021con}
E.~Re, L.~Rottoli and P.~Torrielli, \emph{{Fiducial Higgs and Drell-Yan distributions at N$^3$LL$^\prime$+NNLO with RadISH}},  \href{https://arxiv.org/abs/2104.07509}{{\ttfamily 2104.07509}}.

\bibitem{Camarda:2021ict}
S.~Camarda, L.~Cieri and G.~Ferrera, \emph{{Drell\textendash{}Yan lepton-pair production: qT resummation at N3LL accuracy and fiducial cross sections at N3LO}}, \href{https://doi.org/10.1103/PhysRevD.104.L111503}{\emph{Phys. Rev. D} {\bfseries 104} (2021) L111503} [\href{https://arxiv.org/abs/2103.04974}{{\ttfamily 2103.04974}}].

\bibitem{Ju:2021lah}
W.-L.~Ju and M.~Sch\"onherr, \emph{{The q$_{T}$ and \ensuremath{\Delta}\ensuremath{\phi} spectra in W and Z production at the LHC at N$^{3}$LL'+N$^{2}$LO}}, \href{https://doi.org/10.1007/JHEP10(2021)088}{\emph{JHEP} {\bfseries 10} (2021) 088} [\href{https://arxiv.org/abs/2106.11260}{{\ttfamily 2106.11260}}].

\bibitem{Camarda:2023dqn}
S.~Camarda, L.~Cieri and G.~Ferrera, \emph{{Drell\textendash{}Yan lepton-pair production: qT resummation at N4LL accuracy}}, \href{https://doi.org/10.1016/j.physletb.2023.138125}{\emph{Phys. Lett. B} {\bfseries 845} (2023) 138125} [\href{https://arxiv.org/abs/2303.12781}{{\ttfamily 2303.12781}}].

\bibitem{Neumann:2022lft}
T.~Neumann and J.~Campbell, \emph{{Fiducial Drell-Yan production at the LHC improved by transverse-momentum resummation at N4LLp+N3LO}}, \href{https://doi.org/10.1103/PhysRevD.107.L011506}{\emph{Phys. Rev. D} {\bfseries 107} (2023) L011506} [\href{https://arxiv.org/abs/2207.07056}{{\ttfamily 2207.07056}}].

\bibitem{Moos:2023yfa}
V.~Moos, I.~Scimemi, A.~Vladimirov and P.~Zurita, \emph{{Extraction of unpolarized transverse momentum distributions from the fit of Drell-Yan data at N$^{4}$LL}}, \href{https://doi.org/10.1007/JHEP05(2024)036}{\emph{JHEP} {\bfseries 05} (2024) 036} [\href{https://arxiv.org/abs/2305.07473}{{\ttfamily 2305.07473}}].

\bibitem{Bubanja:2023nrd}
I.~Bubanja et~al., \emph{{The small $k_{\textrm{T}}$region in Drell\textendash{}Yan production at next-to-leading order with the parton branching method}}, \href{https://doi.org/10.1140/epjc/s10052-024-12507-0}{\emph{Eur. Phys. J. C} {\bfseries 84} (2024) 154} [\href{https://arxiv.org/abs/2312.08655}{{\ttfamily 2312.08655}}].

\bibitem{Billis:2024dqq}
G.~Billis, J.K.L.~Michel and F.J.~Tackmann, \emph{{Drell-Yan transverse-momentum spectra at N$^{3}$LL$\prime$ and approximate N$^{4}$LL with SCETlib}}, \href{https://doi.org/10.1007/JHEP02(2025)170}{\emph{JHEP} {\bfseries 02} (2025) 170} [\href{https://arxiv.org/abs/2411.16004}{{\ttfamily 2411.16004}}].

\bibitem{Hautmann:2025fkw}
F.~Hautmann, L.~Keersmaekers, A.~Lelek, S.S.~Barzani and S.~Taheri~Monfared, \emph{{Collinear and TMD distributions with dynamical soft-gluon resolution scale}},  \href{https://arxiv.org/abs/2502.19380}{{\ttfamily 2502.19380}}.

\bibitem{Belle:2005dmx}
{\scshape Belle} collaboration, \emph{{Measurement of azimuthal asymmetries in inclusive production of hadron pairs in e+ e- annihilation at Belle}}, \href{https://doi.org/10.1103/PhysRevLett.96.232002}{\emph{Phys. Rev. Lett.} {\bfseries 96} (2006) 232002} [\href{https://arxiv.org/abs/hep-ex/0507063}{{\ttfamily hep-ex/0507063}}].

\bibitem{Belle:2008fdv}
{\scshape Belle} collaboration, \emph{{Measurement of Azimuthal Asymmetries in Inclusive Production of Hadron Pairs in e+e- Annihilation at s**(1/2) = 10.58-GeV}}, \href{https://doi.org/10.1103/PhysRevD.78.032011}{\emph{Phys. Rev. D} {\bfseries 78} (2008) 032011} [\href{https://arxiv.org/abs/0805.2975}{{\ttfamily 0805.2975}}].

\bibitem{Belle:2019nve}
{\scshape Belle} collaboration, \emph{{Azimuthal asymmetries of back-to-back $\pi^\pm-(\pi^0,\eta,\pi^\pm)$ pairs in $e^+e^-$ annihilation}}, \href{https://doi.org/10.1103/PhysRevD.100.092008}{\emph{Phys. Rev. D} {\bfseries 100} (2019) 092008} [\href{https://arxiv.org/abs/1909.01857}{{\ttfamily 1909.01857}}].

\bibitem{BaBar:2013jdt}
{\scshape BaBar} collaboration, \emph{{Measurement of Collins asymmetries in inclusive production of charged pion pairs in $e^+e^-$ annihilation at BABAR}}, \href{https://doi.org/10.1103/PhysRevD.90.052003}{\emph{Phys. Rev. D} {\bfseries 90} (2014) 052003} [\href{https://arxiv.org/abs/1309.5278}{{\ttfamily 1309.5278}}].

\bibitem{BESIII:2015fyw}
{\scshape BESIII} collaboration, \emph{{Measurement of azimuthal asymmetries in inclusive charged dipion production in $e^+e^-$ annihilations at $\sqrt{s}$ = 3.65 GeV}}, \href{https://doi.org/10.1103/PhysRevLett.116.042001}{\emph{Phys. Rev. Lett.} {\bfseries 116} (2016) 042001} [\href{https://arxiv.org/abs/1507.06824}{{\ttfamily 1507.06824}}].

\bibitem{Boer:2008fr}
D.~Boer, \emph{{Angular dependences in inclusive two-hadron production at BELLE}}, \href{https://doi.org/10.1016/j.nuclphysb.2008.06.011}{\emph{Nucl. Phys. B} {\bfseries 806} (2009) 23} [\href{https://arxiv.org/abs/0804.2408}{{\ttfamily 0804.2408}}].

\bibitem{Metz:2016swz}
A.~Metz and A.~Vossen, \emph{{Parton Fragmentation Functions}}, \href{https://doi.org/10.1016/j.ppnp.2016.08.003}{\emph{Prog. Part. Nucl. Phys.} {\bfseries 91} (2016) 136} [\href{https://arxiv.org/abs/1607.02521}{{\ttfamily 1607.02521}}].

\bibitem{Collins:1993kq}
J.C.~Collins, S.F.~Heppelmann and G.A.~Ladinsky, \emph{{Measuring transversity densities in singly polarized hadron hadron and lepton - hadron collisions}}, \href{https://doi.org/10.1016/0550-3213(94)90078-7}{\emph{Nucl. Phys. B} {\bfseries 420} (1994) 565} [\href{https://arxiv.org/abs/hep-ph/9305309}{{\ttfamily hep-ph/9305309}}].

\bibitem{Bianconi:1999cd}
A.~Bianconi, S.~Boffi, R.~Jakob and M.~Radici, \emph{{Two hadron interference fragmentation functions. Part 1. General framework}}, \href{https://doi.org/10.1103/PhysRevD.62.034008}{\emph{Phys. Rev. D} {\bfseries 62} (2000) 034008} [\href{https://arxiv.org/abs/hep-ph/9907475}{{\ttfamily hep-ph/9907475}}].

\bibitem{Collins:1992kk}
J.C.~Collins, \emph{{Fragmentation of transversely polarized quarks probed in transverse momentum distributions}}, \href{https://doi.org/10.1016/0550-3213(93)90262-N}{\emph{Nucl. Phys. B} {\bfseries 396} (1993) 161} [\href{https://arxiv.org/abs/hep-ph/9208213}{{\ttfamily hep-ph/9208213}}].

\bibitem{Ju:2022wia}
W.-L.~Ju and M.~Sch\"onherr, \emph{{Projected transverse momentum resummation in top-antitop pair production at LHC}}, \href{https://doi.org/10.1007/JHEP02(2023)075}{\emph{JHEP} {\bfseries 02} (2023) 075} [\href{https://arxiv.org/abs/2210.09272}{{\ttfamily 2210.09272}}].

\bibitem{Catani:2017tuc}
S.~Catani, M.~Grazzini and H.~Sargsyan, \emph{{Azimuthal asymmetries in QCD hard scattering: infrared safe but divergent}}, \href{https://doi.org/10.1007/JHEP06(2017)017}{\emph{JHEP} {\bfseries 06} (2017) 017} [\href{https://arxiv.org/abs/1703.08468}{{\ttfamily 1703.08468}}].

\bibitem{Konishi:1979cb}
K.~Konishi, A.~Ukawa and G.~Veneziano, \emph{{Jet Calculus: A Simple Algorithm for Resolving QCD Jets}}, \href{https://doi.org/10.1016/0550-3213(79)90053-1}{\emph{Nucl. Phys. B} {\bfseries 157} (1979) 45}.

\bibitem{Sukhatme:1980vs}
U.P.~Sukhatme and K.E.~Lassila, \emph{{$Q^2$ Evolution of Multi - Hadron Fragmentation Functions}}, \href{https://doi.org/10.1103/PhysRevD.22.1184}{\emph{Phys. Rev. D} {\bfseries 22} (1980) 1184}.

\bibitem{deFlorian:2003cg}
D.~de~Florian and L.~Vanni, \emph{{Two hadron production in e+ e- annihilation to next-to-leading order accuracy}}, \href{https://doi.org/10.1016/j.physletb.2003.10.047}{\emph{Phys. Lett. B} {\bfseries 578} (2004) 139} [\href{https://arxiv.org/abs/hep-ph/0310196}{{\ttfamily hep-ph/0310196}}].

\bibitem{Chen:2022pdu}
H.~Chen, M.~Jaarsma, Y.~Li, I.~Moult, W.J.~Waalewijn and H.X.~Zhu, \emph{{Multi-collinear splitting kernels for track function evolution}}, \href{https://doi.org/10.1007/JHEP07(2023)185}{\emph{JHEP} {\bfseries 07} (2023) 185} [\href{https://arxiv.org/abs/2210.10058}{{\ttfamily 2210.10058}}].

\bibitem{Chen:2022muj}
H.~Chen, M.~Jaarsma, Y.~Li, I.~Moult, W.J.~Waalewijn and H.X.~Zhu, \emph{{Collinear parton dynamics beyond Dokshitzer-Gribov-Lipatov-Altarelli-Parisi framework}}, \href{https://doi.org/10.1103/PhysRevD.111.076021}{\emph{Phys. Rev. D} {\bfseries 111} (2025) 076021} [\href{https://arxiv.org/abs/2210.10061}{{\ttfamily 2210.10061}}].

\bibitem{Majumder:2004wh}
A.~Majumder and X.-N.~Wang, \emph{{The Dihadron fragmentation function and its evolution}}, \href{https://doi.org/10.1103/PhysRevD.70.014007}{\emph{Phys. Rev. D} {\bfseries 70} (2004) 014007} [\href{https://arxiv.org/abs/hep-ph/0402245}{{\ttfamily hep-ph/0402245}}].

\bibitem{Majumder:2004br}
A.~Majumder and X.-N.~Wang, \emph{{Evolution of the parton dihadron fragmentation functions}}, \href{https://doi.org/10.1103/PhysRevD.72.034007}{\emph{Phys. Rev. D} {\bfseries 72} (2005) 034007} [\href{https://arxiv.org/abs/hep-ph/0411174}{{\ttfamily hep-ph/0411174}}].

\bibitem{Ceccopieri:2007ip}
F.A.~Ceccopieri, M.~Radici and A.~Bacchetta, \emph{{Evolution equations for extended dihadron fragmentation functions}}, \href{https://doi.org/10.1016/j.physletb.2007.04.065}{\emph{Phys. Lett. B} {\bfseries 650} (2007) 81} [\href{https://arxiv.org/abs/hep-ph/0703265}{{\ttfamily hep-ph/0703265}}].

\bibitem{Bacchetta:2008wb}
A.~Bacchetta, F.A.~Ceccopieri, A.~Mukherjee and M.~Radici, \emph{{Asymmetries involving dihadron fragmentation functions: from DIS to e+e- annihilation}}, \href{https://doi.org/10.1103/PhysRevD.79.034029}{\emph{Phys. Rev. D} {\bfseries 79} (2009) 034029} [\href{https://arxiv.org/abs/0812.0611}{{\ttfamily 0812.0611}}].

\bibitem{Bacchetta:2011ip}
A.~Bacchetta, A.~Courtoy and M.~Radici, \emph{{First glances at the transversity parton distribution through dihadron fragmentation functions}}, \href{https://doi.org/10.1103/PhysRevLett.107.012001}{\emph{Phys. Rev. Lett.} {\bfseries 107} (2011) 012001} [\href{https://arxiv.org/abs/1104.3855}{{\ttfamily 1104.3855}}].

\bibitem{Courtoy:2012ry}
A.~Courtoy, A.~Bacchetta, M.~Radici and A.~Bianconi, \emph{{First extraction of Interference Fragmentation Functions from $e^+e^-$ data}}, \href{https://doi.org/10.1103/PhysRevD.85.114023}{\emph{Phys. Rev. D} {\bfseries 85} (2012) 114023} [\href{https://arxiv.org/abs/1202.0323}{{\ttfamily 1202.0323}}].

\bibitem{Bacchetta:2012ty}
A.~Bacchetta, A.~Courtoy and M.~Radici, \emph{{First extraction of valence transversities in a collinear framework}}, \href{https://doi.org/10.1007/JHEP03(2013)119}{\emph{JHEP} {\bfseries 03} (2013) 119} [\href{https://arxiv.org/abs/1212.3568}{{\ttfamily 1212.3568}}].

\bibitem{Radici:2015mwa}
M.~Radici, A.~Courtoy, A.~Bacchetta and M.~Guagnelli, \emph{{Improved extraction of valence transversity distributions from inclusive dihadron production}}, \href{https://doi.org/10.1007/JHEP05(2015)123}{\emph{JHEP} {\bfseries 05} (2015) 123} [\href{https://arxiv.org/abs/1503.03495}{{\ttfamily 1503.03495}}].

\bibitem{Cocuzza:2023vqs}
{\scshape Jefferson Lab Angular Momentum (JAM)} collaboration, \emph{{First simultaneous global QCD analysis of dihadron fragmentation functions and transversity parton distribution functions}}, \href{https://doi.org/10.1103/PhysRevD.109.034024}{\emph{Phys. Rev. D} {\bfseries 109} (2024) 034024} [\href{https://arxiv.org/abs/2308.14857}{{\ttfamily 2308.14857}}].

\bibitem{Belle:2017rwm}
{\scshape Belle} collaboration, \emph{{Invariant-mass and fractional-energy dependence of inclusive production of di-hadrons in $e^+e^-$ annihilation at $\sqrt{s}=$ 10.58 GeV}}, \href{https://doi.org/10.1103/PhysRevD.96.032005}{\emph{Phys. Rev. D} {\bfseries 96} (2017) 032005} [\href{https://arxiv.org/abs/1706.08348}{{\ttfamily 1706.08348}}].

\bibitem{Wang:2019bvb}
B.~Wang, J.O.~Gonzalez-Hernandez, T.C.~Rogers and N.~Sato, \emph{{Large Transverse Momentum in Semi-Inclusive Deeply Inelastic Scattering Beyond Lowest Order}}, \href{https://doi.org/10.1103/PhysRevD.99.094029}{\emph{Phys. Rev. D} {\bfseries 99} (2019) 094029} [\href{https://arxiv.org/abs/1903.01529}{{\ttfamily 1903.01529}}].

\bibitem{Daleo:2004pn}
A.~Daleo, D.~de~Florian and R.~Sassot, \emph{{O(alpha**2(s)) QCD corrections to the electroproduction of hadrons with high transverse momentum}}, \href{https://doi.org/10.1103/PhysRevD.71.034013}{\emph{Phys. Rev. D} {\bfseries 71} (2005) 034013} [\href{https://arxiv.org/abs/hep-ph/0411212}{{\ttfamily hep-ph/0411212}}].

\bibitem{Gonsalves:1989ar}
R.J.~Gonsalves, J.~Pawlowski and C.-F.~Wai, \emph{{{QCD} Radiative Corrections to Electroweak Boson Production at Large Transverse Momentum in Hadron Collisions}}, \href{https://doi.org/10.1103/PhysRevD.40.2245}{\emph{Phys. Rev. D} {\bfseries 40} (1989) 2245}.

\bibitem{Ellis:1981nt}
R.K.~Ellis, G.~Martinelli and R.~Petronzio, \emph{{Second Order Corrections to the {Drell-Yan} Process at Large Transverse Momentum}}, \href{https://doi.org/10.1016/0370-2693(81)90851-0}{\emph{Phys. Lett. B} {\bfseries 104} (1981) 45}.

\bibitem{Ellis:1981hk}
R.K.~Ellis, G.~Martinelli and R.~Petronzio, \emph{{Lepton Pair Production at Large Transverse Momentum in Second Order QCD}}, \href{https://doi.org/10.1016/0550-3213(83)90188-8}{\emph{Nucl. Phys. B} {\bfseries 211} (1983) 106}.

\bibitem{Ellis:1980wv}
R.K.~Ellis, D.A.~Ross and A.E.~Terrano, \emph{{The Perturbative Calculation of Jet Structure in e+ e- Annihilation}}, \href{https://doi.org/10.1016/0550-3213(81)90165-6}{\emph{Nucl. Phys. B} {\bfseries 178} (1981) 421}.

\bibitem{Kublbeck:1990xc}
J.~Kublbeck, M.~Bohm and A.~Denner, \emph{{Feyn Arts: Computer Algebraic Generation of Feynman Graphs and Amplitudes}}, \href{https://doi.org/10.1016/0010-4655(90)90001-H}{\emph{Comput. Phys. Commun.} {\bfseries 60} (1990) 165}.

\bibitem{Shtabovenko:2023idz}
V.~Shtabovenko, R.~Mertig and F.~Orellana, \emph{{FeynCalc 10: Do multiloop integrals dream of computer codes?}}, \href{https://doi.org/10.1016/j.cpc.2024.109357}{\emph{Comput. Phys. Commun.} {\bfseries 306} (2025) 109357} [\href{https://arxiv.org/abs/2312.14089}{{\ttfamily 2312.14089}}].

\bibitem{Shtabovenko:2020gxv}
V.~Shtabovenko, R.~Mertig and F.~Orellana, \emph{{FeynCalc 9.3: New features and improvements}}, \href{https://doi.org/10.1016/j.cpc.2020.107478}{\emph{Comput. Phys. Commun.} {\bfseries 256} (2020) 107478} [\href{https://arxiv.org/abs/2001.04407}{{\ttfamily 2001.04407}}].

\bibitem{Shtabovenko:2016sxi}
V.~Shtabovenko, R.~Mertig and F.~Orellana, \emph{{New Developments in FeynCalc 9.0}}, \href{https://doi.org/10.1016/j.cpc.2016.06.008}{\emph{Comput. Phys. Commun.} {\bfseries 207} (2016) 432} [\href{https://arxiv.org/abs/1601.01167}{{\ttfamily 1601.01167}}].

\bibitem{Mertig:1990an}
R.~Mertig, M.~Bohm and A.~Denner, \emph{{FEYN CALC: Computer algebraic calculation of Feynman amplitudes}}, \href{https://doi.org/10.1016/0010-4655(91)90130-D}{\emph{Comput. Phys. Commun.} {\bfseries 64} (1991) 345}.

\bibitem{Shtabovenko:2016whf}
V.~Shtabovenko, \emph{{FeynHelpers: Connecting FeynCalc to FIRE and Package-X}}, \href{https://doi.org/10.1016/j.cpc.2017.04.014}{\emph{Comput. Phys. Commun.} {\bfseries 218} (2017) 48} [\href{https://arxiv.org/abs/1611.06793}{{\ttfamily 1611.06793}}].

\bibitem{Tkachov:1981wb}
F.V.~Tkachov, \emph{{A theorem on analytical calculability of 4-loop renormalization group functions}}, \href{https://doi.org/10.1016/0370-2693(81)90288-4}{\emph{Phys. Lett. B} {\bfseries 100} (1981) 65}.

\bibitem{Chetyrkin:1981qh}
K.G.~Chetyrkin and F.V.~Tkachov, \emph{{Integration by parts: The algorithm to calculate $\beta$-functions in 4 loops}}, \href{https://doi.org/10.1016/0550-3213(81)90199-1}{\emph{Nucl. Phys. B} {\bfseries 192} (1981) 159}.

\bibitem{Laporta:2000dsw}
S.~Laporta, \emph{{High-precision calculation of multiloop Feynman integrals by difference equations}}, \href{https://doi.org/10.1142/S0217751X00002159}{\emph{Int. J. Mod. Phys. A} {\bfseries 15} (2000) 5087} [\href{https://arxiv.org/abs/hep-ph/0102033}{{\ttfamily hep-ph/0102033}}].

\bibitem{Kotikov:1990kg}
A.V.~Kotikov, \emph{{Differential equations method: New technique for massive Feynman diagrams calculation}}, \href{https://doi.org/10.1016/0370-2693(91)90413-K}{\emph{Phys. Lett. B} {\bfseries 254} (1991) 158}.

\bibitem{Remiddi:1997ny}
E.~Remiddi, \emph{{Differential equations for Feynman graph amplitudes}}, \href{https://doi.org/10.1007/BF03185566}{\emph{Nuovo Cim. A} {\bfseries 110} (1997) 1435} [\href{https://arxiv.org/abs/hep-th/9711188}{{\ttfamily hep-th/9711188}}].

\bibitem{Gehrmann:1999as}
T.~Gehrmann and E.~Remiddi, \emph{{Differential equations for two-loop four-point functions}}, \href{https://doi.org/10.1016/S0550-3213(00)00223-6}{\emph{Nucl. Phys. B} {\bfseries 580} (2000) 485} [\href{https://arxiv.org/abs/hep-ph/9912329}{{\ttfamily hep-ph/9912329}}].

\bibitem{Argeri:2007up}
M.~Argeri and P.~Mastrolia, \emph{{Feynman Diagrams and Differential Equations}}, \href{https://doi.org/10.1142/S0217751X07037147}{\emph{Int. J. Mod. Phys. A} {\bfseries 22} (2007) 4375} [\href{https://arxiv.org/abs/0707.4037}{{\ttfamily 0707.4037}}].

\bibitem{Henn:2013pwa}
J.M.~Henn, \emph{{Multiloop integrals in dimensional regularization made simple}}, \href{https://doi.org/10.1103/PhysRevLett.110.251601}{\emph{Phys. Rev. Lett.} {\bfseries 110} (2013) 251601} [\href{https://arxiv.org/abs/1304.1806}{{\ttfamily 1304.1806}}].

\bibitem{Collins:1980ui}
J.C.~Collins, \emph{{INTRINSIC TRANSVERSE MOMENTUM. 1. NONGAUGE THEORIES}}, \href{https://doi.org/10.1103/PhysRevD.21.2962}{\emph{Phys. Rev. D} {\bfseries 21} (1980) 2962}.

\bibitem{Collins:1985ue}
J.C.~Collins, D.E.~Soper and G.F.~Sterman, \emph{{Factorization for Short Distance Hadron - Hadron Scattering}}, \href{https://doi.org/10.1016/0550-3213(85)90565-6}{\emph{Nucl. Phys. B} {\bfseries 261} (1985) 104}.

\bibitem{Bodwin:1984hc}
G.T.~Bodwin, \emph{{Factorization of the Drell-Yan Cross-Section in Perturbation Theory}}, \href{https://doi.org/10.1103/PhysRevD.34.3932}{\emph{Phys. Rev. D} {\bfseries 31} (1985) 2616}.

\bibitem{Dixon:2018qgp}
L.J.~Dixon, M.-X.~Luo, V.~Shtabovenko, T.-Z.~Yang and H.X.~Zhu, \emph{{Analytical Computation of Energy-Energy Correlation at Next-to-Leading Order in QCD}}, \href{https://doi.org/10.1103/PhysRevLett.120.102001}{\emph{Phys. Rev. Lett.} {\bfseries 120} (2018) 102001} [\href{https://arxiv.org/abs/1801.03219}{{\ttfamily 1801.03219}}].

\bibitem{Anastasiou:2002yz}
C.~Anastasiou and K.~Melnikov, \emph{{Higgs boson production at hadron colliders in NNLO QCD}}, \href{https://doi.org/10.1016/S0550-3213(02)00837-4}{\emph{Nucl. Phys. B} {\bfseries 646} (2002) 220} [\href{https://arxiv.org/abs/hep-ph/0207004}{{\ttfamily hep-ph/0207004}}].

\bibitem{Anastasiou:2003yy}
C.~Anastasiou, L.J.~Dixon, K.~Melnikov and F.~Petriello, \emph{{Dilepton rapidity distribution in the Drell-Yan process at NNLO in QCD}}, \href{https://doi.org/10.1103/PhysRevLett.91.182002}{\emph{Phys. Rev. Lett.} {\bfseries 91} (2003) 182002} [\href{https://arxiv.org/abs/hep-ph/0306192}{{\ttfamily hep-ph/0306192}}].

\bibitem{Feng:2012iq}
F.~Feng, \emph{{$\tt{Apart}$: A Generalized Mathematica Apart Function}}, \href{https://doi.org/10.1016/j.cpc.2012.03.025}{\emph{Comput. Phys. Commun.} {\bfseries 183} (2012) 2158} [\href{https://arxiv.org/abs/1204.2314}{{\ttfamily 1204.2314}}].

\bibitem{Lee:2012cn}
R.N.~Lee, \emph{{Presenting LiteRed: a tool for the Loop InTEgrals REDuction}},  \href{https://arxiv.org/abs/1212.2685}{{\ttfamily 1212.2685}}.

\bibitem{Lee:2013mka}
R.N.~Lee, \emph{{LiteRed 1.4: a powerful tool for reduction of multiloop integrals}}, \href{https://doi.org/10.1088/1742-6596/523/1/012059}{\emph{J. Phys. Conf. Ser.} {\bfseries 523} (2014) 012059} [\href{https://arxiv.org/abs/1310.1145}{{\ttfamily 1310.1145}}].

\bibitem{Chavez:2012kn}
F.~Chavez and C.~Duhr, \emph{{Three-mass triangle integrals and single-valued polylogarithms}}, \href{https://doi.org/10.1007/JHEP11(2012)114}{\emph{JHEP} {\bfseries 11} (2012) 114} [\href{https://arxiv.org/abs/1209.2722}{{\ttfamily 1209.2722}}].

\bibitem{Gehrmann:2014bfa}
T.~Gehrmann, A.~von Manteuffel, L.~Tancredi and E.~Weihs, \emph{{The two-loop master integrals for $q\overline{q} \to VV$}}, \href{https://doi.org/10.1007/JHEP06(2014)032}{\emph{JHEP} {\bfseries 06} (2014) 032} [\href{https://arxiv.org/abs/1404.4853}{{\ttfamily 1404.4853}}].

\bibitem{Caola:2014lpa}
F.~Caola, J.M.~Henn, K.~Melnikov and V.A.~Smirnov, \emph{{Non-planar master integrals for the production of two off-shell vector bosons in collisions of massless partons}}, \href{https://doi.org/10.1007/JHEP09(2014)043}{\emph{JHEP} {\bfseries 09} (2014) 043} [\href{https://arxiv.org/abs/1404.5590}{{\ttfamily 1404.5590}}].

\bibitem{Henn:2014lfa}
J.M.~Henn, K.~Melnikov and V.A.~Smirnov, \emph{{Two-loop planar master integrals for the production of off-shell vector bosons in hadron collisions}}, \href{https://doi.org/10.1007/JHEP05(2014)090}{\emph{JHEP} {\bfseries 05} (2014) 090} [\href{https://arxiv.org/abs/1402.7078}{{\ttfamily 1402.7078}}].

\bibitem{Gehrmann:2015ora}
T.~Gehrmann, A.~von Manteuffel and L.~Tancredi, \emph{{The two-loop helicity amplitudes for $ q\overline{q}^{\prime}\to {V}_1{V}_2\to 4 $ leptons}}, \href{https://doi.org/10.1007/JHEP09(2015)128}{\emph{JHEP} {\bfseries 09} (2015) 128} [\href{https://arxiv.org/abs/1503.04812}{{\ttfamily 1503.04812}}].

\bibitem{vanNeerven:1985xr}
W.L.~van Neerven, \emph{{Dimensional Regularization of Mass and Infrared Singularities in Two Loop On-shell Vertex Functions}}, \href{https://doi.org/10.1016/0550-3213(86)90165-3}{\emph{Nucl. Phys. B} {\bfseries 268} (1986) 453}.

\bibitem{Beenakker:1988bq}
W.~Beenakker, H.~Kuijf, W.L.~van Neerven and J.~Smith, \emph{{QCD Corrections to Heavy Quark Production in p anti-p Collisions}}, \href{https://doi.org/10.1103/PhysRevD.40.54}{\emph{Phys. Rev. D} {\bfseries 40} (1989) 54}.

\bibitem{Somogyi:2011ir}
G.~Somogyi, \emph{{Angular integrals in d dimensions}}, \href{https://doi.org/10.1063/1.3615515}{\emph{J. Math. Phys.} {\bfseries 52} (2011) 083501} [\href{https://arxiv.org/abs/1101.3557}{{\ttfamily 1101.3557}}].

\bibitem{Devoto:1984wu}
A.~Devoto, D.W.~Duke, J.D.~Kimel and G.A.~Sowell, \emph{{Analytic Calculation of the Fourth Order Quantum Chromodynamic Contribution to the Nonsinglet Quark Longitudinal Structure Function}}, \href{https://doi.org/10.1103/PhysRevD.30.541}{\emph{Phys. Rev. D} {\bfseries 30} (1984) 541}.

\bibitem{Meyer:2016slj}
C.~Meyer, \emph{{Transforming differential equations of multi-loop Feynman integrals into canonical form}}, \href{https://doi.org/10.1007/JHEP04(2017)006}{\emph{JHEP} {\bfseries 04} (2017) 006} [\href{https://arxiv.org/abs/1611.01087}{{\ttfamily 1611.01087}}].

\bibitem{Meyer:2017joq}
C.~Meyer, \emph{{Algorithmic transformation of multi-loop master integrals to a canonical basis with CANONICA}}, \href{https://doi.org/10.1016/j.cpc.2017.09.014}{\emph{Comput. Phys. Commun.} {\bfseries 222} (2018) 295} [\href{https://arxiv.org/abs/1705.06252}{{\ttfamily 1705.06252}}].

\bibitem{Papadopoulos:2015jft}
C.G.~Papadopoulos, D.~Tommasini and C.~Wever, \emph{{The Pentabox Master Integrals with the Simplified Differential Equations approach}}, \href{https://doi.org/10.1007/JHEP04(2016)078}{\emph{JHEP} {\bfseries 04} (2016) 078} [\href{https://arxiv.org/abs/1511.09404}{{\ttfamily 1511.09404}}].

\bibitem{Canko:2020gqp}
D.D.~Canko and N.~Syrrakos, \emph{{Resummation methods for Master Integrals}}, \href{https://doi.org/10.1007/JHEP02(2021)080}{\emph{JHEP} {\bfseries 02} (2021) 080} [\href{https://arxiv.org/abs/2010.06947}{{\ttfamily 2010.06947}}].

\bibitem{Chen:1977oja}
K.-T.~Chen, \emph{{Iterated path integrals}}, \href{https://doi.org/10.1090/S0002-9904-1977-14320-6}{\emph{Bull. Am. Math. Soc.} {\bfseries 83} (1977) 831}.

\bibitem{Goncharov:1998kja}
A.B.~Goncharov, \emph{{Multiple polylogarithms, cyclotomy and modular complexes}}, \href{https://doi.org/10.4310/MRL.1998.v5.n4.a7}{\emph{Math. Res. Lett.} {\bfseries 5} (1998) 497} [\href{https://arxiv.org/abs/1105.2076}{{\ttfamily 1105.2076}}].

\bibitem{Goncharov:2001iea}
A.B.~Goncharov, \emph{{Multiple polylogarithms and mixed Tate motives}},  \href{https://arxiv.org/abs/math/0103059}{{\ttfamily math/0103059}}.

\bibitem{Duhr:2019tlz}
C.~Duhr and F.~Dulat, \emph{{PolyLogTools \textemdash{} polylogs for the masses}}, \href{https://doi.org/10.1007/JHEP08(2019)135}{\emph{JHEP} {\bfseries 08} (2019) 135} [\href{https://arxiv.org/abs/1904.07279}{{\ttfamily 1904.07279}}].

\bibitem{Bauer:2000cp}
C.W.~Bauer, A.~Frink and R.~Kreckel, \emph{{Introduction to the GiNaC framework for symbolic computation within the C++ programming language}}, \href{https://doi.org/10.1006/jsco.2001.0494}{\emph{J. Symb. Comput.} {\bfseries 33} (2002) 1} [\href{https://arxiv.org/abs/cs/0004015}{{\ttfamily cs/0004015}}].

\bibitem{Liu:2022chg}
X.~Liu and Y.-Q.~Ma, \emph{{AMFlow: A Mathematica package for Feynman integrals computation via auxiliary mass flow}}, \href{https://doi.org/10.1016/j.cpc.2022.108565}{\emph{Comput. Phys. Commun.} {\bfseries 283} (2023) 108565} [\href{https://arxiv.org/abs/2201.11669}{{\ttfamily 2201.11669}}].

\bibitem{Liu:2017jxz}
X.~Liu, Y.-Q.~Ma and C.-Y.~Wang, \emph{{A Systematic and Efficient Method to Compute Multi-loop Master Integrals}}, \href{https://doi.org/10.1016/j.physletb.2018.02.026}{\emph{Phys. Lett. B} {\bfseries 779} (2018) 353} [\href{https://arxiv.org/abs/1711.09572}{{\ttfamily 1711.09572}}].

\bibitem{Peraro:2019svx}
T.~Peraro, \emph{{$\text{FiniteFlow}$: multivariate functional reconstruction using finite fields and dataflow graphs}}, \href{https://doi.org/10.1007/JHEP07(2019)031}{\emph{JHEP} {\bfseries 07} (2019) 031} [\href{https://arxiv.org/abs/1905.08019}{{\ttfamily 1905.08019}}].

\bibitem{PSLQ1}
H.R.P.~Ferguson and D.H.~Bailey, \emph{{A Polynomial Time, Numerically Stable Integer Relation Algorithm}},  RNR Technical Report RNR-91-032 (1992).

\bibitem{PSLQ2}
P.~Bertok, ``\emph{PSLQ Integer Relation Algorithm Implementation}.'' \url{http://library.wolfram.com/infocenter/MathSource/4263/}, 2004.

\bibitem{DeFazio:1999ptt}
F.~De~Fazio and M.~Neubert, \emph{{B ---\ensuremath{>} X(u) lepton anti-neutrino lepton decay distributions to order alpha(s)}}, \href{https://doi.org/10.1088/1126-6708/1999/06/017}{\emph{JHEP} {\bfseries 06} (1999) 017} [\href{https://arxiv.org/abs/hep-ph/9905351}{{\ttfamily hep-ph/9905351}}].

\bibitem{Bosch:2004th}
S.W.~Bosch, B.O.~Lange, M.~Neubert and G.~Paz, \emph{{Factorization and shape function effects in inclusive B meson decays}}, \href{https://doi.org/10.1016/j.nuclphysb.2004.07.041}{\emph{Nucl. Phys. B} {\bfseries 699} (2004) 335} [\href{https://arxiv.org/abs/hep-ph/0402094}{{\ttfamily hep-ph/0402094}}].

\bibitem{Passarino:1978jh}
G.~Passarino and M.J.G.~Veltman, \emph{{One Loop Corrections for e+ e- Annihilation Into mu+ mu- in the Weinberg Model}}, \href{https://doi.org/10.1016/0550-3213(79)90234-7}{\emph{Nucl. Phys. B} {\bfseries 160} (1979) 151}.

\bibitem{Patel:2015tea}
H.H.~Patel, \emph{{Package-X: A Mathematica package for the analytic calculation of one-loop integrals}}, \href{https://doi.org/10.1016/j.cpc.2015.08.017}{\emph{Comput. Phys. Commun.} {\bfseries 197} (2015) 276} [\href{https://arxiv.org/abs/1503.01469}{{\ttfamily 1503.01469}}].

\bibitem{Bardeen:1978yd}
W.A.~Bardeen, A.J.~Buras, D.W.~Duke and T.~Muta, \emph{{Deep Inelastic Scattering Beyond the Leading Order in Asymptotically Free Gauge Theories}}, \href{https://doi.org/10.1103/PhysRevD.18.3998}{\emph{Phys. Rev. D} {\bfseries 18} (1978) 3998}.

\bibitem{Altarelli:1977zs}
G.~Altarelli and G.~Parisi, \emph{{Asymptotic Freedom in Parton Language}}, \href{https://doi.org/10.1016/0550-3213(77)90384-4}{\emph{Nucl. Phys. B} {\bfseries 126} (1977) 298}.

\bibitem{Heller:2021qkz}
M.~Heller and A.~von Manteuffel, \emph{{MultivariateApart: Generalized partial fractions}}, \href{https://doi.org/10.1016/j.cpc.2021.108174}{\emph{Comput. Phys. Commun.} {\bfseries 271} (2022) 108174} [\href{https://arxiv.org/abs/2101.08283}{{\ttfamily 2101.08283}}].

\bibitem{Boehm:2020ijp}
J.~Boehm, M.~Wittmann, Z.~Wu, Y.~Xu and Y.~Zhang, \emph{{IBP reduction coefficients made simple}}, \href{https://doi.org/10.1007/JHEP12(2020)054}{\emph{JHEP} {\bfseries 12} (2020) 054} [\href{https://arxiv.org/abs/2008.13194}{{\ttfamily 2008.13194}}].

\end{thebibliography}\endgroup

\end{document}